\documentclass[hidelinks,onefignum,onetabnum]{siamart250211}
\usepackage{graphicx}
\usepackage{booktabs}
\usepackage{multirow}
\usepackage{makecell}
\usepackage{subfigure}
\usepackage{amsmath}
\allowdisplaybreaks[1]
\usepackage{xr}
\usepackage{lipsum}
\usepackage{amsfonts}
\usepackage{graphicx}
\usepackage{epstopdf}
\usepackage{algorithmic}
\ifpdf
  \DeclareGraphicsExtensions{.eps,.pdf,.png,.jpg}
\else
  \DeclareGraphicsExtensions{.eps}
\fi

\newsiamremark{remark}{Remark}
\newsiamremark{hypothesis}{Hypothesis}
\crefname{hypothesis}{Hypothesis}{Hypotheses}
\newsiamthm{claim}{Claim}
\newsiamremark{fact}{Fact}
\crefname{fact}{Fact}{Facts}

\headers{Tumor-Immune Bistability and Noise-Induced Escape}{M. Tan, S. Chen, C. Wei, and D. Zhou}

\title{Bistability and Noise-Induced Evasion in Tumor-Immune Dynamics with Antigen Accumulation and Immune Escape\thanks{Submitted to the editors DATE.
\funding{This work of the fourth author was supported by the National Natural Science Foundation of China (Grant No. 12471475 and 11971405), the Fundamental Research Funds for the Central Universities in China (Grant No. 20720230023 and 20720240151)}}}

\author{
Mengfan Tan\thanks{School of Mathematical Sciences, Xiamen University, Xiamen,  361005, People's Republic of China, and National Institute for Data Science in Health and Medicine, Xiamen University, Xiamen,  361005, People's Republic of China.   (\email{tanmengfan1205@163.com}). }
\and 
Shaoqing Chen\thanks{School of Mathematical Sciences, Xiamen University, Xiamen,  361005, People's Republic of China. (\email{chenshaoqing@stu.xmu.edu.cn}).}
\and 
Chunjin Wei\thanks{School of Sciences, Jimei University, Xiamen,  361021, People's Republic of China.   (\email{chunjinwei92@163.com}).}  
\and
Da Zhou\thanks{Corresponding author. School of Mathematical Sciences, Xiamen University, Xiamen,  361005, People's Republic of China, and National Institute for Data Science in Health and Medicine, Xiamen University, Xiamen,  361005, People's Republic of China. (\email{zhouda@xmu.edu.cn}).}  
}

\usepackage{amsopn}

\ifpdf
\hypersetup{
  pdftitle={Tumor-Immune with Immune Escape},
  pdfauthor={M. Tan, S. Chen, C. Wei, and D. Zhou}
}
\fi

\begin{document}

\maketitle

\begin{abstract}
Tumor-immune interactions are shaped by both antigenic heterogeneity and stochastic perturbations in the tumor microenvironment, yet the mathematical mechanisms underlying immune phase transitions remain poorly understood. We propose a four-compartment dynamical model that incorporates antigen accumulation and immune escape mutations. Bifurcation analysis reveals bistability between immune surveillance and immune escape states, providing a mechanistic explanation for heterogeneous immune outcomes during tumor progression. In the multistable regime, the stable manifold of a saddle point partitions the state space into distinct basins of attraction, determining the long-term fate of the system. We further analyze how stochastic fluctuations in the tumor microenvironment perturb these separatrices, potentially triggering irreversible state transitions. By characterizing the critical noise intensity and estimating the tipping time, we establish a mathematical framework for assessing noise-induced transitions. The model further predicts that increasing tumor cell death can improve system resilience to stochastic perturbations, whereas stronger immune pressure may facilitate immune escape—highlighting the nonlinear and non-monotonic nature of tumor-immune dynamics.
\end{abstract}



\section{Introduction}
Cancer remains one of the leading causes of mortality worldwide, with an estimated 20 million new cases and 9.7 million deaths in 2022 alone \cite{bray2024global}. Despite significant advances in cancer research, many aspects of tumor progression remain poorly understood. The immune system, as the body's natural defense, plays a critical role in recognizing and eliminating malignant cells \cite{depillis2014modeling}. A central question in cancer biology is how the immune system dynamically responds to tumor development and how tumors, in turn, evade immune control \cite{de2006mixed}.

The theory of cancer immunoediting provides a foundational framework for understanding tumor-immune coevolution. This process is typically characterized by three dynamic phases: elimination, equilibrium, and escape \cite{dunn2004three,schreiber2011cancer}. During elimination, immune cells detect and destroy antigen-expressing tumor cells \cite{gonzalez2018roles}. However, prolonged immune pressure can drive the selection of immune-evasive phenotypes, leading to tumor escape. This escape is facilitated by various mechanisms, including reduced neoantigen visibility due to low tumor mutational burden \cite{hellmann2018tumor,ready2019first}, impaired antigen presentation through HLA-I loss or silencing \cite{aptsiauri2018transition,anderson2021hla}, T cell exhaustion driven by PD-L1 overexpression \cite{tang2020mechanisms,iwai2017cancer}, and the formation of an immunosuppressive microenvironment \cite{beatty2015immune,wherry2015molecular}. These adaptations highlight the complexity and plasticity of tumor-immune interactions and underscore the need for mathematical models to explore their dynamical behavior.

Immune escape remains one of the central challenges in cancer treatment. Immunotherapies—aimed at enhancing antitumor immune responses—have emerged as a transformative strategy and are now regarded as the fourth pillar of cancer treatment alongside surgery, radiotherapy, and chemotherapy \cite{schmidt2017benefits,wraith2017future,zhang2021neoantigen}. Despite their clinical success in certain cancers, immunotherapies exhibit highly variable efficacy across patients \cite{riaz2017tumor,alsaab2017pd,deng2024patterns,grasso2018genetic,sade2017resistance}. Intratumor heterogeneity (ITH), driven by genetic instability and selective immune pressure, is a major contributor to this variability \cite{su2025single,aguade2020tumour,zuo2025intratumor}. Recent studies have suggested that negative frequency-dependent selection (NFDS) accelerates phenotypic diversification and promotes adaptive resistance to immune checkpoint blockade \cite{chen2024frequency}. Additionally, the tumor microenvironment can suppress immune activity through non-genetic factors such as inflammation, nutrient deprivation, or local immunosuppression \cite{sinha2025tumor}. While these phenomena are well documented experimentally, their dynamic and mechanistic implications remain poorly understood. In particular, the interplay between antigenic evolution, immune pressure, and stochastic perturbations in the tumor microenvironment has not been systematically quantified.

Mathematical modeling provides a powerful framework for exploring tumor immune dynamics, allowing researchers to investigate nonlinear behavior, multistability, and treatment response under perturbations. Classical models often abstract tumor-immune interactions as predator-prey systems and employ Lotka-Volterra-type equations to study oscillatory dynamics or tumor elimination \cite{wang2024bistability,dehingia2023modelling,li2021complex}. Other efforts have focused on capturing the multistage nature of cancer immunoediting. For example, Alvarez et al. \cite{alvarez2019nonlinear} incorporated multiple immunogenic phenotypes into a nonlinear ODE model to analyze phase transitions between elimination and escape. Delay differential equations have been used to characterize immune regulation \cite{zhang2023bifurcation}, and hybrid cellular automata have captured transient dormancy states \cite{lopez2017dynamics}. More recent models account for tumor heterogeneity, such as the use of integro-differential equations to model phenotypic distributions \cite{kaid2023phenotype}, or agent-based frameworks to study how antigen diversity affects immune control \cite{leschiera2022mathematical}.

While these models have deepened our understanding of tumor-immune interactions, most treat antigenic variation and immune escape as static traits or deterministic inputs. In particular, the coupled dynamics of antigen accumulation and immune escape under stochastic microenvironment have not been fully addressed. A unified model that captures both evolutionary dynamics and stochastic processes is essential for understanding how immune surveillance transitions to immune escape.

These observations motivate the following key questions: \\
(1) How does the dynamic accumulation of antigenic mutations influence transitions in the cancer immunoediting process? \\
(2) What are the underlying mechanisms that lead to divergent immune outcomes such as surveillance or escape? \\
(3) How do stochastic perturbations in the tumor microenvironment trigger irreversible shifts toward immune escape?

To address these questions, we develop a minimal four-compartment model that captures the coupled dynamics of antigen accumulation and immune escape in a fluctuating microenvironment. This framework allows us to explore multistability, bifurcation structures, and noise-induced transitions. By quantifying the tipping thresholds and escape times, our model provides theoretical insights into the emergence of heterogeneous immune responses and the loss of tumor control.

The remainder of this paper is organized as follows. Section 2 provides a detailed description of deterministic and stochastic models, and their long-term dynamic behaviors are analyzed in section 3. Section 4 performs a bifurcation analysis of key parameters. Additionally, the separatrix dividing the two attraction regions is plotted using the stable manifold of the saddle point. Furthermore, we  characterize noise-induced transitions and estimate the critical noise intensity and average escape time. Section 5 concludes the paper with a summary and discussion.

\section{Mathematical model}
\begin{figure}[H]
    \centering
    \includegraphics[width=0.8\linewidth]{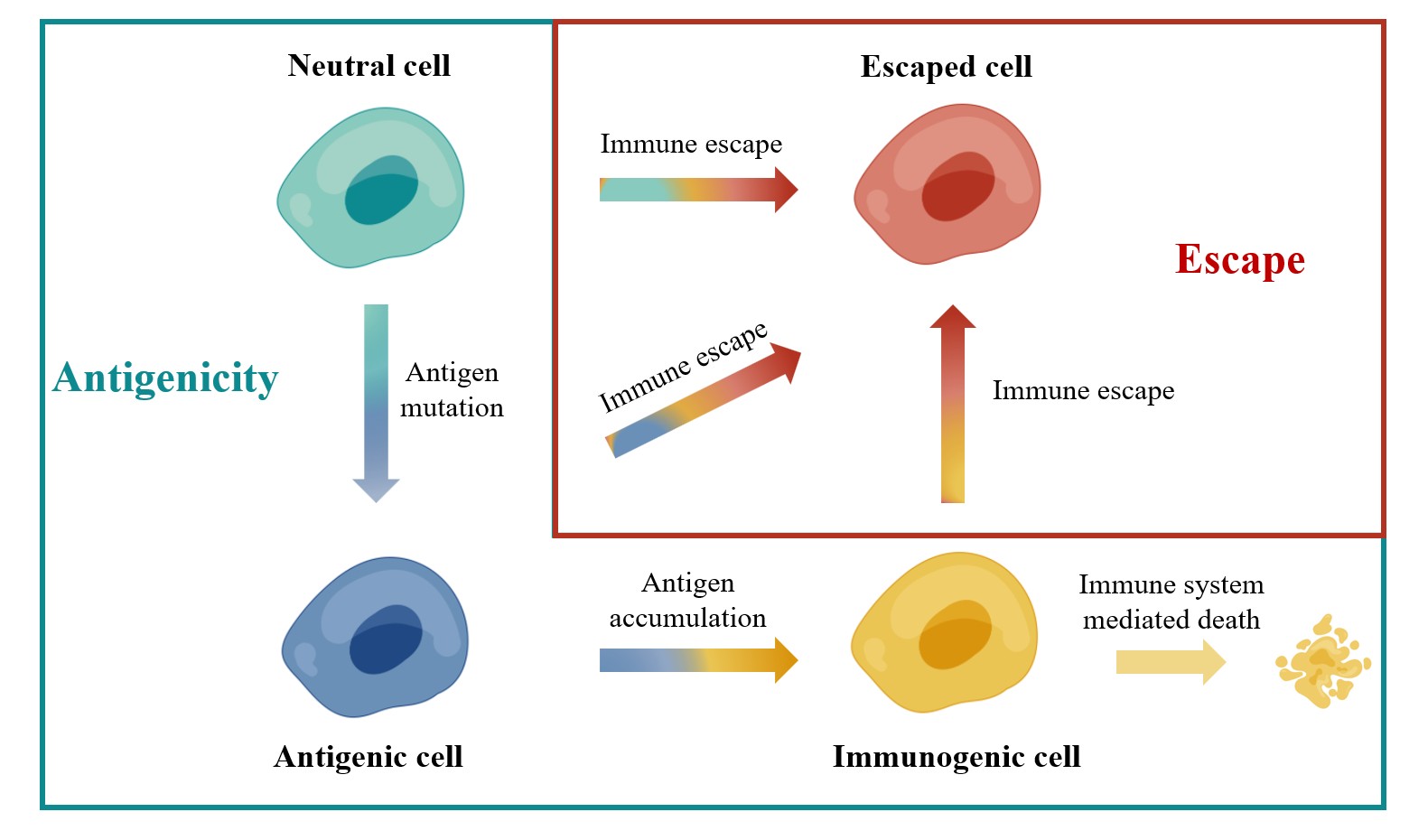}
    \caption{Schematic illustration of the local interactions among the four tumor cell subpopulations considered in the model.}
    \label{figsch}
\end{figure}

In modeling the heterogeneous growth of tumors, we consider antigen mutation, antigen accumulation, and immune escape mutations in tumor cells. Based on these processes, tumor cells are classified into four subpopulations: antigenically neutral cells ($\bar{N}$), antigenic cells ($\bar{A}$), immunogenic cells ($\bar{I}$), and immune-escaped cells ($\bar{E}$). The main transition relationships among these subpopulations are summarized in Fig. \ref{figsch}. To establish a tractable model, we make the following assumptions:

\textbf{(A1)} Factors such as angiogenesis, immune cells, and other cytokines are not considered; the focus is solely on the interplay among these subpopulations.

\textbf{(A2)} Antigenic mutation, antigen accumulation, and immune escape are assumed to occur continuously during tumor growth.

\textbf{(A3)} Each of these processes occurs with a fixed probability throughout tumor growth.

\textbf{(A4)} Immune cells are assumed to recognize and attack tumor cells carrying high-frequency antigen mutations, while cells with low-frequency antigens can evade immune detection.

Consequently, based on these assumptions, we formulate the following word equations for each compartment:

\begin{equation*} \begin{aligned} &(\text{rate of change of cell density})= \\ &(\text{proliferation}) \pm (\text{transformation between the other compartment}) - (\text{apoptosis}). \\ \end{aligned} \end{equation*}

\textbf{Proliferation.} Tumor cell growth in the absence of immune response is modeled by a logistic function, with intrinsic growth rates $r_N$, $r_A$, $r_I$, and $r_E$ for $\bar{N}$, $\bar{A}$, $\bar{I}$, and $\bar{E}$, respectively, and a common carrying capacity $K$. This formulation captures the competition among different tumor cell subpopulations for limited resources such as nutrients and space.

\textbf{Transformation.} In accordance with assumptions (A2) and (A3), linear terms describe the inter-compartmental transitions. Specifically, antigenically neutral cells mutate at rate $p$ to become antigenic cells; antigenic cells accumulate antigens at rate $s$ and convert into immunogenic cells, which can then be recognized by the immune system; moreover, antigenically neutral, antigenic, and immunogenic cells may all undergo immune escape mutations at rate $q$, giving rise to immune-escaped cells.

\textbf{Apoptosis.} Cell death arises from two sources: natural apoptosis and immune-mediated killing. Natural apoptosis is modeled as a linear removal term. Immune-mediated killing follows assumption (A4), acting specifically on immunogenic cells and described by the nonlinear saturation term $\frac{\bar{m}\bar{I}}{\bar{\theta}+\bar{I}}$. This term reflects that when the number of immunogenic cells is low, the immune response increases rapidly, while for large $\bar{I}$ the killing capacity approaches the saturation level $\bar{m}$ once a threshold is exceeded.

Based on the above biological processes, the model can be formulated as follows, with the biological meanings of the parameters in system~(\ref{model}) summarized in Table~\ref{tab1}:
\begin{small}
\begin{eqnarray}\label{model}
\left\{
\begin{aligned}
&\frac{\mathrm{d} \bar{N}}{\mathrm{d} t}=\underbrace{ r_N\bar{N}\left(1-\frac{\bar{N}+\bar{A}+\bar{I}+\bar{E}}{K}\right)}_{\text{proliferation}}-\underbrace{ d_N\bar{N}}_{\text{apoptosis}}-\underbrace{p\bar{N}}_{\substack{\text{antigenic} \\ \text{mutations}}}-\underbrace{q\bar{N}}_{\substack{\text{immune} \\ \text{escape}}},\\
&\frac{\mathrm{d} \bar{A}}{\mathrm{d}t}=\underbrace{ r_A\bar{A}\left(1-\frac{\bar{N}+\bar{A}+\bar{I}+\bar{E}}{K}\right)}_{\text{proliferation}}-\underbrace{ d_A\bar{A}}_{\text{apoptosis}}+\underbrace{p\bar{N}}_{\substack{\text{antigenic} \\ \text{mutations}}}-\underbrace{q\bar{A}}_{\substack{\text{immune} \\ \text{escape}}}-\underbrace{s\bar{A}}_{\substack{\text{antigen} \\ \text{accumulation}}},\\
&\frac{\mathrm{d} \bar{I}}{\mathrm{d} t}=\underbrace{ r_I\bar{I}\left(1-\frac{\bar{N}+\bar{A}+\bar{I}+\bar{E}}{K}\right)}_{\text{proliferation}}+\underbrace{s\bar{A}}_{\substack{\text{antigen} \\ \text{accumulation}}}-\underbrace{d_I\bar{I}}_{\text{apoptosis}}-\underbrace{q\bar{I}}_{\substack{\text{immune} \\ \text{escape}}}-\underbrace{\frac{\tilde{m}\bar{I}}{\tilde{\theta} + \bar{I}}}_{\substack{\text{immune} \\ \text{surveillance}}},\\
&\frac{\mathrm{d} \bar{E}}{\mathrm{d} t}=\underbrace{ r_E\bar{E}\left(1-\frac{\bar{N}+\bar{A}+\bar{I}+\bar{E}}{K}\right)}_{\text{proliferation}}-\underbrace{d_E\bar{E}}_{\text{apoptosis}}+\underbrace{q(\bar{N}+\bar{A}+\bar{I})}_{\text{immune escape}},\\
\end{aligned}
\right.
\end{eqnarray}
\end{small}
with initial conditions
\begin{equation*}
    \bar{N}(0)=\bar{N}_0\geq 0,~\bar{A}(0)=\bar{A}_0\geq 0,~\bar{I}(0)=\bar{I}_0\geq 0,~\bar{E}(0)=\bar{E}_0\geq 0.
\end{equation*}


\begin{table}[ht]
\caption{\centering Summary of parameters used for model (\ref{model})}\label{tab1}
\scalebox{0.76}{
\begin{tabular}{ccl}
\toprule
Parameter & Description  & Unit\\
\midrule
$r_N$    & Intrinsic growth rate of the antigenically neutral
cells     &  $\text{day}^{-1}$  \\
$K$    &  Carrying capacity     &  \text{cell}  \\
$d_N$    & Natural mortality rate of the antigenically neutral cells      & $\text{day}^{-1}$   \\
$p$    & Antigenic mutation rate      & $\text{day}^{-1}$   \\
$q$    & Immune escape rate      & $\text{day}^{-1}$   \\
$r_A$    & Intrinsic growth rate of the antigenic cells      & $\text{day}^{-1}$   \\
$d_A$    & Natural mortality rate of the antigenic cells     & $\text{day}^{-1}$   \\
$s$    & Transformation rate form the antigenic cells to the immunogenic cells      & $\text{day}^{-1}$   \\
$r_I$    & Intrinsic growth rate of the immunogenic cells       & $\text{day}^{-1}$   \\
$d_I$    & Natural mortality rate of the immunogenic cells     & $\text{day}^{-1}$   \\
$\tilde{m}$    & Immune cell mediated death rate of the immunogenic cells      & $\text{cell}\cdot\text{day}^{-1}$   \\
$\tilde{\theta}$    & Half saturation constant       &  \text{cell}  \\
$r_E$    & Intrinsic growth rate of the immune-escaped cells      & $\text{day}^{-1}$   \\
$d_E$    &  Natural mortality rate of the immune-escaped cells    & $\text{day}^{-1}$   \\
\hline
\end{tabular}}
\end{table}

To begin our analysis, we scale the original variables $N$, $A$, $I$, and $E$ by the carrying capacity $K$. Specifically, we set $N=\frac{\bar{N}}{K}$, $A=\frac{\bar{A}}{K}$, $I=\frac{\bar{I}}{K}$, and $E=\frac{\bar{E}}{K}$, which simplifies system~(\ref{model}) to
\begin{eqnarray}\label{nonmodel}
\left\{
\begin{aligned}
&\frac{\mathrm{d} N}{\mathrm{d} t}= r_NN\left[1-(N+A+I+E)\right]- d_NN-pN-qN,\\
&\frac{\mathrm{d} A}{\mathrm{d}t}= r_AA\left[1-(N+A+I+E)\right]- d_AA+pN-qA-sA,\\
&\frac{\mathrm{d} I}{\mathrm{d} t}= r_II\left[1-(N+A+I+E)\right]+sA-d_II-qI-\frac{mI}{\theta + I},\\
&\frac{\mathrm{d} E}{\mathrm{d} t}= r_EE\left[1-(N+A+I+E)\right]-d_EE+q(N+A+I),\\
\end{aligned}
\right.
\end{eqnarray}
where $m=\frac{\tilde{m}}{K}$, $\theta=\frac{\tilde{\theta}}{K}$. All parameters in the aforementioned model are nonnegative throughout this work.

To account for the effects of environmental fluctuations, we construct a stochastic model by introducing multiplicative white noise into the tumor-immune system. Specifically, we assume that environmental stochasticity primarily affects the mortality of antigenically neutral, antigenic, immunogenic, and immune-escaped cells. Accordingly,
\begin{align*}
    -d_N &\rightarrow  -d_N+\sigma_1\mathrm{d}B_1(t),\quad
    -d_A \rightarrow  -d_A+\sigma_2\mathrm{d}B_2(t),\\
    -d_I &\rightarrow  -d_I+\sigma_3\mathrm{d}B_3(t),\quad
    -d_E \rightarrow  -d_E+\sigma_4\mathrm{d}B_4(t),\\
\end{align*}
where $B_i(t)$ $(i=1,2,3,4)$ are independent standard Brownian motions with intensities $\sigma_i$. Throughout this work, we assume that $B_i(t)$ are defined on a complete probability space $(\Omega , \mathcal{F}, \mathbb{P})$ with filtration ${\mathcal{F}t}{t\geq 0}$ satisfying the usual conditions (i.e., right-continuity and completeness, with $\mathcal{F}_0$ containing all $\mathbb{P}$-null sets). Under these settings, the stochastic model is given by:
\begin{small}
\begin{eqnarray}\label{Smodel}
\left\{
\begin{aligned}
&\mathrm{d} N=\left[ r_NN\left(1-N-A-I-E\right)-d_NN-pN-qN\right]\mathrm{d} t + \sigma_1N\mathrm{d}B_1(t),\\
&\mathrm{d} A=\left[r_AA\left(1-N-A-I-E\right)-d_AA+pN-qA-sA\right]\mathrm{d} t + \sigma_2A\mathrm{d}B_2(t),\\
&\mathrm{d} I=\left[r_II\left(1-N-A-I-E\right)+sA-d_II-qI-\frac{mI}{\theta + I}\right]\mathrm{d} t + \sigma_3I\mathrm{d}B_3(t),\\
&\mathrm{d} E=\left[r_EE\left(1-N-A-I-E\right)-d_EE+q(N+A+I)\right]\mathrm{d} t + \sigma_4E\mathrm{d}B_4(t),\\
\end{aligned}
\right.
\end{eqnarray}
\end{small}

\section{Model analysis}
In this section, we establish threshold conditions governing the long-term dynamics of both the deterministic and stochastic models. For clarity, the proofs of all theorems are provided in the supplementary material.

\subsection{Long-term dynamics of the deterministic model (\ref{nonmodel})}

We analyze the equilibria and stability of the deterministic system to verify that the model is biologically consistent. Since the right-hand side functions of system~(\ref{nonmodel}) are continuously differentiable, the existence of solutions follows from standard theory of ordinary differential equations.

As a preliminary step, we show that the system neither yields negative cell populations nor exhibits unbounded growth over time, but instead remains within biologically realistic bounds.

\begin{theorem}\label{thm1}
     For any nonnegative initial value $(N(0), A(0), I(0), E(0))\in \mathbb{R}^4_+$, the solution $(N(t), A(t), I(t), E(t))$ of system (\ref{nonmodel}) originating from $\mathbb{R}^4_+$, remains positive and bounded within the invariant set $\Gamma = \{(N(t), A(t), I(t), E(t)) \in \mathbb{R}^4_+ ~|~0 \leq N(t) + A(t) + I(t) + E(t) \leq \Upsilon  \}$, where $\Upsilon =\max\left\{N(0)+A(0)+I(0)+E(0), \frac{\Lambda}{\varrho}\right\}$, $\Lambda= \frac{r_N+r_A+r_I+r_E}{4}$, and $\varrho=\min\{d_N,d_A,d_I,d_E\}$.
\end{theorem}

The proof of Theorem \ref{thm1} is given in \ref{positive} of the supplementary material.

In view of their biological significance, we focus on the steady states of system~(\ref{nonmodel}) within $\Gamma$, including trivial, semi-trivial, and positive equilibria. For convenience, we further impose the following assumptions:\\
\begin{align*}
  (H_1)~~&r_N<d_N+p+q,~~r_A<d_A+q+s,\\
    &r_I<d_I+q+\frac{m}{\theta},~~r_E<d_E.
\end{align*}

\begin{align*}
  (H_2)~~&\frac{r_N}{r_E}<\frac{d_N+p+q}{d_E},\quad
    \frac{r_A}{r_E}<\frac{d_A+q+s}{d_E},\quad
    \frac{r_I}{r_E}<\frac{d_I+q+\frac{m}{\theta}}{d_E}.
\end{align*}

\begin{align*}
  (H_3)~~&r_N\left(1-I_2-E_2\right)<d_N+p+q,\quad
    r_A\left(1-I_2-E_2\right)<d_A+q+s,\\
    &r_II_2+\frac{qI_2}{E_2}+r_EE_2>\frac{mI_2}{(\theta+I_2)^2},\quad
    qr_II_2\frac{I_2+E_2}{E_2}>\frac{mI_2}{(\theta+I_2)^2}\left(\frac{qI_2}{E_2}+r_EE_2\right).
\end{align*}

\begin{theorem}\label{sxi0}
    Tumor extinction state $\Xi_0=(0,0,0,0)$ always exists. If condition ($H_1$) holds, the equilibrium  $\Xi_0$ is locally asymptotically stable; otherwise, it is unstable.
\end{theorem}

The proof of Theorem \ref{sxi0} is given in  \ref{psxi0}.

\begin{theorem}\label{sxi1}
   The immune escape dominant state $\Xi_1=(0,0,0,E_{1})$ exists when $r_E>d_E$, where $E_{1}=1-\frac{d_E}{r_E}$. If condition ($H_2$) holds, the equilibrium  $\Xi_1$ is locally asymptotically stable; otherwise, it is unstable.
\end{theorem}

The proof of Theorem \ref{sxi1} is given in  \ref{psxi1}.

\begin{theorem}\label{sxi2}
    Suppose the immune surveillance state $\Xi_2=(0,0,I_{2},E_{2})$ exists. If condition ($H_3$) holds, the equilibrium  $\Xi_2$ is locally asymptotically stable; otherwise, it is unstable.
\end{theorem}

The proof of Theorem \ref{sxi2} is given in  \ref{psxi2}.

Define
\begin{align*}
    \mathcal{W}=&\frac{r_I}{r_A}(d_A+q+s)-s-d_I-q,\quad
    \mathcal{U}=s\frac{\mathcal{X}}{r_A}\frac{\mathcal{V}}{\mathcal{V}-q},\\
    \mathcal{V}=&r_E\frac{d_A+q+s}{r_A}-d_E,\quad
    \mathcal{X}=r_A-d_A-q-s.
\end{align*}
Antigenic accumulation transition state $\Xi_3=(0,A_3,I_3,E_3)$, where $E_3=\frac{q\mathcal{X}}{r_A(q-\mathcal{V})}$, $A_3=\frac{ \mathcal{U} }{s}-I_3$, and $I_3$ satisfies the following quadratic equation

\begin{equation}\label{E3}
    \mathcal{W}I^2+(\theta \mathcal{W}+\mathcal{U}-m)I+\theta \mathcal{U}=0.
\end{equation}

rom the properties of quadratic equations, it follows that equation~(\ref{E3}) has at most two positive roots. For details regarding the number of positive roots and the associated conditions, we refer the reader to \ref{Xi3}. In summary, the existence conditions and the corresponding number of equilibria $\Xi_3$ are presented in Table~\ref{tab2}.

\begin{table}[!htp]
    \renewcommand{\arraystretch}{2}
    \centering
    \caption{\centering Summary of the number and existence conditions for equilibrium $\Xi_3$} \label{tab2}
    \scalebox{0.73}{
    \begin{tabular}{ccccc}
        \toprule
        \multicolumn{3}{c}{Conditions}   & Numbers of  $\Xi_3$  & Cases\\
        \midrule
        \multirow{3}{*}{$\mathcal{X}>0,\mathcal{V}<0$} & $\mathcal{W}>\frac{s^2\theta}{\mathcal{U}}$ & \makecell{$\theta \mathcal{W}+\mathcal{U}+2\sqrt{\theta\mathcal{WU}}<m<(s+\mathcal{W})(\theta+\frac{\mathcal{U}}{s})$}  & 2 & Case(i)\\
        \cline{2-5}
         & - & $m>(\mathcal{W}+s)(\theta+\frac{\mathcal{U}}{s})$ & 1 & \makecell{Case(ii) \& Case(iii) \\ \& Case(v)} \\
         \cline{2-5}
         & $\mathcal{W}<-\frac{s^2\theta}{\mathcal{U}}$ & $m=\theta\mathcal{W}+\mathcal{U}>0$ & 1 & Case(iv)\\
         \cline{2-5}
         & $\mathcal{W}>\frac{s^2\theta}{\mathcal{U}}$ & $m=\theta\mathcal{W}+\mathcal{U}+2\sqrt{\theta\mathcal{WU}}$ & 1 & Case(vi)\\
         \cline{1-5}
         \multicolumn{3}{c}{Other conditions} & 0 & Case(vii)\\
        \hline
    \end{tabular}}
\end{table}

\begin{theorem}\label{sxi3}
    Suppose the state $\Xi_3=(0,A_3,I_3,E_3)$ exists. If condition $\frac{r_N}{r_A}<\frac{d_N+p+q}{d_A+q+s}$,
    $\Upsilon_{1\Xi_{3}},\Upsilon_{3\Xi_{3}}>0$, and $\Upsilon_{1\Xi_{3}}\Upsilon_{2\Xi_{3}}>\Upsilon_{3\Xi_{3}}$ hold, the equilibrium  $\Xi_3$ is locally asymptotically stable, where $\Upsilon_{1\Xi_{3}}, \Upsilon_{2\Xi_{3}}, \Upsilon_{3\Xi_{3}}$ are given in (\ref{T}).
\end{theorem}

The proof of Theorem \ref{sxi3} is given in \ref{psxi3}.

Define
\begin{align*}
    \bar{G}=&\frac{p}{p+\mathcal{Q}},\quad
    \bar{H}=\frac{1}{r_N}\frac{\mathcal{SR}}{\mathcal{S}+q},\quad
    \mathcal{P}=\frac{r_I}{r_N}(d_A+q+s)-s\bar{G}-d_I-q,\\
    \mathcal{Q}=&q+s+d_A-\frac{r_A(d_N+p+q)}{r_N},\quad
    \mathcal{R}=r_N-d_N-p-q,\\
    \mathcal{S}=&d_E-\frac{r_E(d_N+p+q)}{r_N}.
\end{align*}
Adaptive heterogeneity state $\Xi^*=(N^*,A^*,I^*,E^*)$, where $N^*=\frac{\mathcal{Q}}{p}A^*$, $A^*=\bar{G}(\bar{H}-I^*)$, $E^*=\frac{q\mathcal{R}}{r_N(q+\mathcal{S})}$ and $I^*$ satisfies the following quadratic equation

\begin{equation}\label{E*}
    \mathcal{P}I^{*2}+(\theta \mathcal{P}+s\bar{G}\bar{H}-m)I^*+\theta s\bar{G}\bar{H}=0.
\end{equation}

From the properties of quadratic equations, equation~(\ref{E*}) can have at most two positive roots. For a detailed analysis of the number of positive roots and the corresponding conditions, we refer the reader to \ref{xi*}.
In summary, the existence conditions and the corresponding number of equilibria $\Xi^*$ are listed in Table~\ref{tab3}.

\begin{table}[!htp]
    \renewcommand{\arraystretch}{2}
    \centering
    \caption{\centering Summary of the number and existence conditions for equilibrium $\Xi^*$} \label{tab3}
    \scalebox{0.7}{
    \begin{tabular}{ccccc}
        \toprule
        \multicolumn{3}{c}{Conditions}   & Numbers of  $\Xi_3$  & Cases\\
        \midrule
        \multirow{3}{*}{$\mathcal{Q,R,S}>0$} & $\mathcal{P}>\frac{s\theta\bar{G}}{\bar{H}}$ & \makecell{$\theta \mathcal{P}+s\bar{G}\bar{H}+2\sqrt{s\theta\mathcal{P}\bar{G}\bar{H}}<m< (\mathcal{P}+s\bar{G})
        (\theta+\bar{H})$}  & 2 & Case(i)\\
        \cline{2-5}
         & - & $m>(\mathcal{P}+s\bar{G})(\theta+\bar{H})$ & 1 & \makecell{Case(ii) \& Case(iii)\\ \& Case(iv)}\\
         \cline{2-5}
         & $\mathcal{P}<-\frac{s\theta\bar{G}}{\bar{H}}$ & $m=\theta\mathcal{P}+s\bar{G}\bar{H}>0$ & 1 & Case(v)\\
         \cline{2-5}
         & $\mathcal{P}>\frac{s\theta\bar{G}}{\bar{H}}$ & $m=\theta\mathcal{P}+s\bar{G}\bar{H}+2\sqrt{s\theta\mathcal{P}\bar{G}\bar{H}}$ & 1 & Case(vi)\\
         \hline
         \multicolumn{3}{c}{Other conditions} & 0 & Case(vii)\\
        \hline
    \end{tabular}}
\end{table}



\begin{theorem}\label{sxi}
    Suppose the equilibrium $\Xi^*$ exists. If condition $\Upsilon_{1\Xi^*},\Upsilon_{4\Xi^*}>0$,  $\Upsilon_{1\Xi^*}\Upsilon_{2\Xi^*}-\Upsilon_{3\Xi^*}>0$, and $\Upsilon_{3\Xi^*}(\Upsilon_{1\Xi^*}\Upsilon_{2\Xi^*}-\Upsilon_{3\Xi^*})-\Upsilon_{1\Xi^*}^2\Upsilon_{4\Xi^*}>0$ hold, the equilibrium  $\Xi^*$ is locally asymptotically stable, where $\Upsilon_{1\Xi^*},\Upsilon_{2\Xi^*},\Upsilon_{3\Xi^*},\Upsilon_{4\Xi^*}$ are given in (\ref{ts}).
\end{theorem}

The proof of Theorem \ref{sxi} is given in \ref{psxi}.

\begin{remark}
By analyzing the existence and stability conditions of the equilibria of system~(\ref{nonmodel}), we observe the following:
\begin{enumerate}
    \item If the immune escape-dominant state $\Xi_1$ exists, the tumor extinction state $\Xi_0$ is unstable.
    \item The immune escape-dominant state $\Xi_1$ is unstable whenever the antigenic accumulation transition state $\Xi_3$ or the adaptive heterogeneity state $\Xi^*$ exists.
    \item The antigenic accumulation transition state $\Xi_3$ is unstable if the adaptive heterogeneity state $\Xi^*$ exists.
    \item Bistability in system~(\ref{nonmodel}) can only occur in one of the following cases: (a) $\Xi_1$ and $\Xi_2$, (b) $\Xi_2$ and $\Xi_3$, or (c) $\Xi_2$ and $\Xi^*$.
\end{enumerate}
\end{remark}

\subsection{Long-term dynamics of the stochastic model (\ref{Smodel})}

To study the dynamical behavior of the stochastic model, a primary concern is whether its solution is global and positive. The following result addresses the existence and uniqueness of a global positive solution, which is a prerequisite for analyzing the long-term behavior of system~(\ref{Smodel}).

\begin{theorem}\label{unique}
    (Existence and uniqueness of the positive solution) For any initial value $(N_0, A_0, I_0, E_0 )\in \mathbb{R}^4_+$, there is a unique positive solution $(N(t), A(t), I(t), E(t))$ of system (\ref{Smodel}) on $t \geq 0$ and the solution will remain in $\mathbb{R}^4_+$ with probability one.
\end{theorem}

The proof of Theorem \ref{unique} is given in \ref{punique}.

\begin{theorem}\label{moment}
(Moment boundedness) Let $N(t),A(t),I(t),E(t)$ be solution of the stochastic system (\ref{Smodel}) satisfying the initial condition $(N_0, A_0, I_0, E_0) \in  \mathbb{R}^4_+$, then
$$
\limsup\limits_{t\rightarrow\infty}\mathbb{E}[(N(t)+A(t)+I(t)+E(t))^{\theta}]\leq L_1(\theta),
$$
where $L(\theta)$ is a positive constant dependent on $\theta$, which is defined by (\ref{L(theta)}).
\end{theorem}

The proof of Theorem \ref{moment} is given in \ref{pmoment}.

In the following, we derive sufficient conditions for the existence and uniqueness of an ergodic stationary distribution for the positive solutions of system~(\ref{Smodel}).

Let $X(t)$ be a regular time-homogeneous Markov process in $\mathbb{R}^d$ described by the stochastic differential equation  $\mathrm{d}X (t ) = f (X (t ))\mathrm{d}t + g(X (t ))\mathrm{d}B(t)$.  The diffusion matrix of the process $X(t)$ is defined as follows  $\mathcal{A}(x) = (a_{ij}(x)), a_{ij}(x) = g^i(x)g^j(x)$.

\begin{lemma}\cite{Has}\label{SD}
The Markov process $X(t)$ has a unique ergodic stationary distribution $\pi(\cdot)$ if there exists a bounded open domain $D\subset \mathbb{R}^d$ with regular boundary, having the following properties:  \\
(A1)~ there is a positive number $H$ such that $\sum_{i,j=1}^da_{ij}(x)\xi_i\xi_j\geq H|\xi|^2$, $x\in D$, $\xi\in\mathbb{R}^d$.  \\
(A2)~ there exists a nonnegative $C^2$-function $V$ such that $LV$ is negative for any $\mathbb{R}^d \backslash D$.
\end{lemma}

We first make some notations, denote
    \begin{equation*}
        \begin{aligned}
        R^{*}_s=&\frac{psqd_E(d_N+p+q+\sigma_1^2)}{\mathcal{D}(d_A+q+s+\frac{\sigma_2^2}{2})(d_I+q+\frac{\sigma_3^2}{2}+\frac{m}{\theta})(d_E+\frac{\sigma_4^2}{2})}, \\
    \mathcal{D}=&\sup\limits_{(N,A,I,E)\in \mathbb{R}^4_+}\left\{
    -\min\{r_N,r_A,r_I,r_E\}(N+A+I+E)^2+\left(\max\{r_N,r_A,r_I,r_E\}\right.\right.\\
    &\left.\left.+c_1r_A+c_2r_I+c_3r_E
    \right)(N+A+I+E)\right\},
        \end{aligned}
    \end{equation*}
where  $$c_1=\frac{psqd_E(d_N+p+q+\sigma_1^2)}{(d_A+q+s+\frac{\sigma_2^2}{2})^2(d_I+q+\frac{\sigma_3^2}{2}+\frac{m}{\theta})(d_E+\frac{\sigma_4^2}{2})},$$
$$c_2=\frac{psqd_E(d_N+p+q+\sigma_1^2)}{(d_A+q+s+\frac{\sigma_2^2}{2})(d_I+q+\frac{\sigma_3^2}{2}+\frac{m}{\theta})^2(d_E+\frac{\sigma_4^2}{2})},$$
$$c_3=\frac{psqd_E(d_N+p+q+\sigma_1^2)}{(d_A+q+s+\frac{\sigma_2^2}{2})(d_I+q+\frac{\sigma_3^2}{2}+\frac{m}{\theta})(d_E+\frac{\sigma_4^2}{2})^2}.$$

\begin{theorem}\label{stationary}
 (Existence of an ergodic stationary distribution)   Assume that $R^{*}_s> 1$, then for any initial value  $(N_0,A_0,I_0,E_0)\in \mathbb{R}^4_+$, system (\ref{Smodel}) has a unique stationary distribution  and it has the ergodic property.
\end{theorem}

The proof of Theorem \ref{stationary} is given in \ref{pstationary}.


\begin{lemma}\label{L1}
     Let $(N(t),A(t),I(t),E(t))$ be the solution of system (\ref{Smodel}) with any initial value $(N_0,A_0,I_0,E_0)\in \mathbb{R}^4_+$. Then
     \begin{equation}\label{17}
         \lim_{t\rightarrow\infty}\frac{N(t)}{t}=0,~\lim_{t\rightarrow\infty}\frac{A(t)}{t}=0,~\lim_{t\rightarrow\infty}\frac{I(t)}{t}=0,~\lim_{t\rightarrow\infty}\frac{E(t)}{t}=0~~a.s.
     \end{equation}
     Furthermore,
     \begin{equation}\label{18}
     \begin{aligned}
         &\lim_{t\rightarrow\infty}\frac{1}{t}\int_0^tN(s)\mathrm{d}B_1(s)=0,~~\lim_{t\rightarrow\infty}\frac{1}{t}\int_0^t A(s)\mathrm{d}B_2(s)=0,\\
         &\lim_{t\rightarrow\infty}\frac{1}{t}\int_0^t I(s)\mathrm{d}B_3(s)=0,~~\lim_{t\rightarrow\infty}\frac{1}{t}\int_0^t E(s)\mathrm{d}B_4(s)=0~~a.s.
    \end{aligned}
    \end{equation}
\end{lemma}

The proof of Lemma \ref{L1} proceeds along the same lines as the proofs of Lemmas 2.1 and 2.2 in \cite{zhao2014}, and is omitted for brevity.

\begin{theorem}\label{extinction}
(Extinction)    Let $(N(t),A(t),I(t),E(t))$ be the solution of system (\ref{Smodel}) with any initial value $(N_0,A_0,I_0,E_0)\in \mathbb{R}^4_+$. If $R_0^s<1$, where $R_0^s=\frac{4\max\{r_N,r_A,r_I,r_E\}}{\min\{d_N+\frac{\sigma_1^2}{2},d_A+\frac{\sigma_2^2}{2},d_I+\frac{\sigma_3^2}{2},d_E+\frac{\sigma_4^2}{2}\}}$,
    then
$$
\lim_{t\rightarrow\infty}N(t)=\lim_{t\rightarrow\infty}A(t)=\lim_{t\rightarrow\infty}I(t)=\lim_{t\rightarrow\infty}E(t)=0~~a.s.
$$
\end{theorem}

The proof of Theorem \ref{extinction} is given in \ref{pextinction}.


\section{Results}

To address the key questions concerning the dynamics of cancer immunoediting raised in the Introduction, we systematically investigate the interplay of deterministic and stochastic factors in driving phase transitions among tumor-immune states.
Our results show that the kinetics of antigenic mutations (Question 1) act as a critical threshold mechanism, governing phase transitions in the immunoediting process. Bifurcation analysis demonstrates how mutation and accumulation reshape the system's bifurcation structure, shifting dynamics from a multicellular coexistence state to an immune surveillance state, and ultimately giving rise to a bistable regime.
We further uncover the mechanisms underlying divergent immune outcomes (Question 2) through two complementary paradigms: (i) the reshaping of the state-space topology by the stable manifold of a saddle point, which partitions attraction basins between surveillance and escape; and (ii) intervention-induced reprogramming of immune states, whereby elevated cell mortality suppresses escape dominance.
Finally, we assess the role of stochastic perturbations (Question 3) in irreversible fate transitions. We show that environmental noise above a critical intensity destabilizes the immune surveillance state, driving a phase transition toward immune escape dominance.

In the following subsections, we present these results in detail.

\subsection{Key parameters driving immunoediting phase transitions and regulating tumor progression}

Bifurcation analysis provides mathematical insight into the dynamical regulation of tumor progression, systematically elucidating how parameter thresholds govern immunoediting phase transitions and revealing state shifts across the three stages of immunoediting. We perform single-parameter bifurcation analysis to quantify the critical thresholds of antigenic mutation rate ($p$), antigen accumulation rate ($s$), and immune escape rate ($q$). When these parameters cross their respective thresholds, the system undergoes transcritical or saddle-node bifurcations, leading to irreversible phase transitions from multistable regimes toward either immune surveillance states ($\Xi_2$) or immune escape-dominant states ($\Xi_1$).

\subsubsection{Single-parameter bifurcation analysis of the antigenic mutation rate \texorpdfstring{$p$}{p}}
\begin{figure}[hbt]
    \centering
\subfigure[]{\includegraphics[width=0.45\linewidth]{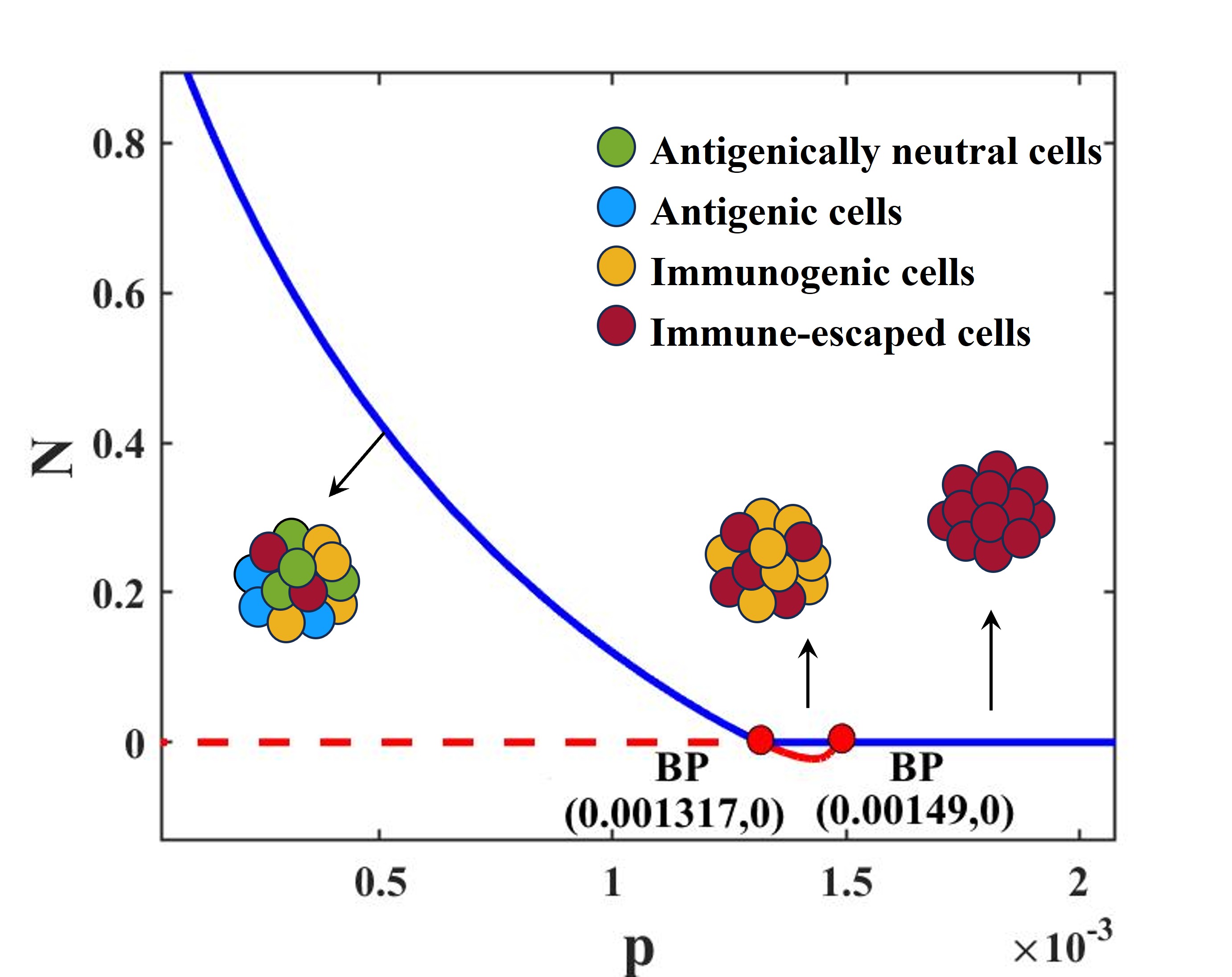}}	
\subfigure[]{\includegraphics[width=0.45\linewidth]{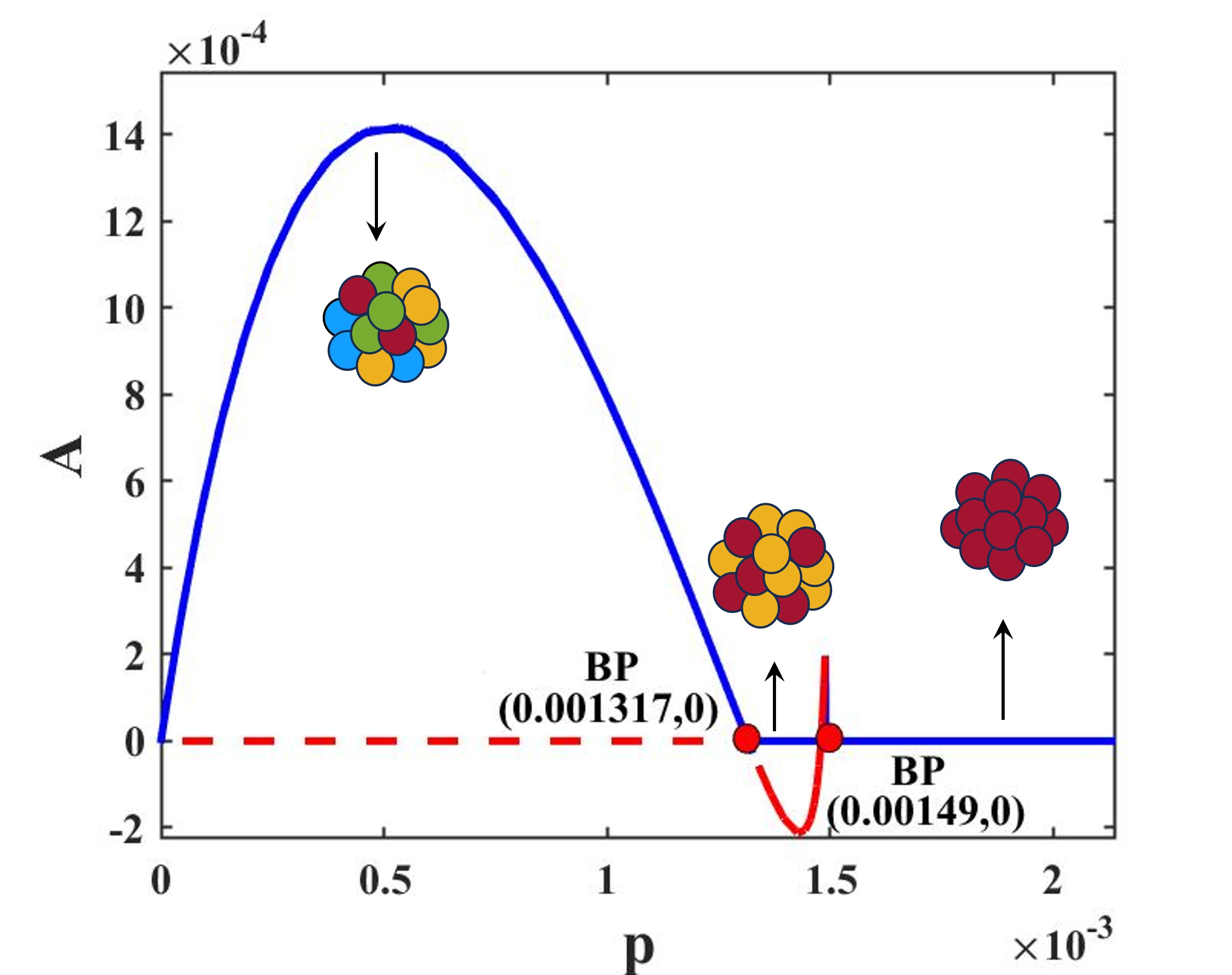}}\\
\subfigure[]{\includegraphics[width=0.45\linewidth]{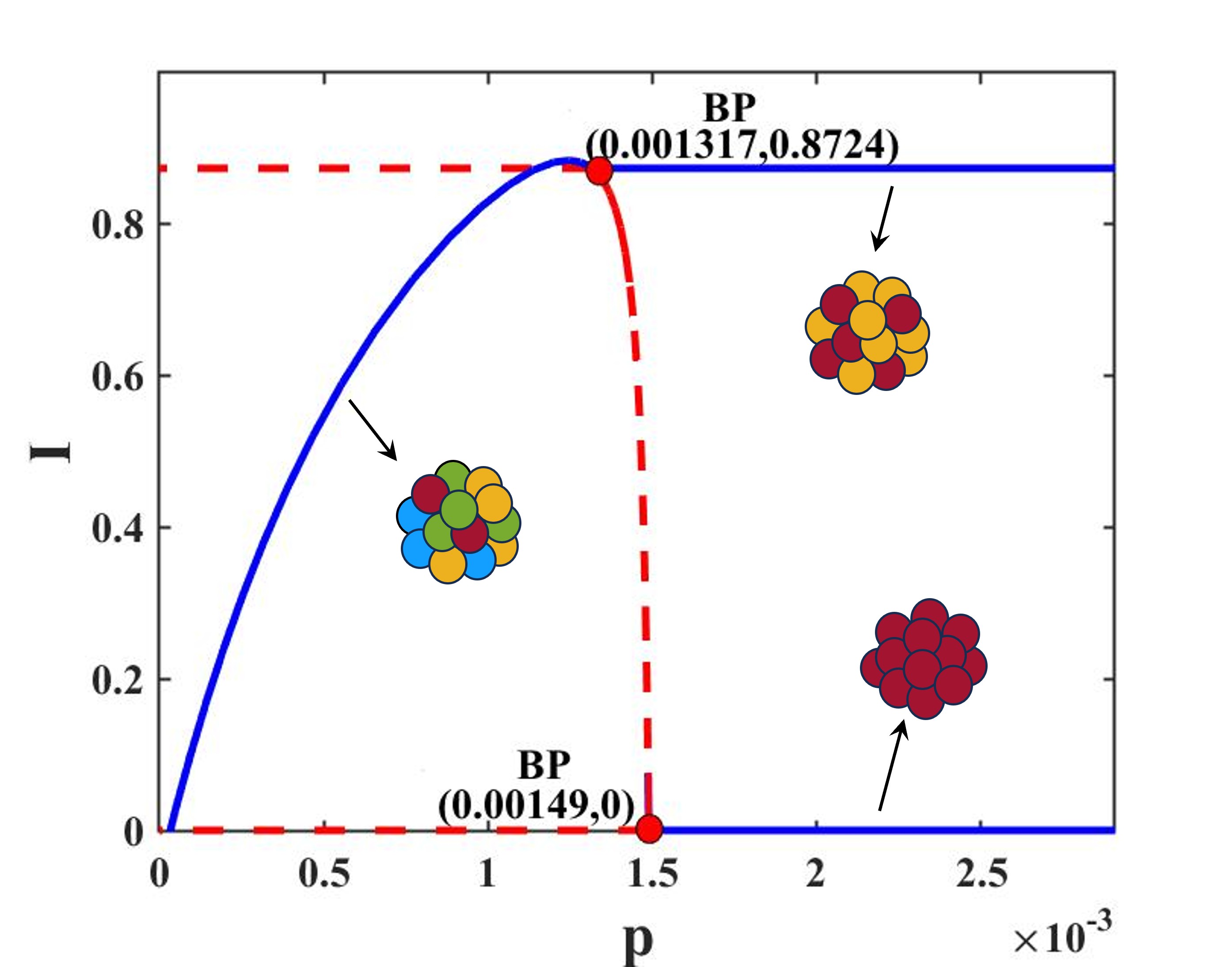}}	
\subfigure[]{\includegraphics[width=0.45\linewidth]{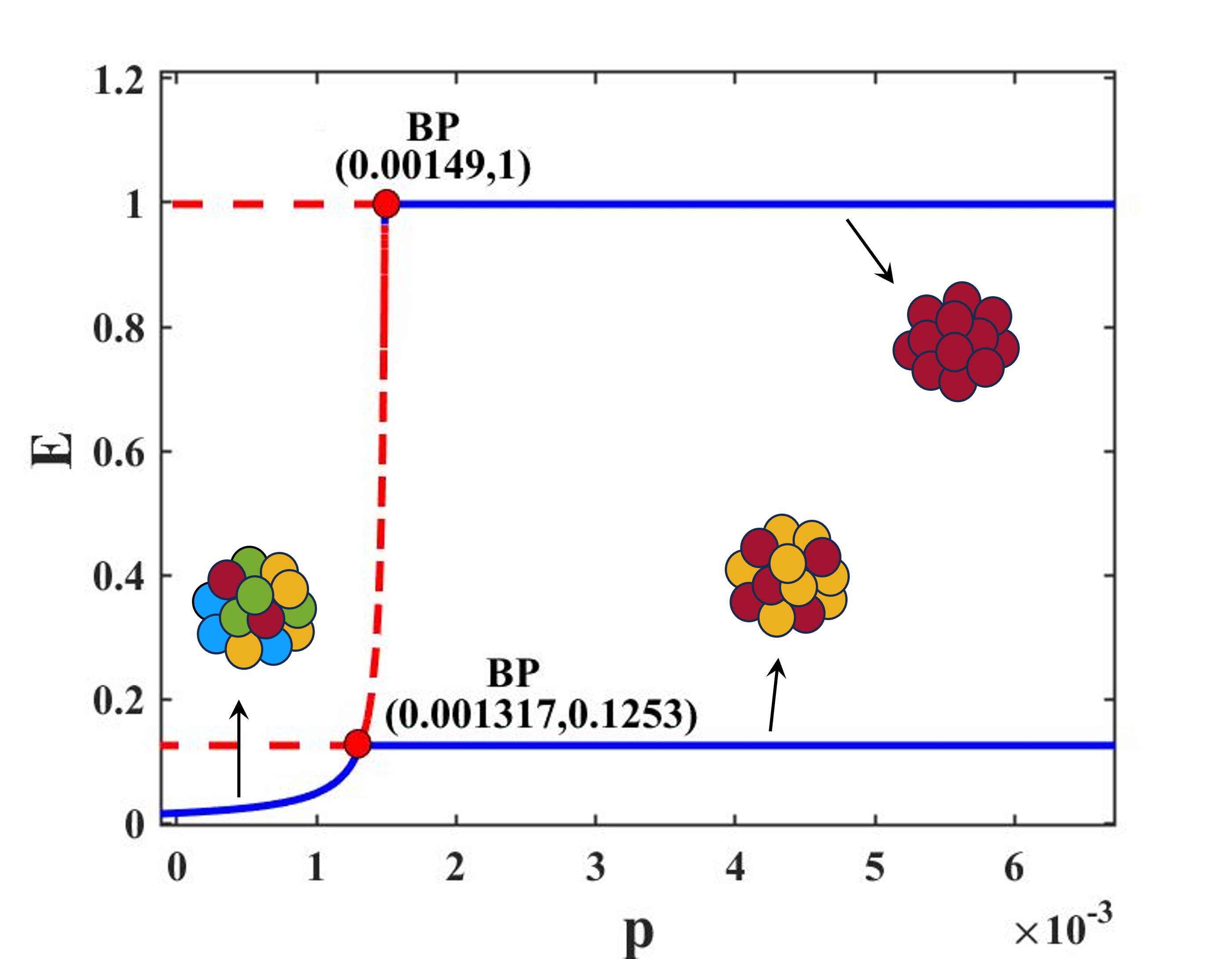}}\\
    \caption{ One parameter bifurcation diagrams with varying antigenic mutation rate $p$. The solid blue curves represent stable equilibria and dashed red curves represent unstable equilibria. Green, blue, yellow, and red spheres denote antigenically neutral cells, antigenic cells, immunogenic cells, and immune-escaped cells, respectively. "BP" means transcritical bifurcation and the other parameters are given in Table \ref{smtab}.}
    \label{fig3}
\end{figure}

To examine how antigenic mutations influence tumor dynamics, we focus on parameter $p$ and investigate the system's behavior across different intervals. Two transcritical bifurcation points are identified ($BP_1 \approx 0.001317$, $BP_2 \approx 0.00149$), which partition the parameter space into three distinct regions: $0<p<BP_1$, $BP_1<p<BP_2$, and $p>BP_2$ (Fig.~\ref{fig3}). This highlights the critical role of antigenic mutations in shaping the interplay between tumor cells and the immune system.

When $p<BP_1$, the system remains in a multicellular coexistence state where antigenically neutral cells $N$, antigenic cells $A$, immunogenic cells $I$, and immune-escaped cells $E$ maintain a dynamic balance. Moreover, nonlinear responses emerge among cell populations:
(i) As $p$ increases, more neutral cells $N$ undergo mutation, reducing the antigen-neutral population.
(ii) The antigenic cells $A$ first increase and then decline, since moderate $p$ produces more mutants while higher $p$ accelerates their conversion into immunogenic or immune-escaped cells.
(iii) Both $I$ and $E$ increase with $p$, with $I$ rising more steeply.

When $p>BP_1$, the system transitions into an immune surveillance state where only $I$ and $E$ persist. In this regime, the immune system effectively controls $I$ via the saturating killing term $\tfrac{mI}{\theta+I}$, while the growth of $E$ depends on immune escape $q(N+A+I)$.

Once $p>BP_2$, the system enters a bistable region in which the immune surveillance state $\Xi_2$ (dominated by $I$) coexists with the immune escape state $\Xi_1$ (dominated by $E$). The eventual outcome depends on the initial conditions. This bistability supports the irreversibility of the "escape phase" in immunoediting theory.

Importantly, the stability of $\Xi_1$ and $\Xi_2$ does not depend on $p$, and their equilibrium expressions are independent of $p$. This indicates that once the system settles into either state, it becomes highly robust, requiring external perturbations to escape from the established homeostasis.

\subsubsection{Single-parameter bifurcation analysis of the transformation rate from antigenic to immunogenic cells \texorpdfstring{$s$}{s}}
We investigate the effect of the antigenic accumulation rate $s$ on tumor immune escape dynamics. The system exhibits two transcritical bifurcation points ($BP_1\approx0.000153$ and $BP_2\approx0.00024$), indicating a threshold-driven mechanism underlying antigen accumulation, immune surveillance, and immune escape. When $s<BP_1$, antigenic cells $A$ dominate the system (state $\Xi_3$) through proliferation and mutation, while the densities of immunogenic cells $I$ and immune-escaped cells $E$ remain low. Once $s$ exceeds $BP_1$, the conversion from $A$ to $I$ becomes sufficient to activate immune clearance ($\frac{mI}{\theta+I}$), leading to a transition to the immune surveillance state $\Xi_2$, where $I$ and $E$ constitute the dynamic equilibrium. Further increasing $s$ beyond $BP_2$ drives the system into a bistable regime, where $\Xi_2$ (immune surveillance) coexists with $\Xi_1$ (immune escape dominance). This bistability arises from the synergistic effect of $s$ and the immune escape parameter $q$: while $s$ accelerates the $A\rightarrow I$ transition, $q$ facilitates the $I\rightarrow E$ escape, ultimately enabling $E$ proliferation to surpass apoptosis and immune clearance. Notably, the stability of $\Xi_1$ and $\Xi_2$ themselves is not directly affected by $s$.

These results highlight the unique role of antigen accumulation in shaping immune escape kinetics and suggest that therapeutic strategies aimed at reducing $s$ (e.g., inhibiting antigen presentation pathways) may complement approaches targeting immune enhancement or escape inhibition.

\begin{figure}[H]
    \centering
\subfigure[]{\includegraphics[width=0.45\linewidth]{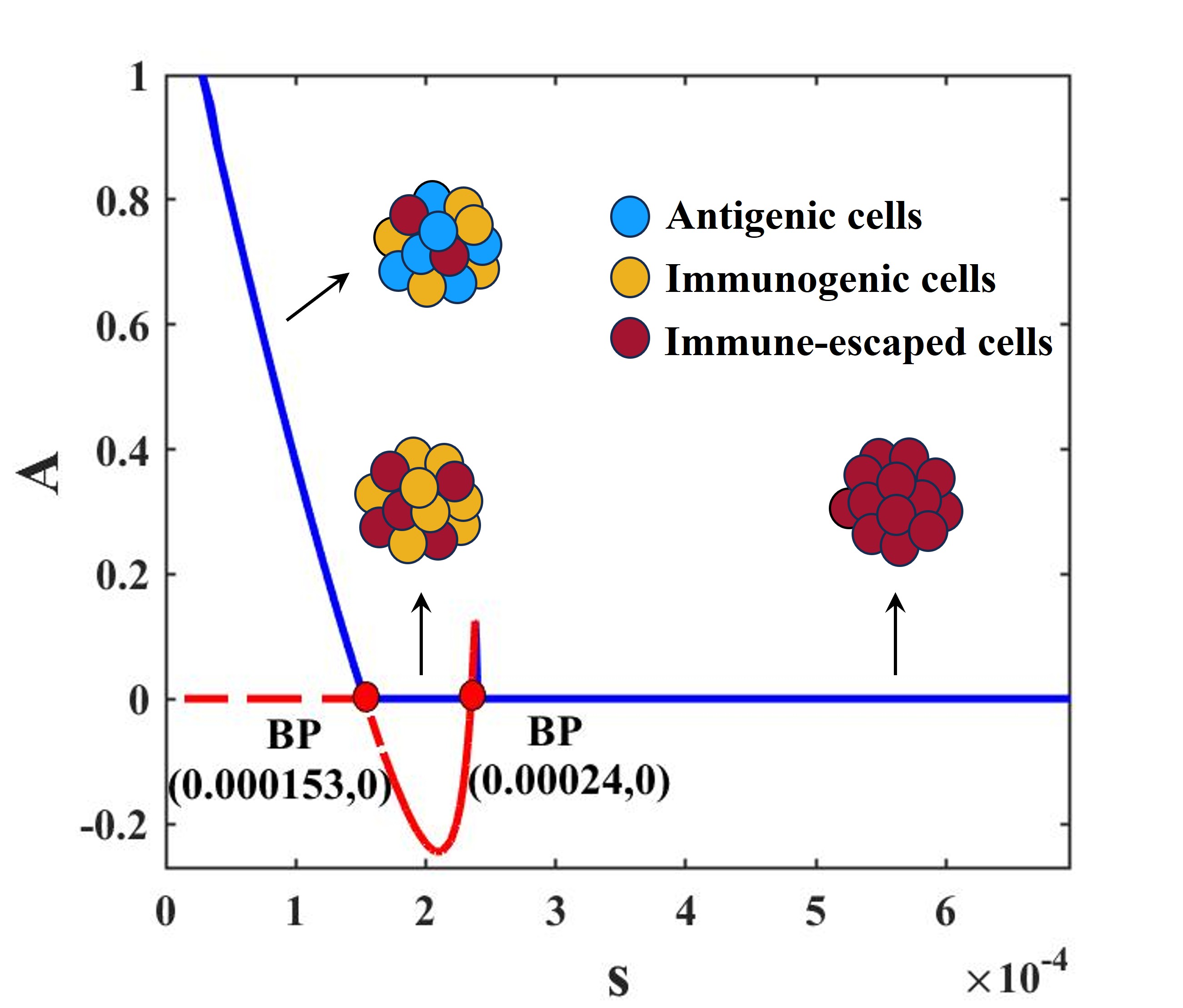}}
\subfigure[]{\includegraphics[width=0.45\linewidth]{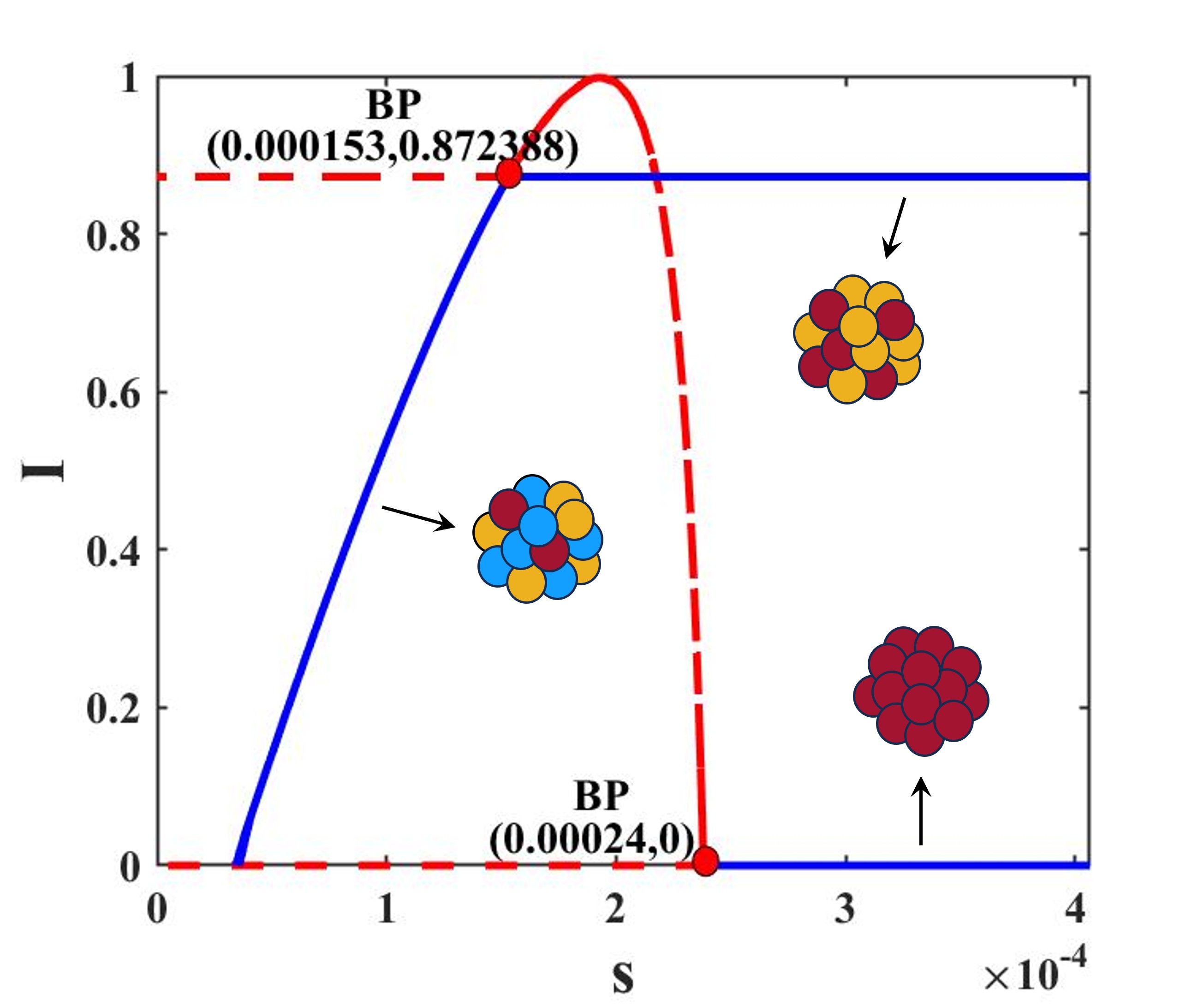}}\\
\subfigure[]{\includegraphics[width=0.45\linewidth]{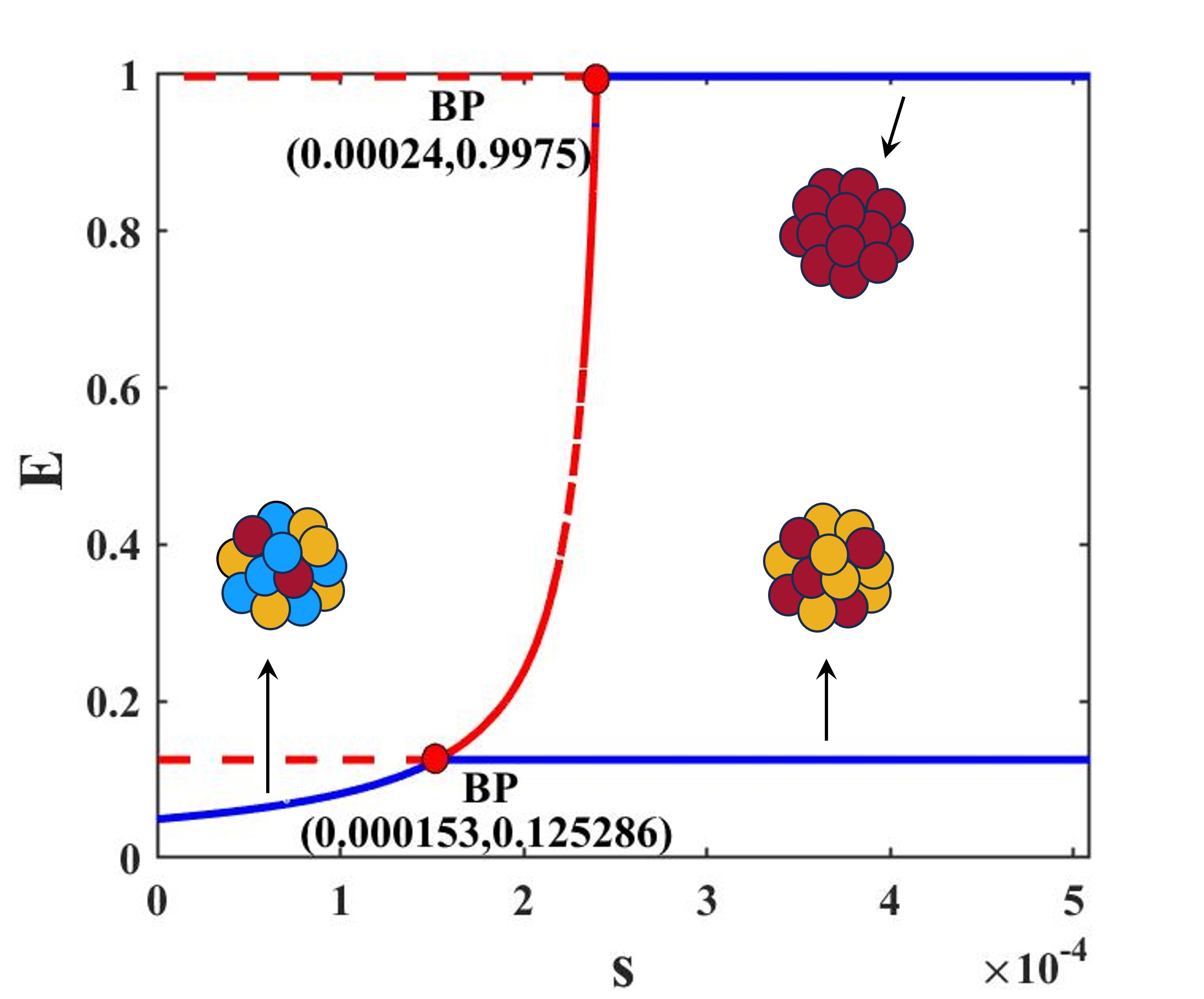}}	
    \caption{ One parameter bifurcation diagrams with varying transformation rate form the antigenic cells to the immunogenic cells $s$. The solid blue curves represent stable equilibria and dashed red curves represent unstable equilibria. Blue, yellow, and red spheres denote antigenic cells, immunogenic cells, and immune-escaped cells, respectively. "BP" means transcritical bifurcation and the other parameters are given in Table \ref{smtab}}
    \label{fig4}
\end{figure}

\subsubsection{Single-parameter bifurcation analysis of the immune escape rate \texorpdfstring{$q$}{q}}
\begin{figure}[H]
    \centering
\subfigure[]{\includegraphics[width=0.45\linewidth]{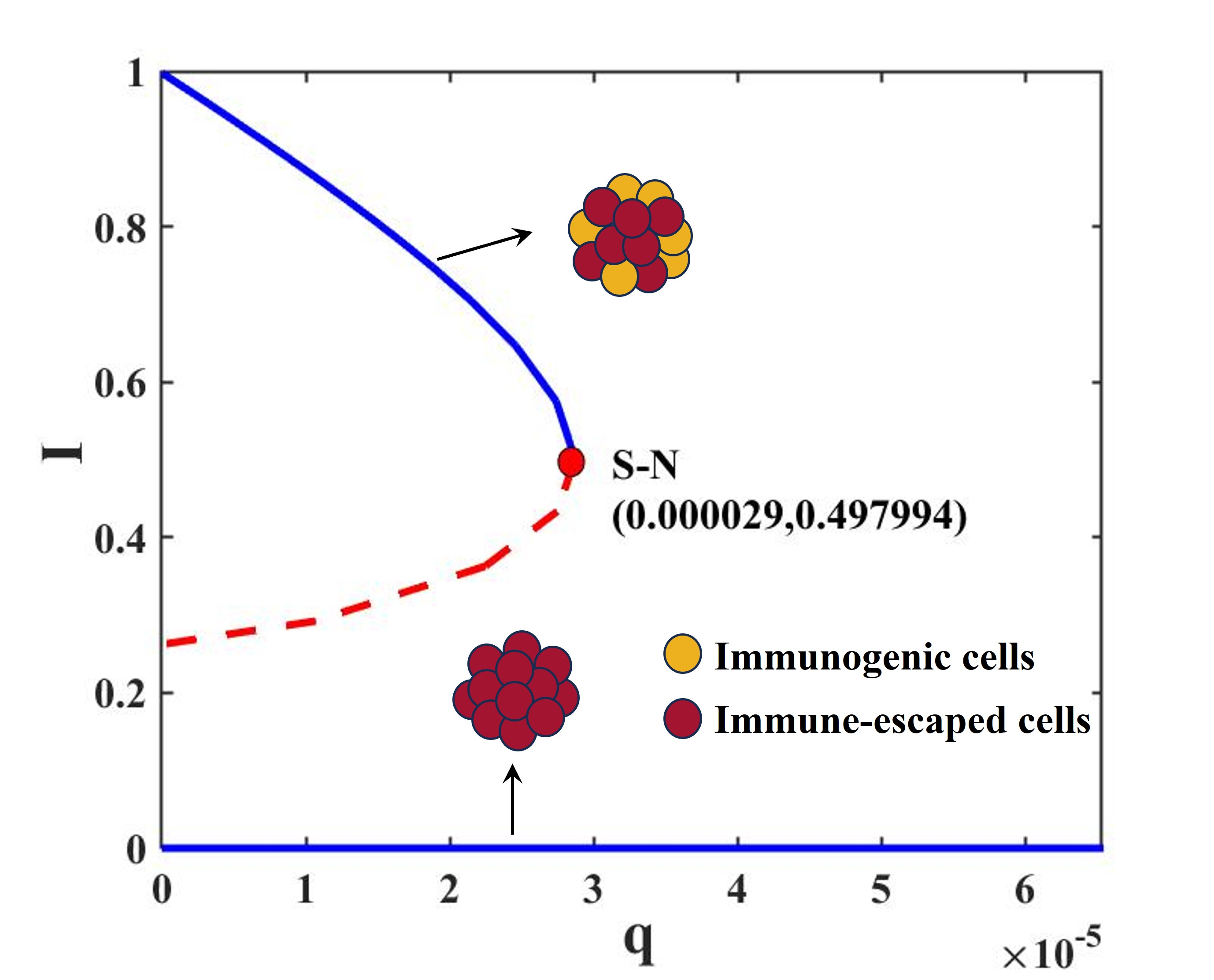}}
\subfigure[]{
\includegraphics[width=0.45\linewidth]{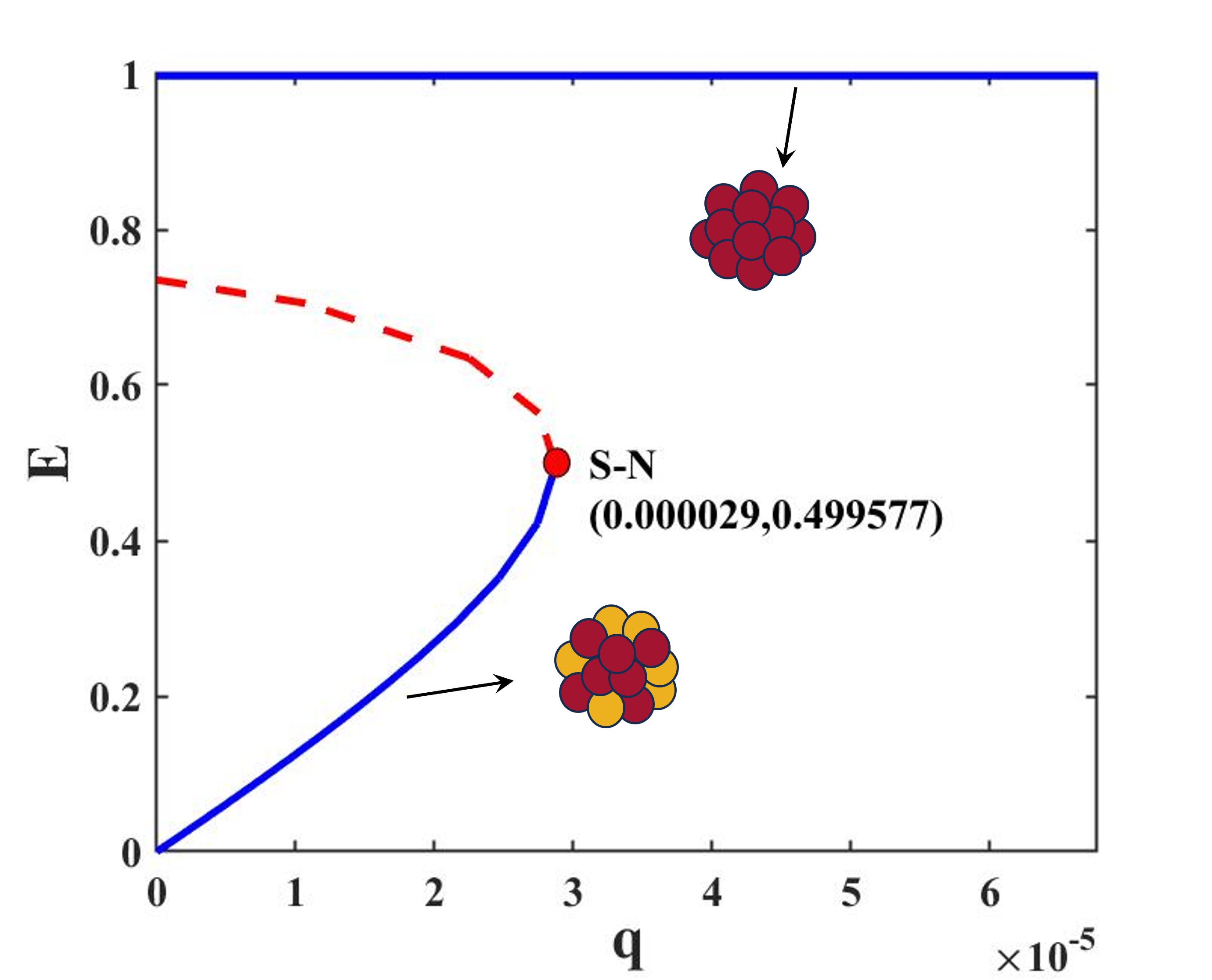}}
    \caption{ One parameter bifurcation diagrams with varying immune escape rate $q$. The solid blue curves represent stable equilibria and dashed red curves represent unstable equilibria. Yellow and red spheres indicate immunogenic cells and immune-escaped cells, respectively. "S-N" means saddle-node bifurcation and the other parameters are given in Table \ref{smtab}.}
    \label{fig2}
\end{figure}

The influence of immune escape mutation rates on tumor dynamics is illustrated in Fig.~\ref{fig2}. As the immune escape rate $q$ increases, both the existence and stability of equilibria change. Stable equilibria are indicated by blue solid curves, unstable equilibria by red dashed curves, and saddle-node bifurcations by the label ``S-N.'' The immune escape dominant state $\Xi_1$ always exists and remains stable.

Fig.~\ref{fig2} demonstrates that the system exhibits threshold-dependent bistability. When $q$ is below the saddle-node bifurcation threshold ($\text{S-N}\approx 0.000029$), the system admits two stable equilibria: the immune surveillance state $\Xi_2$ (dominated by immunogenic cells $I$) and the immune escape state $\Xi_1$ (dominated by immune-escaped cells $E$). The long-term outcome depends on initial conditions: the system converges to $\Xi_2$ if immune surveillance ($\tfrac{mI}{\theta+I}$) effectively suppresses $I$, or to $\Xi_1$ if immune escape ($q(N+A+I)$) dominates.

Once $q$ exceeds the bifurcation threshold, $\Xi_2$ disappears, leaving only the immune escape state $\Xi_1$. This transition is irreversible and corresponds to the ``escape phase'' in the theory of tumor immunoediting, highlighting the critical role of $q$ as a tipping point for immune evasion.

It is noteworthy that $\Xi_1$ is structurally stable, supported by the predominance of the proliferation term $r_EE$ and escape term $q(N+A+I)$ over the apoptotic term $d_EE$. This quantitative analysis suggests that maintaining $q$ below the bifurcation threshold is essential for preserving bistability and sustaining the possibility of immune surveillance.

\subsection{Elevated cell mortality weakens the dominance of immune escape and reprograms immune response states}
The single-parameter bifurcation analysis above illustrates the regulatory role of key parameters in phase transitions of immunoediting. However, the influence of multi-parameter coupling effects on tumor subpopulation interactions requires further investigation. To this end, we perform a global sensitivity analysis. Following the numerical framework described by Marino et al.~\cite{marino2008}, the Latin hypercube sampling (LHS) method is employed to generate 1000 parameter sets, and the partial rank correlation coefficient (PRCC) of the proportions of four cell types is computed. Detailed results of this sensitivity analysis are provided in \ref{PSA}.

Within the bistable region at an antigenic mutation rate of $p=0.15$, sensitivity analysis reveals a marked reduction in system sensitivity to conventional regulatory parameters (e.g., immune system-mediated mortality $m$ and immune escape rate $q$), accompanied by a substantial increase in sensitivity to cell-autonomous mortality rates ($d_I, d_E$), as shown in Figs.~\ref{fig5} and \ref{fig51} of the supplementary material. Building on this, we investigate system dynamics using a two-parameter ($q$-$m$) analysis and introduce a high-mortality condition to simulate chemotherapeutic intervention.

\begin{figure}[hbt]
    \centering
\includegraphics[width=1\linewidth]{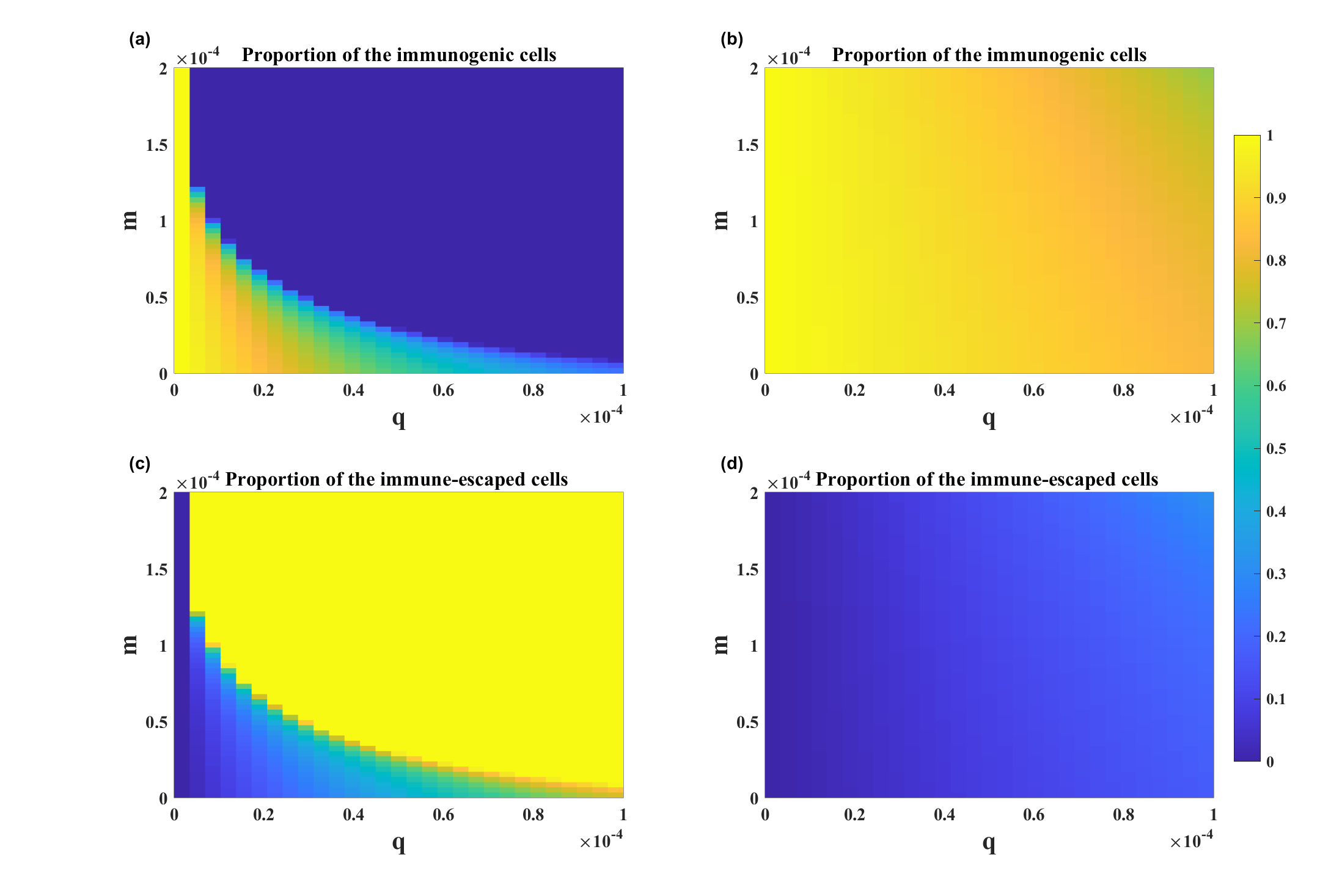}
    \caption{Proportion of immune-escaped cells and immunogenic cells to the total number of cells when the immune escape rate $q$ and immune system-mediated immunogenic cell death $m$ are varied. (a) (c) $d_N=d_A=d_I=d_E=0.001$, (b) (d) $d_N=d_A=d_I=d_E=0.005$, and other parameters are shown in Table \ref{smtab}.}
    \label{fig}
\end{figure}

Specifically, when the mortality rates of all four cell types are simultaneously elevated ($d_N = d_A = d_I = d_E = 0.005$), the proliferative advantage of immune-escaped cells $E$ is diminished, effectively restricting their proportion to below $30\%$. At the same time, the proportion of immunogenic cells $I$ significantly increases, driving the system toward an irreversible transition from the immune escape-dominant state ($\Xi_1$) to the immune surveillance state ($\Xi_2$) (see Fig.~\ref{fig}). By contrast, increasing only the immune system-mediated mortality $m$ promotes the likelihood of immune escape (see Fig.~\ref{fig}(c)).

These findings suggest that interventions targeting tumor immune escape require multidimensional strategies. In particular, chemotherapy (increasing $d_E$) combined with immune regulation (e.g., enhancing $r_I$ or inhibiting $r_E$) is likely more effective than relying on a single mechanism.

\subsection{The stable manifold of the saddle point as the boundary between different tumor fates}
The single-parameter bifurcation analysis in the previous section shows that the same parameter set can correspond to multiple behavioral modes. The eventual trajectory of the system depends on the initial conditions, specifically on which attractor basin they lie in. The boundary between these basins, called the separatrix, captures the system's sensitivity in critical states: small perturbations near the separatrix may trigger a qualitative shift in the trajectory.

From a dynamical systems perspective, the separatrix typically coincides with a stable invariant manifold. When this manifold is one-dimensional, it can be tracked by integrating time-reversed trajectories starting near the saddle point \cite{ku1998}. For two-dimensional manifolds, one may compute orbital segments using numerical continuation methods \cite{krauskopf2005survey}, which solve associated two-point boundary value problems. However, our immune-escape kinetic model involves four state variables, making direct visualization of the invariant manifold infeasible. In such cases, analytical approximation combined with projection into three dimensions is used.

We employ the local quadratic approximation method proposed in \cite{venkatasubramanian1997numerical} to compute the approximate $(n-1)$-dimensional stable manifold of the unstable equilibrium lying on the stability boundary. The resulting manifold is described by the approximate quadratic form $h_2(N,A,I,E) = 0$ (see \ref{App1}). To highlight the interactions among antigenically neutral cells ($N$), immunogenic cells ($I$), and immune-escaped cells ($E$), we fix $A$ at its equilibrium value. We then plot the reduced manifold $h_2(N,I,E \mid A=0) = 0$, which partitions the state space into two distinct regions of attraction.

\begin{figure}[htb]
    \centering
\subfigure[]{\includegraphics[width=0.48\linewidth]{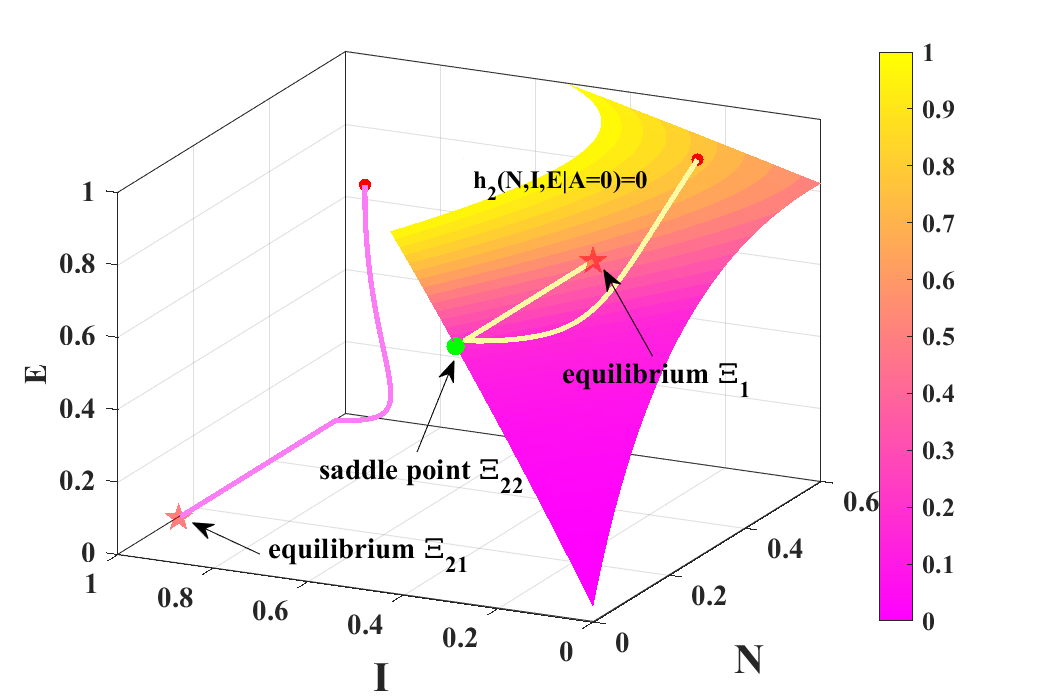}}	
\subfigure[]{
\includegraphics[width=0.48\linewidth]{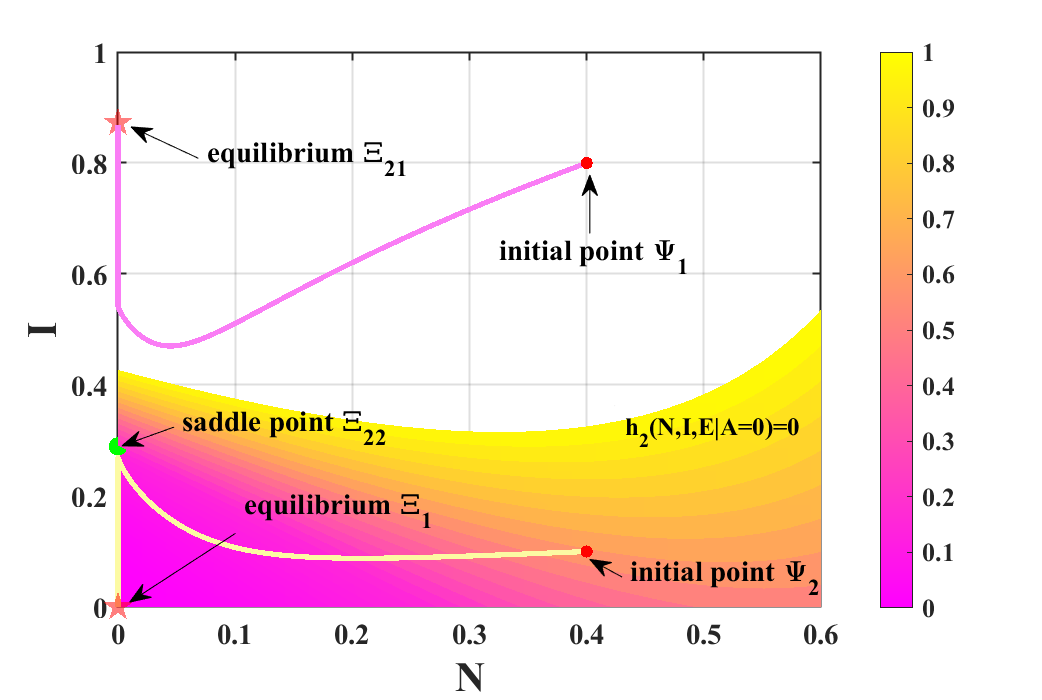}}
    \caption{The regions of attraction for high and low immune-escaped cells are delineated by an approximate stable manifold  at saddle point, as shown by the surface $h_2( N , I, E~|~A= 0) = 0$. Simulated trajectories from different regions of attraction stay within the region and converge to the corresponding equilibrium. (a) 3-dimensional illustration of the surface $h_2( N , I, E~|~A = 0) = 0$, (b) 2-dimensional projection of the surface $h_2( N , I, E~|~A = 0) = 0$ into the N-I plane, where the two red pentagrams denote the stable equilibria $\Xi_1=(0,0,0,0.9975)$ and $\Xi_{21}=(0,0,0.8724 ,0.1253)$, green dots denote unstable saddle point $\Xi_{22}=(0,0,0.2895,0.7081)$, and two red dots denote initial points taken from the two domains of attraction $\Psi_1=(0.4,0,0.8,0.8)$ and $\Psi_2=(0.4,0,0.1,1)$, respectively. All parameters are taken from Table \ref{smtab}.}
    \label{fig1}
\end{figure}

Specifically, the region below the manifold corresponds to the immune-escape dominant state $\Xi_1$, while the region above it corresponds to the immune-surveillance state $\Xi_2$. As illustrated in Fig.~\ref{fig1}, trajectories starting in different regions (purple vs. yellow curves) converge to the corresponding stable equilibria.

As shown in Fig.\ref{fig1}, the system converges to a low immune-escape steady state ($E=0.1289$), where immune surveillance effectively suppresses the clonal expansion of immune-escaped cells. In contrast, when $I$ falls below this critical value (within the surface projection, see Fig.~\ref{fig1}), the system reaches a high immune-escape steady state ($E=0.9975$), suggesting that excessive depletion of immunogenic cells by the immune system can drive the system toward an immune escape scenario.

This bistable behavior provides a theoretical basis for the threshold phenomenon observed in treatment response, indicating that the efficacy of tumor immunotherapy depends not only on drug potency but also on the initial densities of each cell type within the tumor.

\subsection{Environmental noise irreversibly induces a phase transition from immune surveillance state to immune escape dominant state}

\begin{figure}[htb]
    \centering
\subfigure[The initial value is $(0.1,0.001,0.1,0.8)$]{\includegraphics[width=0.45\linewidth]{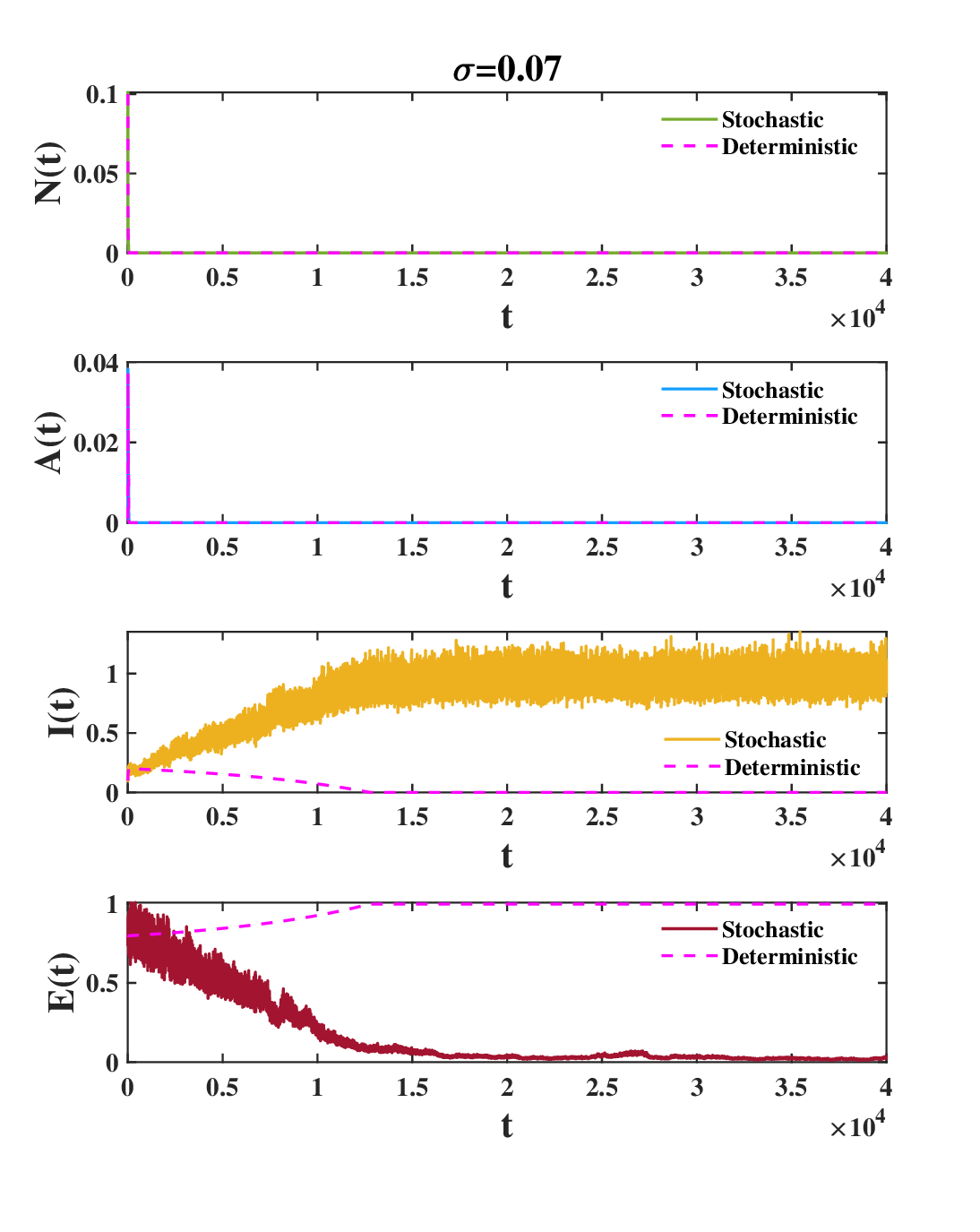}}
\subfigure[The initial value is $(0.8,0.1,0.01,0.01)$]{\includegraphics[width=0.45\linewidth]{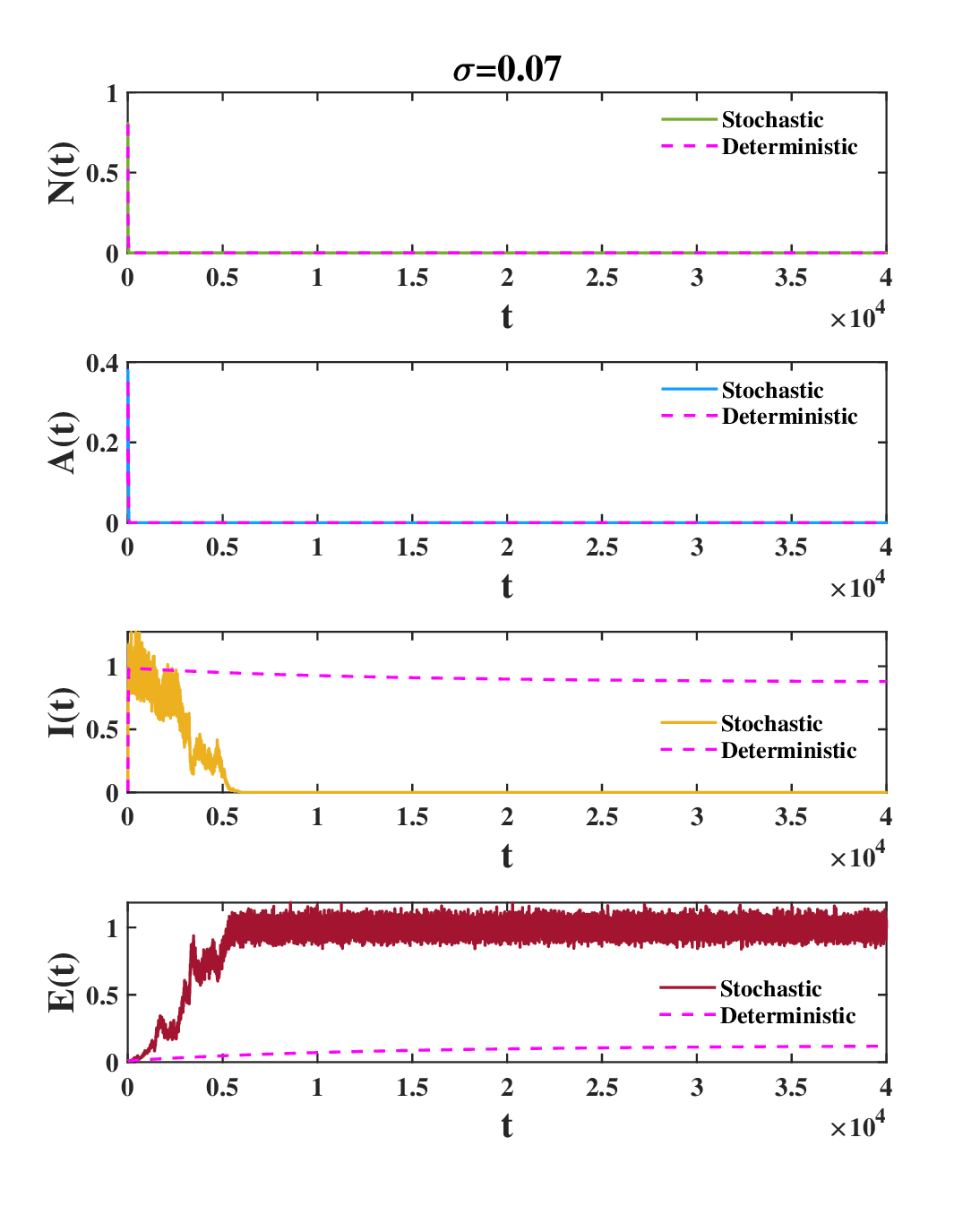}}\\
    \caption{Trajectories of stochastic system (\ref{Smodel}) and its corresponding deterministic system (\ref{nonmodel}) with different initial values. Here the noise intensity is $\sigma=0.007$ and the other parameters are given in Table \ref{smtab}.}
    \label{fig6}
\end{figure}

It has been established that the deterministic model (\ref{nonmodel}) exhibits parameter-dependent bistable regions ($\Xi_1$ and $\Xi_2$). In this subsection, our main focus is to investigate the noise-induced state transition from the immune surveillance state $\Xi_2$  to the immune escape dominant state $\Xi_1$. We also explore how the environmental stochasticity affects the dynamical behavior of the system, as shown in Fig. \ref{fig6}.

Fig. \ref{fig6} illustrates that the dynamical behavior of the deterministic system (\ref{nonmodel}) within the bistable region is highly sensitive to initial conditions. Specifically, the trajectory converges to the immune escape dominant state $\Xi_1=(0,0,0,0,0.9975)$ when initialized at $(0.1,0.001,0.1,0.1,0.8)$ (dashed line in Fig. \ref{fig6}(a)). In contrast, with the initial value $(0.8,0.1,0.01,0.01)$, the trajectory converges to the immune surveillance state $\Xi_2=(0,0,0.8724,0.1253)$ (dashed line in Fig. \ref{fig6}(b)).

However, when environmental perturbation with noise intensity $\sigma = 0.07$ is incorporated into the system, the stochastic trajectories exhibit opposite convergence patterns. The trajectory starting from $(0.1,0.001,0.1,0.8)$ fluctuates around $\Xi_2$ (solid line in Fig. \ref{fig6}(a)), whereas the trajectory starting from $(0.8,0.1,0.01,0.01)$ converges to $\Xi_1$ (solid line in Fig. \ref{fig6}(b)). This indicates that environmental perturbations can induce transitions between immune surveillance and immune escape states by disrupting the deterministic system's steady-state attractor domains.

Furthermore, we numerically estimate the critical noise intensity at which the system transitions from the immune surveillance state $\Xi_2$ to the immune escape dominant state $\Xi_1$.

\subsubsection{Numerical estimation of critical noise intensity}

When the deterministic model (\ref{nonmodel}) is perturbed by stochastic fluctuations, transitions between different attractors may occur. We now examine how the dynamics of the stochastic model (\ref{Smodel}) change under varying noise intensities. Specifically, we consider $\sigma = 0.012$ and $\sigma = 0.05$ to illustrate the effect of noise. The initial condition is set at the equilibrium point $\Xi_2 = (0,0,0.8724,0.1253)$.

\begin{figure}[htb]
    \centering
\subfigure{\includegraphics[width=0.48\linewidth]{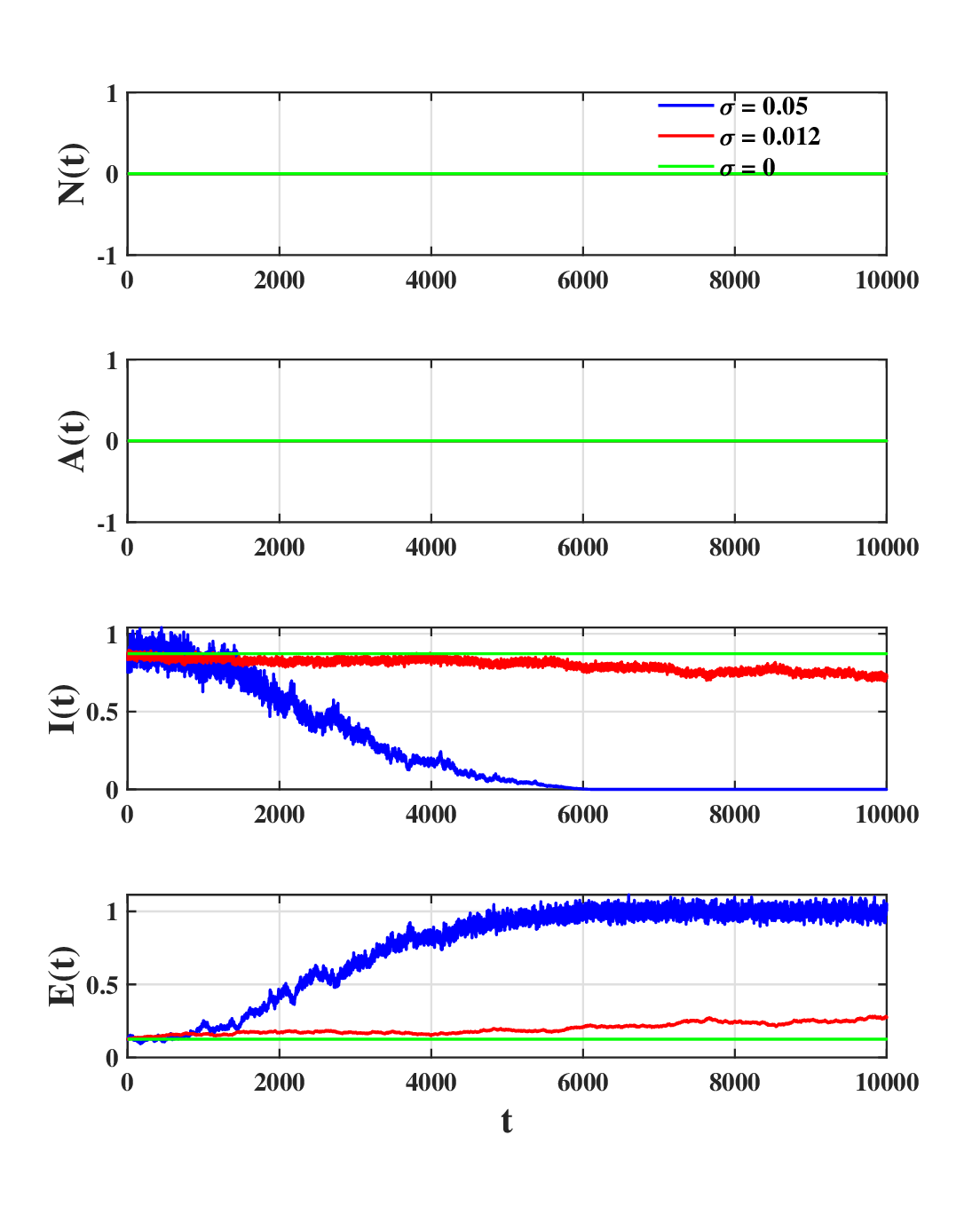}}
\subfigure{\includegraphics[width=0.48\linewidth]{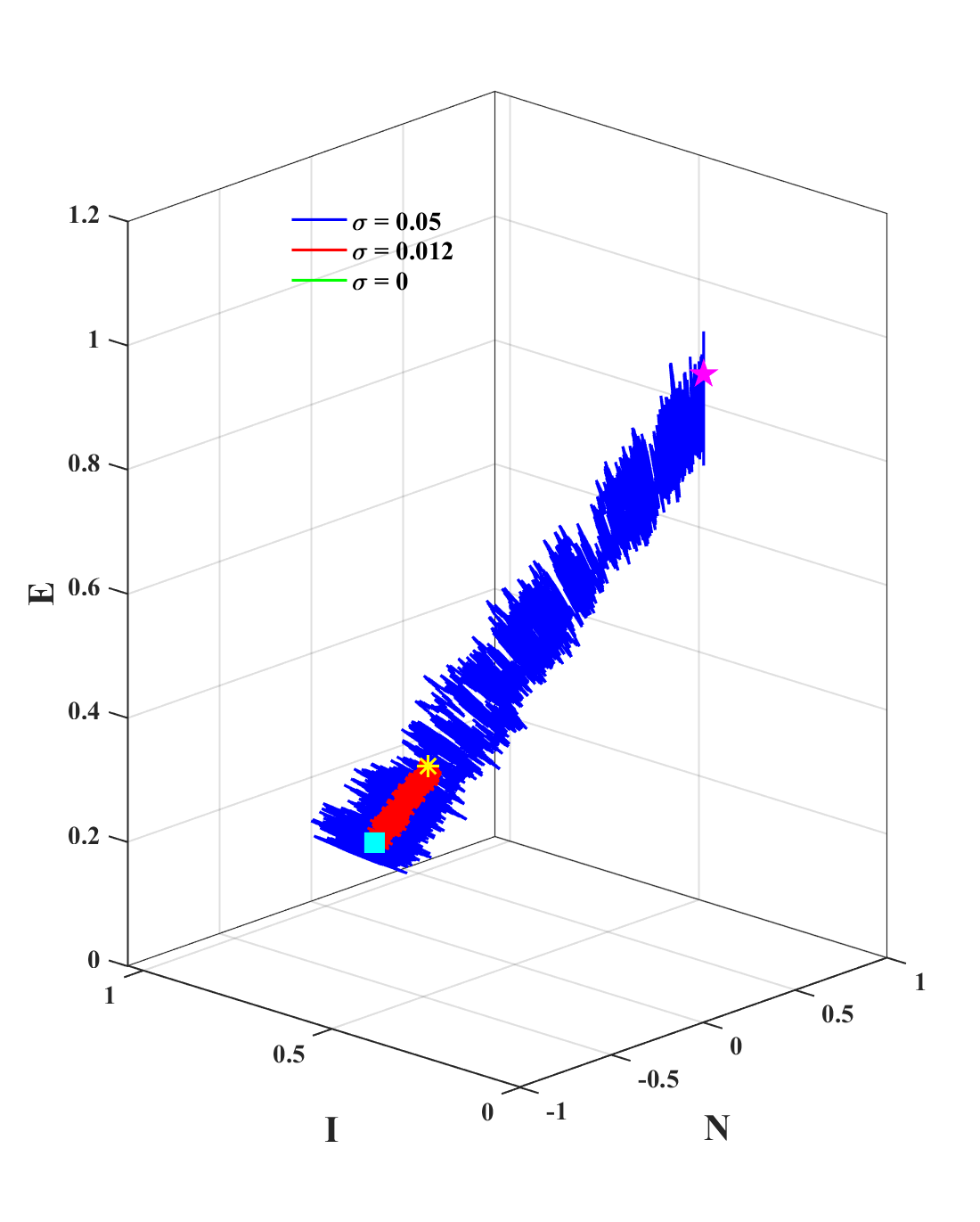}}\\
    \caption{The time series (left) and phase trajectories (right) for stochastic model (\ref{Smodel}) with initial value $\Xi_2=(0,0,0.8724,0.1253)$ and different noise intensity $\sigma=0$ (green), $\sigma=0.012$ (red), $\sigma=0.05$ (blue). The cyan square indicates the initial point, the yellow asterisk and the magenta pentagram indicate the position of the trajectory  at time $t = 10^4$ with noise intensity of 0.05 and 0.012, respectively.}
    \label{fig8}
\end{figure}

Fig. \ref{fig8} shows three sample trajectories of model (\ref{Smodel}) under different noise intensities. When the noise is weak ($\sigma = 0.012$), the stochastic trajectories (red) fluctuate around $\Xi_2$. However, with stronger noise ($\sigma = 0.05$), the stochastic trajectories (blue) converge towards the immune escape dominant state $\Xi_1$, as seen in the left panel of Fig. \ref{fig8}. This indicates that sufficiently strong noise can induce a transition from the immune surveillance state $\Xi_2$ to the immune escape dominant state $\Xi_1$, which is further confirmed by the phase diagram in the right panel of Fig. \ref{fig8}. For comparison, the deterministic solution (green) for the same parameters is also plotted in the left panel.

\begin{figure}[htb]
    \centering
\subfigure{\includegraphics[width=0.48\linewidth]{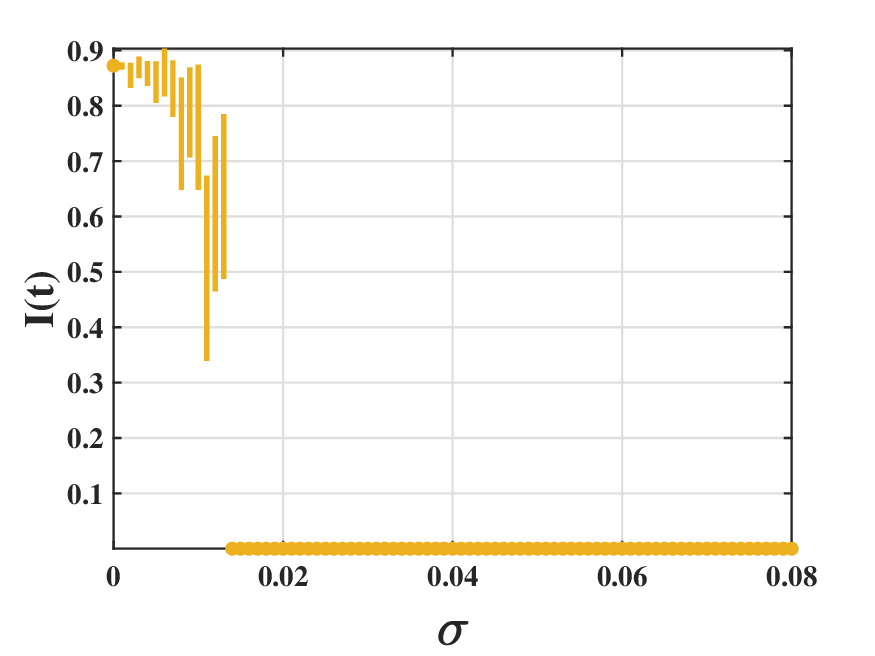}}
\subfigure{\includegraphics[width=0.48\linewidth]{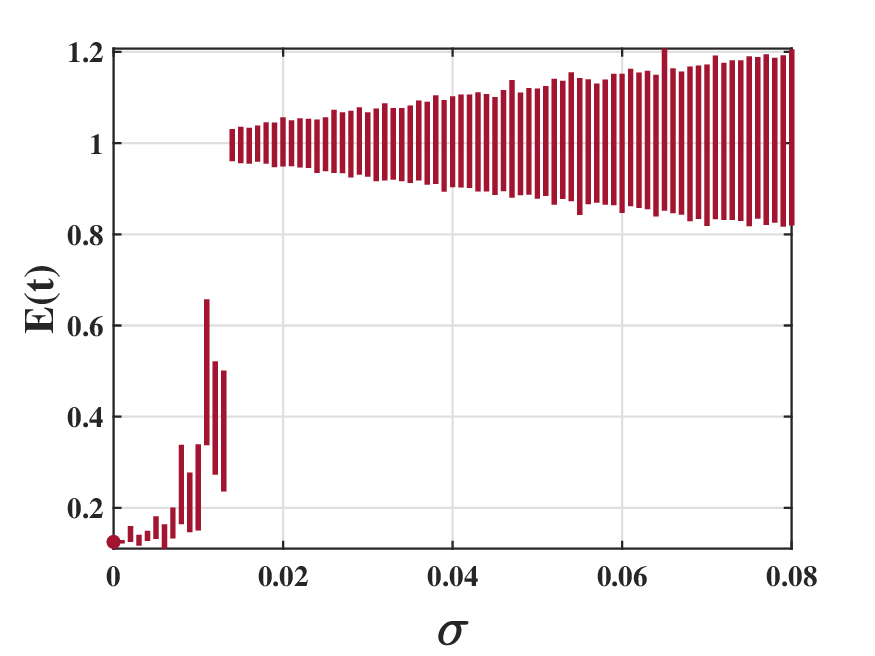}}
\caption{Dispersion of random states of stochastic model (\ref{Smodel}) with initial value $\Xi_2=(0,0,0.8724,0.1253)$ for different noise intensities.}
\label{fig9}
\end{figure}

Fig. \ref{fig9} illustrates the dispersion of the random states $I(t)$ and $E(t)$ in model (\ref{Smodel}) as a function of noise intensity $\sigma$. In the figure, the random state $I(t)$ is shown in yellow and $E(t)$ in red. A critical noise intensity can be identified, approximately $\sigma \approx 0.014$. When $\sigma$ is below this threshold, the dispersion of random states around the immune surveillance state $\Xi_2$ increases monotonically with $\sigma$, and the stochastic trajectories remain confined within the basin of attraction of $\Xi_2$. In contrast, once $\sigma$ exceeds $0.014$, the dispersion of random states shifts to being distributed around the immune escape state $\Xi_1$. In this regime, the stochastic trajectories leave the basin of attraction of $\Xi_2$ and enter that of $\Xi_1$. Consequently, the dimensionless immunogenic cell density $I(t)$ decreases markedly, and the stochastic trajectories exhibit strong oscillations around the immune escape state $\Xi_1$.

\subsubsection{The tipping probability and time from the immune surveillance state to the immune escape dominant state}

The preceding analysis indicates that an appropriate noise intensity can induce transitions of the system from the immune surveillance state $\Xi_2$ to the immune escape dominant state $\Xi_1$. From a biological perspective, this raises a natural question regarding the regulatory role of environmental stochasticity in driving such transitions. To rigorously characterize these dynamics, we employ the concepts of tipping time and tipping probability introduced by Feng et al. \cite{feng2024tipping} for critical transitions in nonlinear systems.

We simulate the stochastic model (\ref{Smodel}) using the Euler-Maruyama scheme with a time step $\triangle t=10^{-3}$. The tipping probability is defined as the likelihood that a trajectory, initialized within the attraction domain of $\Xi_2$, crosses the separatrix for the first time during the interval $[0,T]$, and subsequently enters the attraction domain of $\Xi_1$. It is estimated by the frequency-based relation
\begin{equation}\label{tipping}
P_{tipping}=\frac{n}{M},
\end{equation}
where $M=1000$ denotes the number of independent realizations, $n$ is the number of realizations for which immunogenic cell extinction ($I=0$) occurs, and $T=10^5$. The corresponding time elapsed during such transitions is referred to as the tipping time.

\begin{figure}[htb]
    \centering
\subfigure{\includegraphics[width=0.5\linewidth]{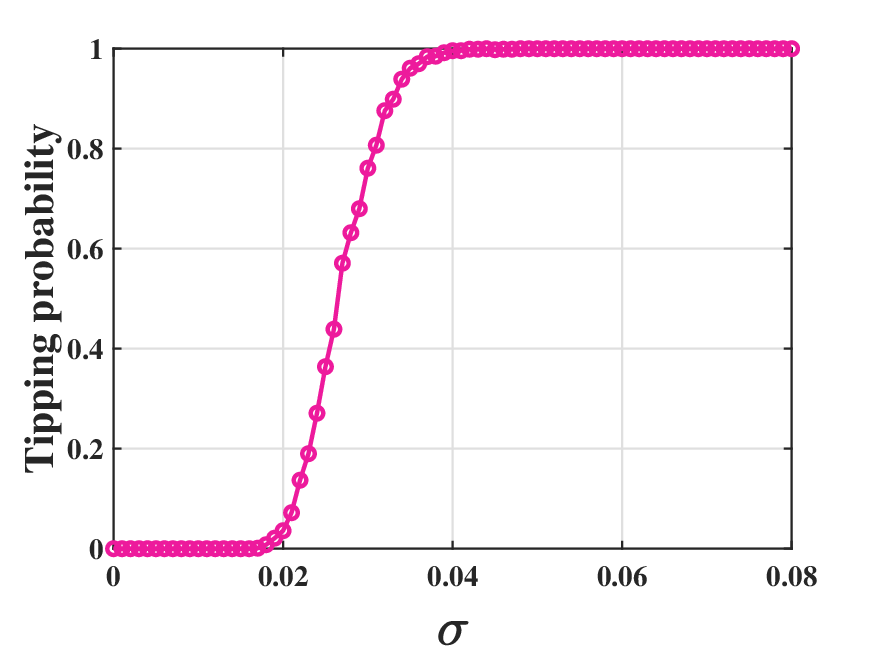}}
    \caption{Tipping probability from the immune surveillance state $\Xi_2$ to the immune escape dominant state $\Xi_1$ of the stochastic system (\ref{Smodel}) under different noise intensities.}
    \label{fig10}
\end{figure}

\begin{figure}[htb]
    \centering
\subfigure{\includegraphics[width=0.48\linewidth]{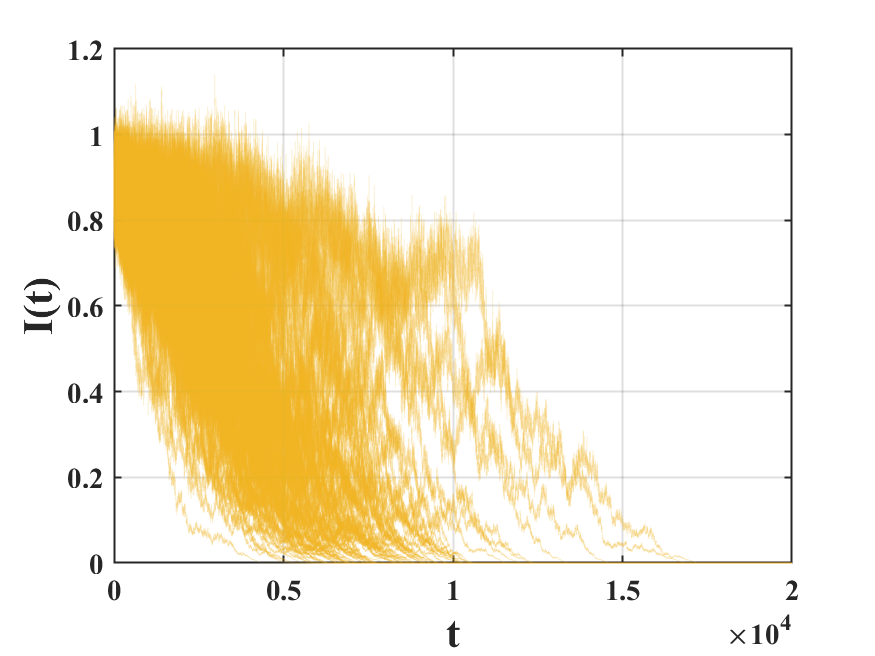}}
\subfigure{\includegraphics[width=0.48\linewidth]{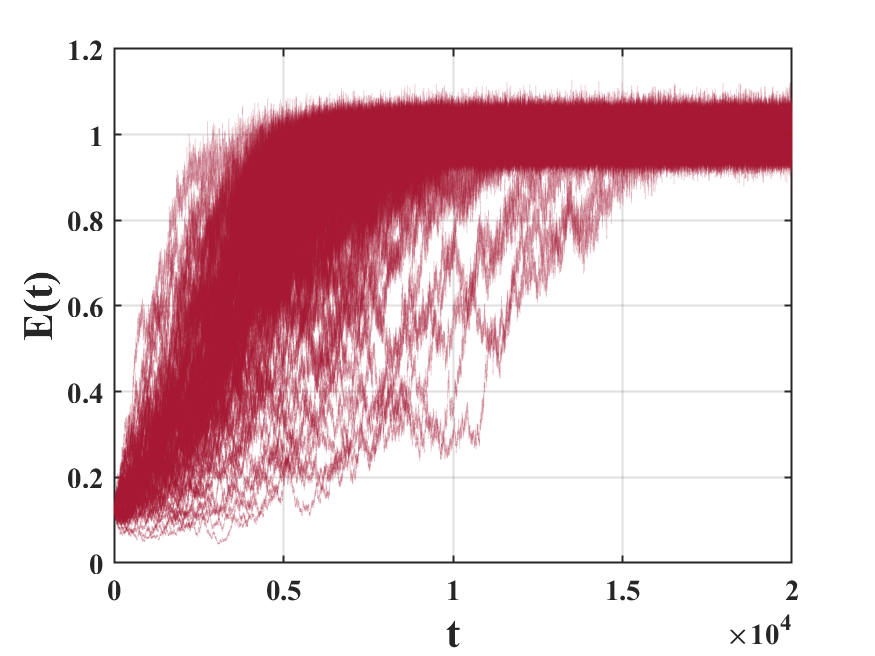}}
    \caption{100 trajectories of variables $I$ and $E$ for the stochastic model (\ref{Smodel}) with noise intensity $\sigma = 0.044$.}
    \label{fig11}
\end{figure}

The results of the previous section show that when the noise intensity is relatively weak, the stochastic model (\ref{Smodel}) does not exhibit noise-induced tipping. To further investigate, we compute the tipping probability under different noise intensities using (\ref{tipping}). As shown in Fig. \ref{fig10}, the tipping probability approaches $1$ once $\sigma \geq 0.044$. Taking $\sigma = 0.044$ as an illustrative case, Fig. \ref{fig11} presents the trajectories of $I(t)$ and $E(t)$ across 100 independent simulations. Consequently, when examining the relationship between noise intensity and average tipping time, we restrict attention to $\sigma \geq 0.044$ to ensure that tipping events occur with certainty.

\begin{figure}[htb]
    \centering
\subfigure{\includegraphics[width=0.7\linewidth]{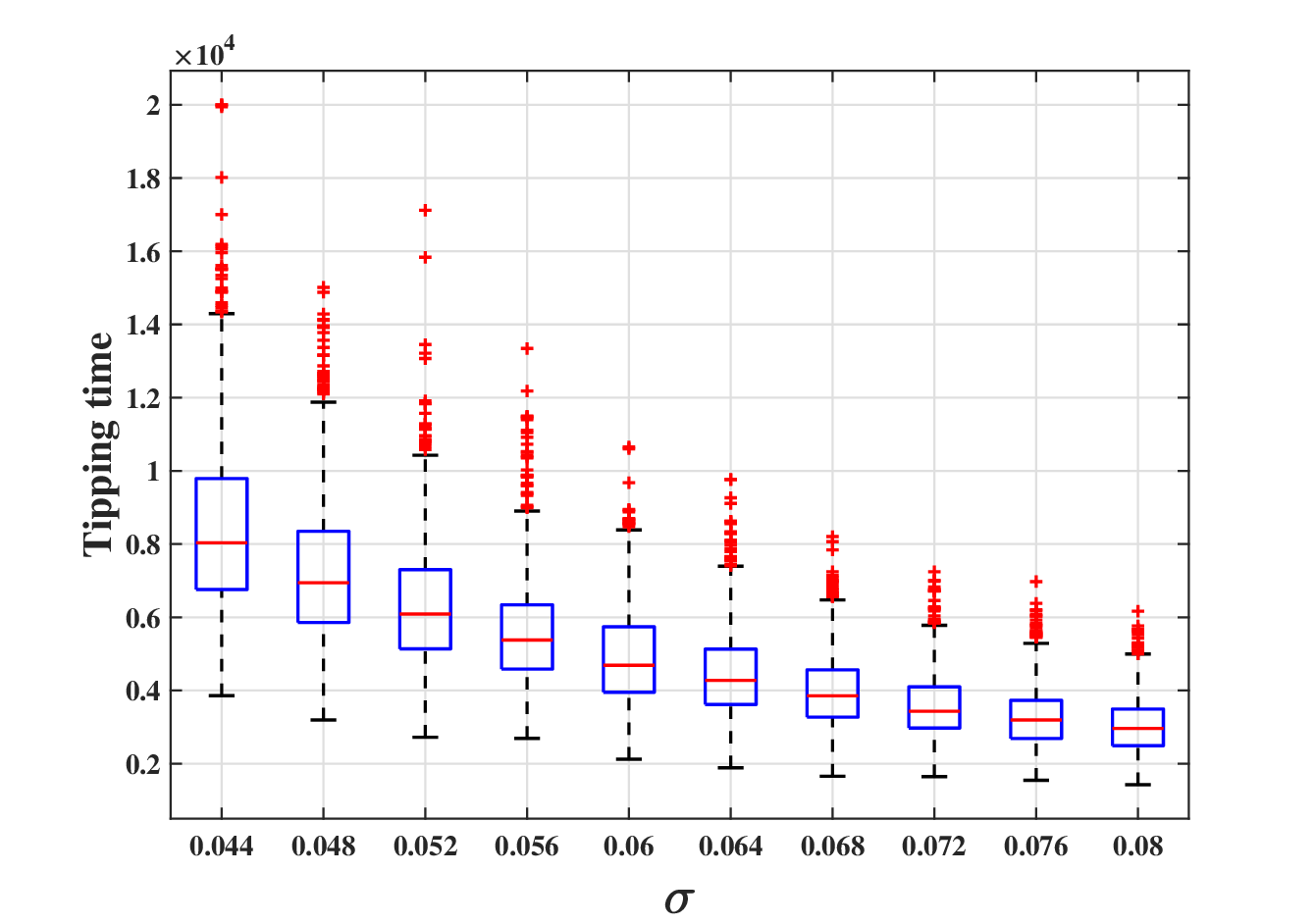}}
    \caption{Tipping time from the immune surveillance state $\Xi_2$ to the immune escape dominant state $\Xi_1$ of the stochastic system (\ref{Smodel}) under different noise intensities. The red horizontal line in the box-plot indicates the median of the tipping time at each noise intensity, and the height of the box represents the 25th and 75th quartiles of the tipping time distribution, respectively. The red symbol ``+" indicates outliers.}
    \label{fig12}
\end{figure}

Numerically, to characterize the tipping time from the steady state $\Xi_2$ (immune surveillance state) to the steady state $\Xi_1$ (immune escape dominant state) in the stochastic model, we define the tipping time as the duration required for the variable $I$ to first enter a prescribed small neighborhood of $I = 0$, starting from the initial condition. Fig. \ref{fig12} illustrates the relationship between the tipping time and the noise intensity based on 1000 simulations.
Preliminary observations indicate that as the noise intensity increases, the tipping occurs faster and the tipping time becomes shorter. At relatively low levels of environmental stochasticity, the tipping time decreases rapidly in the early stage, accompanied by marked fluctuations. In addition, when the disturbance intensity exceeds a certain threshold, the sensitivity of the tipping time to further variations in $\sigma$ diminishes. In other words, as the noise intensity grows, the average tipping time changes from a steep decline to a more gradual flattening trend.


\subsubsection{Nonlinear regulation of tipping time and critical noise intensity by mortality \texorpdfstring{$d$}{d} and immune pressure \texorpdfstring{$m$}{m}}

\begin{figure}[htb]
    \centering
\subfigure[]{\includegraphics[width=0.48\linewidth]{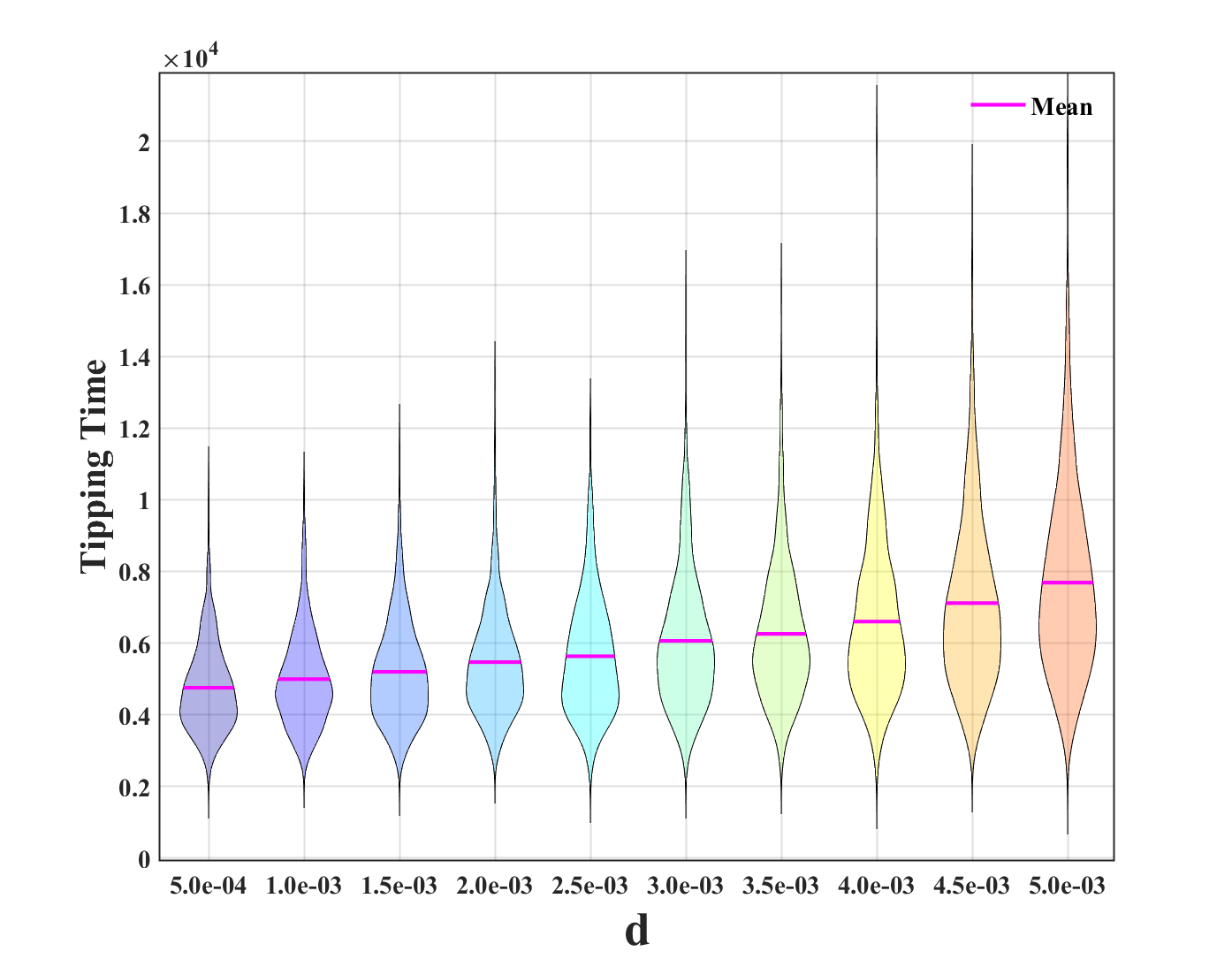}}
\subfigure[]{\includegraphics[width=0.48\linewidth]{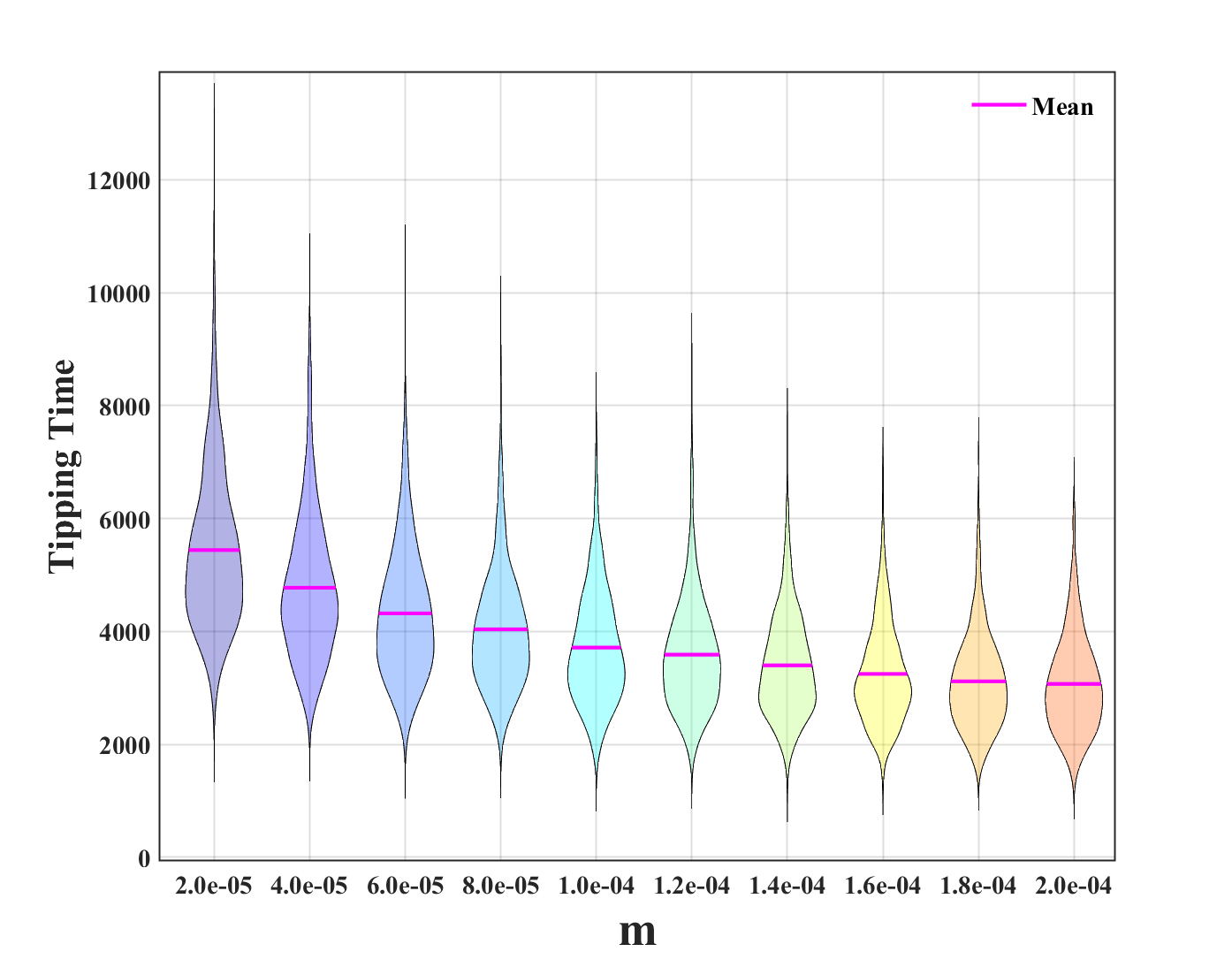}}

    \caption{Tipping time from the immune surveillance state $\Xi_2$ to the immune escape dominant state $\Xi_1$ for the stochastic system (\ref{Smodel}) under a fixed noise intensity $\sigma=0.06$. (a) The relationship between different cell mortality rates and tipping times, where $d_N=d_A=d_I=d_E=d$.  (b) The relationship between immune system-mediated cell death $m$ and tipping time. And other parameters are shown in Table \ref{smtab}. }
    \label{fig13}
\end{figure}

As shown in Fig. \ref{fig13}, for a fixed noise intensity $\sigma = 0.06$, increasing the mortality rate of all four cell types ($d_N, d_A, d_I, d_E$) significantly prolongs the time required for the system to enter the immune escape dominant state $\Xi_1$. In contrast to conventional expectations, enhancing immune system-mediated killing ($m$) accelerates the escape process (Fig. \ref{fig13}(b)). These results reveal a therapeutic paradox: under persistent environmental perturbations in the tumor microenvironment, simply increasing immune pressure may produce counterproductive effects. By comparison, chemotherapy (modeled as elevated $d$) delays escape by expanding the therapeutic window.

\begin{figure}[htb]
    \centering
\subfigure[]{\includegraphics[width=0.48\linewidth]{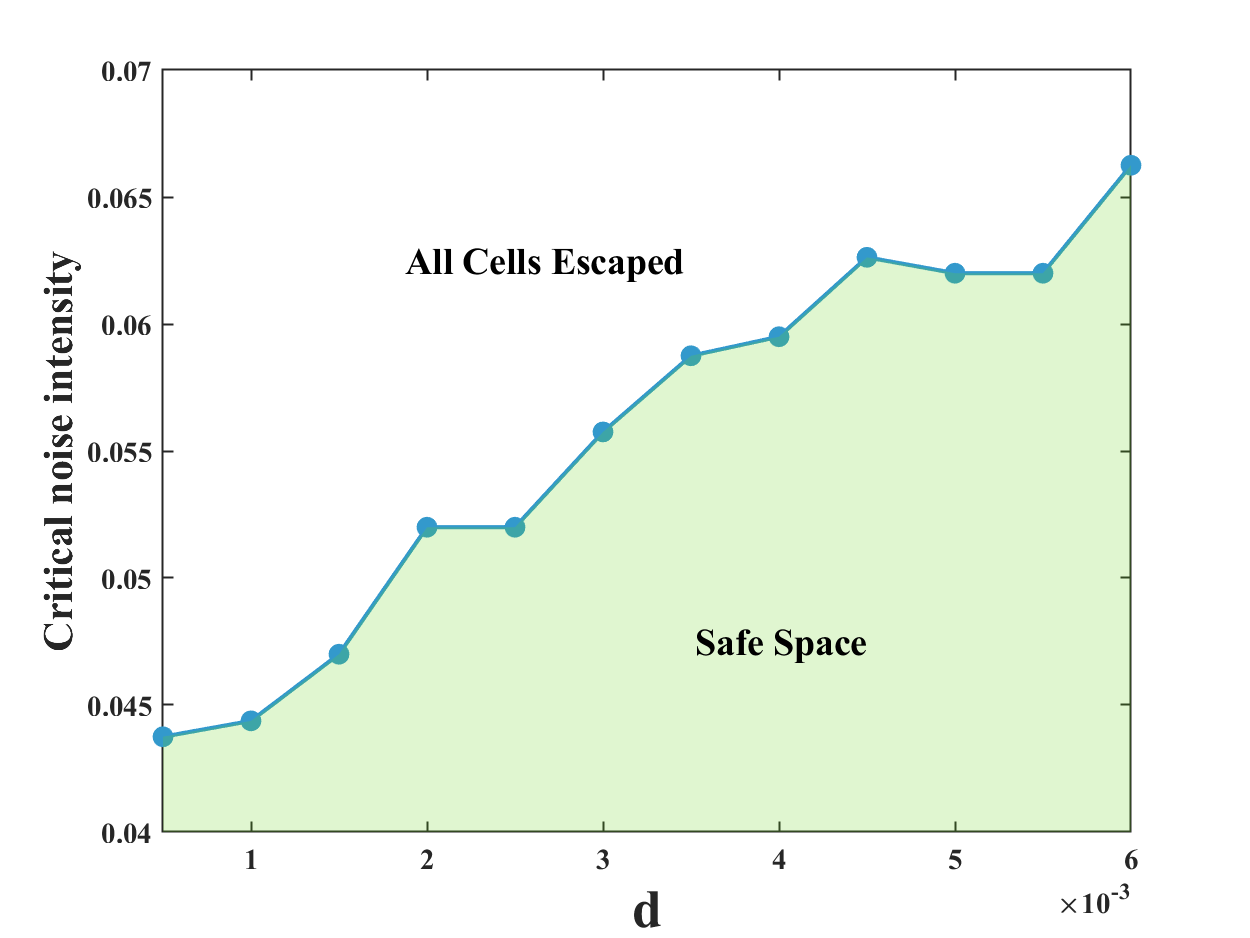}}
\subfigure[]{\includegraphics[width=0.48\linewidth]{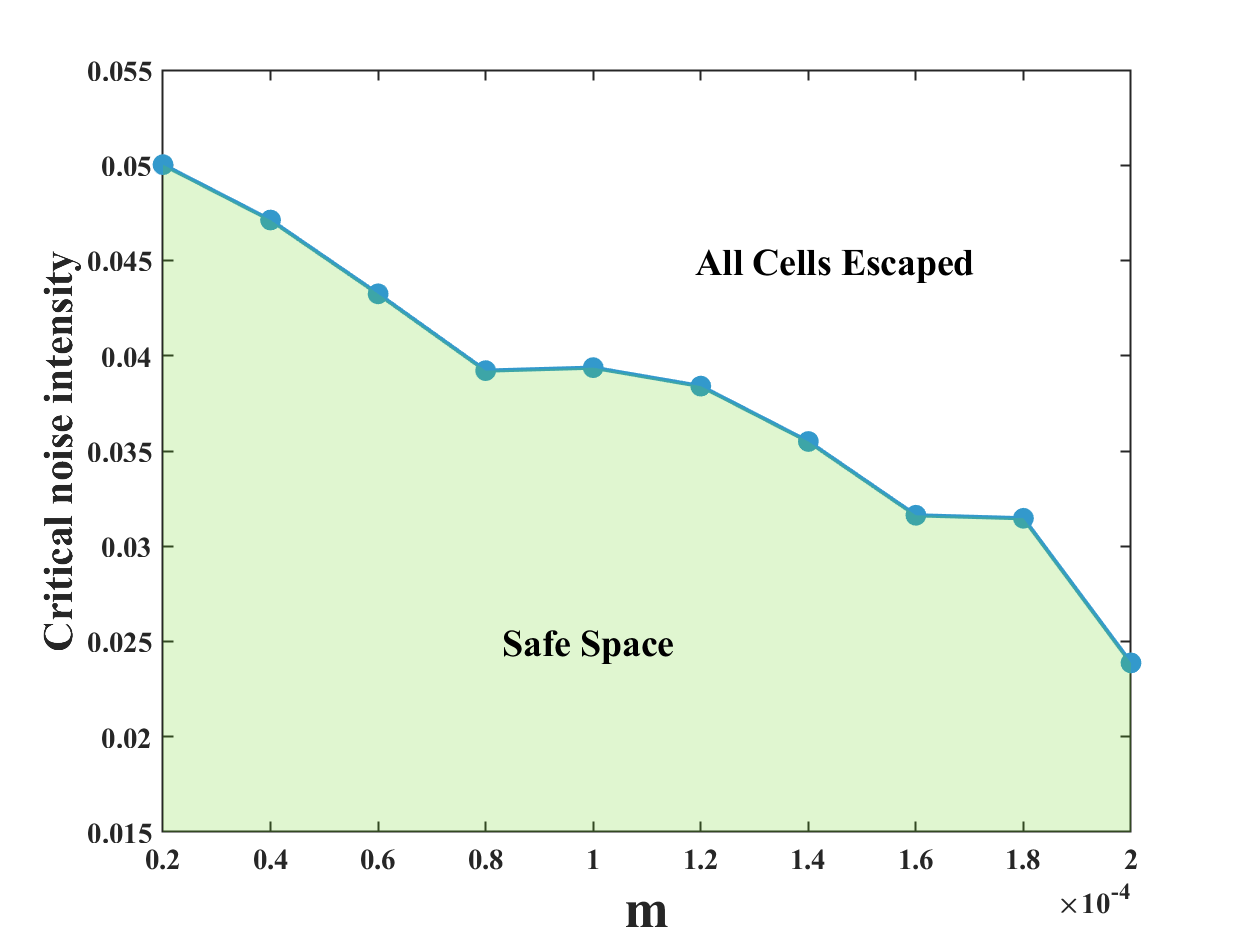}}
    \caption{Critical noise intensity and safe space for the transition from immune surveillance state $\Xi_2$ to immune escape dominant state $\Xi_1$ in the stochastic system (\ref{Smodel}). (a) The relationship between different cell mortality rates and critical noise intensity, where $d_N=d_A=d_I=d_E=d$.  (b) The relationship between immune system-mediated cell death and critical noise intensity. And other parameters are shown in Table \ref{smtab}. }
    \label{fig14}
\end{figure}

Fig. \ref{fig14} further characterizes how mortality ($d$) and immune killing power ($m$) regulate the critical noise intensity required for transition into the immune escape dominant state $\Xi_1$. The light green shaded region defines the safety zone, whose boundary (fold line) marks the phase transition threshold from immune surveillance ($\Xi_2$) to immune escape ($\Xi_1$). This zone also represents the stability domain of $\Xi_2$.

Simulation results indicate that higher cell mortality $d$ increases the minimum noise strength required for critical transition (Fig. \ref{fig14}(a)), thereby broadening the safety zone and enhancing robustness against stochastic perturbations. Conversely, immune killing intensity $m$ exhibits a negative correlation with the critical noise intensity (Fig. \ref{fig14}(b)): as $m$ increases, the safe zone contracts, making the system more susceptible to noise-induced escape.

This nonlinear interplay highlights a therapeutic trade-off. Chemotherapy (increasing $d$) stabilizes the system by expanding the safety zone, though at the cost of non-specific immune cell depletion. Immunotherapy (increasing $m$) strengthens direct immune suppression of tumor cells but, under noisy microenvironments, paradoxically compresses the safety zone and accelerates escape.

\section{Discussion}


We proposed and analyzed a dynamical model to investigate the coupled evolution of tumor antigenicity and immune escape, under both deterministic and stochastic conditions. By introducing two orthogonal axes of phenotypic variation—antigen expression and immune evasion—we derived a minimal yet biologically motivated framework that captures core tumor–immune dynamics through nonlinear interactions among four distinct tumor subpopulations. While inspired by cancer immunology, the structure of the model reflects a broader class of systems characterized by trait-driven selection and nonlinear feedback, such as eco-evolutionary predator–prey systems with evolving defense traits.

Our theoretical analysis identified up to five biologically relevant equilibria and established the conditions for their existence and stability. We demonstrated that the system undergoes multiple bifurcations, including transcritical and saddle-node bifurcations, with respect to key biological parameters such as antigen mutation rate ($p$), antigen accumulation rate ($s$), and immune escape rate ($q$). Of particular importance is the emergence of bistability between an immune surveillance state and an immune escape state. This bistability offers a generic dynamical explanation for divergent immune outcomes observed in tumor progression and reflects a broader class of systems in which state-dependent feedback and nonlinear conversion rates give rise to multistability.

We further characterized the separatrix dividing the basins of attraction of the two stable equilibria using the stable manifold of the saddle point. This analysis revealed how initial conditions, particularly the relative proportions of immunogenic and immune-escaped cells, dictate long-term outcomes. Such sensitivity to initial conditions underpins the observed heterogeneity in tumor-immune responses and provides a mechanistic justification for therapeutic strategies that shift tumor composition toward immunogenic phenotypes. For instance, immunotherapies such as checkpoint blockade can be viewed as perturbations that reshape the system's initial condition or parameter regime to favor immune surveillance \cite{wang2025autophagy,auslander2018robust}.

To model treatment-induced perturbations, we studied the effect of increasing tumor cell mortality—representing cytotoxic chemotherapy—on the steady-state tumor composition. Our results show that elevated death rates favor immunogenic subpopulations over immune-escaped ones, effectively shifting the system toward the immune surveillance basin. This finding supports the hypothesis that chemotherapy may have immunostimulatory effects, and it provides a theoretical foundation for the synergy observed in chemo-immunotherapy combinations \cite{sakai2013effects,pol2015trial}.

Beyond deterministic dynamics, we examined how stochastic fluctuations in the tumor microenvironment affect the stability and transitions between immune states. Using numerical simulations, we quantified noise-induced tipping events and estimated the critical noise intensity that triggers escape from immune surveillance. We found that even in parameter regimes where immune control is deterministically stable, stochastic perturbations can induce transitions to immune escape. This behavior aligns with metastability phenomena commonly studied in stochastic dynamical systems and highlights the importance of accounting for microenvironmental variability in treatment design.

Interestingly, we observed that the tipping time—the expected time for transition from surveillance to escape—decreases with increasing noise intensity but is also modulated by the immune killing rate and cell death rate. Higher cell death delays escape, whereas increased immune clearance paradoxically accelerates tipping under noise. To systematically quantify robustness, we defined a noise-dependent “safe zone” in parameter space. Within this zone, the system remains resilient to stochastic perturbations. Our results suggest a compensatory mechanism: chemotherapy-like death effects expand this safe region, while intensified immune pressure may destabilize it—revealing the dual role of immunotherapy as both a control mechanism and a source of dynamical fragility.

To facilitate mathematical tractability, our model adopts a coarse-grained structure that aggregates complex immunological processes into key compartments and transitions. Despite these simplifications, the framework provides analytical access to nonlinear behaviors such as bistability, bifurcation cascades, and stochastic transitions. From a modeling standpoint, this study contributes to the theoretical understanding of trait-mediated population dynamics with feedback-driven transitions between metastable states. From an application perspective, it offers testable predictions about the conditions under which immune escape may be reversed or delayed, thereby providing a mathematical foundation for rational intervention strategies in cancer immunotherapy.

\bibliographystyle{siamplain}
\bibliography{references}
\end{document}


\maketitle

\section{Proof of Theorem \ref{thm1}}\label{positive}

Before proving Theorem \ref{thm1}, we first state a significant lemma.
\begin{lemma}\label{lemma1}
    The positive cone $\mathbb{R}_+^n$ remains invariant under the flow induced by the differential equation,  $\frac{dv}{dt} = f(v)$,  if and only if the function $f(v)$ is quasi-positive. This means that for each $i = 1, 2, 3, \cdots, n$, the inequality  $$f_i (v_1, \dots , 0, \dots, v_n ) \geq 0$$  must hold, where the $i$-th entry is zero, and $v_j \geq 0$ for all $j \ne i$.
\end{lemma}
We will now begin to prove Theorem \ref{thm1}.
\begin{proof}
    By applying Lemma \ref{lemma1} to the model (\ref{nonmodel}), we obtain for $n=4$ and 

    \begin{align*}
    f_1(0,A,I,E)&=0,\quad
    f_2(N,0,I,E)=pN,\\
    f_3(N,A,0,E)&=sA,\quad
    f_4(N,A,I,0)=q(N+A+I).
    \end{align*}

    Due to all parameter values are positive and nonnegative initial conditions, it imply that all $f_i\geq 0$. Therefore, according to Lemma \ref{lemma1}, we can show that all the solutions of system (\ref{nonmodel}) remain positive for any $t\geq 0$.

    Define $V(t)=N(t)+A(t)+I(t)+E(t)$, then

    \begin{align*}
     \frac{d V(t)}{d t}=&(r_NN+r_AA+r_II+r_EE)\left[1-(N+A+I+E)\right]-d_NN-d_AA-d_II\\
     &-d_EE-\frac{mI}{\theta +I}\\  
     \leq &\frac{r_N+r_A+r_I+r_E}{4}-\min\{d_N,d_A,d_I,d_E\}V\\
     :=&\Lambda- \varrho V.
    \end{align*}

Employing the comparison theorem, we can yield that $$V(t) \leq V(0)e^{-\varrho t} + \frac{\Lambda}{\varrho}(1-e^{-\varrho t}).$$ Thus, all the solutions of the system (\ref{nonmodel}) with initial value $(N(0),A(0),I(0),E(0))\in \mathbb{R}_+^4$ are confined to the invariant set $\Gamma$.

\end{proof}

\section{The proof of Theorem \ref{sxi0}}\label{psxi0}
\begin{proof}
   The Jacobi matrix of system (\ref{nonmodel}) at $\Xi_0$ is 
   \begin{eqnarray*}
J|_{\Xi_0}=\begin{pmatrix}
r_N-d_N-p-q & 0  & 0  & 0\\
p & r_A-d_A-q-s  & 0  & 0\\
0 & s  & r_I-d_I-q-\frac{m}{\theta}  & 0\\
q & q  & q  & r_E-d_E\\
\end{pmatrix},
\end{eqnarray*}
The eigenvalues of the matrix $J|_{\Xi_0}$ are
\begin{align*}
    &\lambda_1=r_N-d_N-p-q,~~\lambda_2=r_A-d_A-q-s,\\
    &\lambda_3=r_I-d_I-q-\frac{m}{\theta},~~\lambda_4=r_E-d_E. 
\end{align*}
Therefore, the equilibrium $\Xi_0$ is locally asymptotically stable if condition ($H_1$) holds.
\end{proof}

\section{The proof of Theorem \ref{sxi1}}\label{psxi1}
\begin{proof}
   The Jacobi matrix of system (\ref{nonmodel}) at $\Xi_1$ is 
   \begin{eqnarray*}
J|_{\Xi_1}=\begin{pmatrix}
r_N\frac{d_E}{r_E}-d_N-p-q & 0  & 0  & 0\\
p & r_A\frac{d_E}{r_E}-d_A-q-s  & 0  & 0\\
0 & s  & r_I\frac{d_E}{r_E}-d_I-q-\frac{m}{\theta}  & 0\\
-r_E+d_E+q & -r_E+d_E+q  & -r_E+d_E+q  & -r_E+d_E\\
\end{pmatrix},
\end{eqnarray*}
the eigenvalues of the matrix $J|_{\Xi_1}$ are
\begin{align*}
    &\lambda_1=r_N\frac{d_E}{r_E}-d_N-p-q,~~\lambda_2=r_A\frac{d_E}{r_E}-d-A-q-s,\\
    &\lambda_3=r_I\frac{d_E}{r_E}-d_I-q-\frac{m}{\theta},~~\lambda_4=-r_E+d_E<0. 
\end{align*}
Therefore, the equilibrium $\Xi_1$ is locally asymptotically stable if condition ($H_2$) holds.
\end{proof}

\section{The proof of Theorem \ref{sxi2}}\label{psxi2}

\begin{proof}
   The Jacobi matrix of system (\ref{nonmodel}) at $\Xi_2$ is 
   \begin{eqnarray*}
J|_{\Xi_2}=\begin{pmatrix}
r_N\left(1-I_2-E_2\right)-d_N-p-q & 0  & 0  & 0\\
p & r_A\left(1-I_2-E_2\right)-d_A-q-s  & 0  & 0\\
-r_II_2 & -r_II_2+s  & -r_II_2+\frac{mI_2}{(\theta+I_2)^2}  & -r_II_2\\
-r_EE_2+q & -r_EE_2+q  & -r_EE_2+q  & -\frac{qI_2}{E_2}-r_EE_2\\
\end{pmatrix}.
\end{eqnarray*}
Therefore, the characteristic equation for the matrix $J|_{\Xi_2}$ is
\begin{equation}\label{CE2}
    (\lambda-\lambda_1)(\lambda-\lambda_2)(\lambda^2+\Upsilon_{1\Xi_{2}}\lambda+\Upsilon_{2\Xi_{2}})=0,
\end{equation}
where
\begin{align*}
    &\lambda_1=r_N\left(1-I_2-E_2\right)-d_N-p-q,~~\lambda_2=r_A\left(1-I_2-E_2\right)-d_A-q-s,\\
    &\Upsilon_{1\Xi_{2}}= r_II_2+\frac{qI_2}{E_2}+r_EE_2-\frac{m I_2}{(\theta+I_2)^2},\\
    &\Upsilon_{2\Xi_{2}}= qr_II_2\frac{I_2+E_2}{E_2}-\frac{mI_2}{(\theta+I_2)^2}\left(\frac{qI_2}{E_2}+r_EE_2\right).
\end{align*}
If condition ($H_3$) holds, then the eigenvalues of equation (\ref{CE2}) have negative real parts, and the equilibrium $\Xi_2$ is locally asymptotically stable.
\end{proof}

\section{The number of positive roots of equation (\ref{E3})}\label{Xi3}
In this section, we discuss in detail the number of positive roots of equation (\ref{E3}) for different combinations of parameters.

(i) When condition $\mathcal{W}>\frac{s^2\theta}{\mathcal{U}}$, $\mathcal{X}>0$, $\mathcal{V}<0$, and $\theta \mathcal{W}+\mathcal{U}+2\sqrt{\theta\mathcal{WU}}<m<(\mathcal{W}+s)(\theta+\frac{\mathcal{U}}{s})$ hold, equation (\ref{E3}) has two positive roots and satisfies $E_3 > 0$ and $I_3 > 0$. That is, under these conditions, the system has two antigenically neutral cells free equilibrium $\Xi_{31}=(0,A_{31},I_{31},E_3)$ and $\Xi_{32}=(0,A_{32},I_{32},E_3)$, where $E_3=\frac{q\mathcal{X}}{r_A(q-\mathcal{V})}$, 
$$
A_{31}=\frac{ \mathcal{U} }{s}-I_{31},~~ I_{31}=\frac{(m-\theta\mathcal{W}-\mathcal{U})-\sqrt{(m-\theta\mathcal{W}-\mathcal{U})^2-4\theta\mathcal{W}\mathcal{U}}}{2\mathcal{W}}, 
$$
$$
A_{32}=\frac{ \mathcal{U} }{s}-I_{32},~~I_{32}=\frac{(m-\theta\mathcal{W}-\mathcal{U})+\sqrt{(m-\theta\mathcal{W}-\mathcal{U})^2-4\theta\mathcal{W}\mathcal{U}}}{2\mathcal{W}}.
$$

(ii)  When condition $\mathcal{W}> 0$, $\mathcal{X}>0$, $\mathcal{V}<0$, and $m>(\mathcal{W}+s)(\theta+\frac{\mathcal{U}}{s})$ hold, equation (\ref{E3}) has a unique positive root and satisfies $E_3 > 0$ and $I_3 > 0$. That is, under these conditions, the system has a unique antigenically neutral cells free equilibrium $\Xi_{31}=(0,A_{31},I_{31},E_3)$, where  $A_{31}=\frac{ \mathcal{U} }{s}-I_{31},~ I_{31}=\frac{(m-\theta\mathcal{W}-\mathcal{U})-\sqrt{(m-\theta\mathcal{W}-\mathcal{U})^2-4\theta\mathcal{W}\mathcal{U}}}{2\mathcal{W}}$, and $E_3=\frac{q\mathcal{X}}{r_A(q-\mathcal{V})}$.

(iii) When condition $\mathcal{W}<0$, $\mathcal{X}>0$, $\mathcal{V}<0$, and $m>(\mathcal{W}+s)(\theta+\frac{\mathcal{U}}{s})$ hold, equation (\ref{E3}) has a unique positive root and satisfies $E_3 > 0$ and $I_3 > 0$. That is, under these conditions, the system has a unique antigenically neutral cells free equilibrium $\Xi_{31}=(0,A_{31},I_{31},E_3)$, where  $A_{31}=\frac{ \mathcal{U} }{s}-I_{31},~ I_{31}=\frac{(m-\theta\mathcal{W}-\mathcal{U})-\sqrt{(m-\theta\mathcal{W}-\mathcal{U})^2-4\theta\mathcal{W}\mathcal{U}}}{2\mathcal{W}}$, and $E_3=\frac{q\mathcal{X}}{r_A(q-\mathcal{V})}$.

(iv) When consitions $\mathcal{W}<-\frac{s^2\theta}{\mathcal{U}}$, $\mathcal{X}>0$, $\mathcal{V}<0$, and $m=\theta\mathcal{W}+\mathcal{U}>0$ hold, equation (\ref{E3}) has a unique positive root and satisfies $E_3 > 0$ and $I_3 > 0$. That is, under these conditions, the system has a unique antigenically neutral cells free equilibrium $\Xi_{31}=(0,A_{31},I_{31},E_3)$, where  $A_{31}=\frac{ \mathcal{U} }{s}-I_{31},~ I_{31}=\sqrt{-\frac{\theta \mathcal{U}}{\mathcal{W}}}$, and $E_3=\frac{q\mathcal{X}}{r_A(q-\mathcal{V})}$.

(v) When consitions $\mathcal{W}=0$, $\mathcal{X}>0$, $\mathcal{V}<0$, and $m>s\theta+\mathcal{U}$ hold, equation (\ref{E3}) has a unique positive root and satisfies $E_3 > 0$ and $I_3 > 0$. That is, under these conditions, the system has a unique antigenically neutral cells free equilibrium $\Xi_{31}=(0,A_{31},I_{31},E_3)$, where  $A_{31}=\frac{ \mathcal{U} }{s}-I_{31},~ I_{31}=\frac{\theta \mathcal{U}}{m-\mathcal{U}}$, and $E_3=\frac{q\mathcal{X}}{r_A(q-\mathcal{V})}$.

(vi) When consitions $\mathcal{W}>\frac{s^2\theta}{\mathcal{U}}$, $\mathcal{X}>0$, $\mathcal{V}<0$, and $m=\theta\mathcal{W}+\mathcal{U}+2\sqrt{\theta\mathcal{WU}}$ hold, equation (\ref{E3}) has a unique positive root and satisfies $E_3 > 0$ and $I_3 > 0$. That is, under these conditions, the system has a unique antigenically neutral cells free equilibrium $\Xi_{31}=(0,A_{31},I_{31},E_3)$, where  $A_{31}=\frac{ \mathcal{U} }{s}-I_{31},~ I_{31}=\sqrt{\frac{\theta \mathcal{U}}{\mathcal{W}}}$, and $E_3=\frac{q\mathcal{X}}{r_A(q-\mathcal{V})}$.

(vii) The system does not exist equilibrium $\Xi_3=(0,A_3,I_3,E_3)$ for all conditions except for the three parameter combination cases mentioned above.

\section{The proof of Theorem \ref{sxi3}}\label{psxi3}
\begin{proof}
    We calculate the Jacobi matrix evaluated around steady state $\Xi_3$, which is expressed as
\begin{eqnarray*}
J|_{\Xi_3}=
\begin{pmatrix}
r_N\frac{d_A+q+s}{r_A}-d_N-p-q & 0  & 0  & 0\\
-r_AA_3+p & -r_AA_3  & -r_AA_3  & -r_AA_3\\
-r_II_3 & -r_II_3+s  & -r_II_3-s\frac{A_3}{I_3}+\frac{mI_3}{(\theta+I_3)^2}  & -r_II_3\\
-r_EE_3+q & -r_EE_3+q  & -r_EE_3+q  & -\frac{q(A_3+I_3)}{E_3}-r_EE_3\\
\end{pmatrix}.
\end{eqnarray*}
Therefore, the characteristic equation for the matrix $J|_{\Xi_3}$ is

\begin{equation}\label{CE3}
    (\lambda-\lambda_1)(\lambda^3+\Upsilon_{1\Xi_{3}}\lambda^2+\Upsilon_{2\Xi_{3}}\lambda+\Upsilon_{3\Xi_{3}})=0,
\end{equation}
where
\begin{equation}\label{T}
\begin{aligned}
    \lambda_1=&r_N\frac{d_A+q+s}{r_A}-d_N-p-q,\\
    \Upsilon_{1\Xi_{3}}=& r_AA_3-\frac{mI_3}{(\theta+I_3)^2}+\frac{sA_3}{I_3}+r_II_3+\frac{q(A_3+I_3)}{E_3}+r_EE_3,\\
    \Upsilon_{2\Xi_{3}}=& -\frac{mI_3}{(\theta+I_3)^2}\left(r_AA_3+\frac{q(A_3+I_3)}{E_3}+r_EE_3\right)+r_AA_3\left(\frac{sA_3}{I_3}+\frac{q(A_3+I_3)}{E_3}+q\right.\\
    &\left.+s\right)+\frac{sA_3}{I_3}\left(\frac{q(A_3+I_3)}{E_3}+r_EE_3\right)+r_II_3\left(\frac{q(A_3+I_3)}{E_3}+q\right),\\
    \Upsilon_{3\Xi_{3}}=&r_AA_3\left(s\frac{A_3}{I_3}-\frac{mI_3}{(\theta+I_3)^2}\right)\left(\frac{q(A_3+I_3)}{E_3}+q\right)+\frac{r_AqsA_3(A_3+I_3)}{E_3}+sqr_AA_3 .
\end{aligned}
\end{equation}
Hence, According to the Routh-Hurwitz criterion, we know that all roots of (\ref{CE3}) have negative real parts if and only if the conditions $\Upsilon_{1\Xi_{3}},\Upsilon_{3\Xi_{3}}>0$, and $\Upsilon_{1\Xi_{3}}\Upsilon_{2\Xi_{3}}>\Upsilon_{3\Xi_{3}}$ are satisfied. 
That is, the equilibrium $\Xi_3$ is locally asymptotically stable under the conditions $\Upsilon_{1\Xi_{3}},\Upsilon_{3\Xi_{3}}>0$, and $\Upsilon_{1\Xi_{3}}\Upsilon_{2\Xi_{3}}>\Upsilon_{3\Xi_{3}}$.
\end{proof}

\section{The number of positive roots of equation (\ref{E*})}\label{xi*}
In this section, we discuss in detail the number of positive roots of equation (\ref{E*}) for different combinations of parameters.

(i) When condition $\mathcal{P}>\frac{s\theta\bar{G}}{\bar{H}}$, $\mathcal{Q}>$, $\mathcal{R}>0$, $\mathcal{S}>0$, and $\theta \mathcal{P}+s\bar{G}\bar{H}+2\sqrt{s\theta\mathcal{P}\bar{G}\bar{H}}<m<(\mathcal{P}+s\bar{G})(\theta+\bar{H})$ hold, equation (\ref{E*}) has two positive roots and satisfies $N^*>0$, $A^* > 0$ and $E^* > 0$. That is, under these conditions, the system has two coexistence equilibrium $\Xi^*_1=(N^*_1,A^*_1,I^*_1,E^*)$ and $\Xi^*_2=(N^*_2,A^*_2,I^*_2,E^*)$, where $E^*=\frac{q\mathcal{R}}{r_N(q+\mathcal{S})}$, 
\begin{align*}
&N^*_i=\frac{\mathcal{Q} }{p}A^*_i,~~ A^*_i=\bar{G}(\bar{H}-I^*_i)  ~~~(i=1,2),\\
&I^*_1=\frac{(m-\theta\mathcal{P}-s\bar{G}\bar{H})-\sqrt{(m-\theta\mathcal{P}-s\bar{G}\bar{H})^2-4s\theta \mathcal{P}\bar{G}\bar{H}}}{2\mathcal{P}}, \\
&I^*_2=\frac{(m-\theta\mathcal{P}-s\bar{G}\bar{H})+\sqrt{(m-\theta\mathcal{P}-s\bar{G}\bar{H})^2-4s\theta \mathcal{P}\bar{G}\bar{H}}}{2\mathcal{P}}.
\end{align*}

(ii) When condition $\mathcal{P}>0$, $\mathcal{Q}>0$, $\mathcal{R}>0$, $\mathcal{S}>0$, and $m>(\mathcal{P}+s\bar{G})(\theta+\bar{H})$ hold, equation (\ref{E*}) has a unique positive roots and satisfies $N^*>0$, $A^* > 0$ and $E^* > 0$. That is, under these conditions, the system has a unique coexistence equilibrium $\Xi^*_1=(N^*_1,A^*_1,I^*_1,E^*)$, where  
$N^*_1=\frac{\mathcal{Q} }{p}A^*_1$, $A^*_1=\bar{G}(\bar{H}-I^*_1)$,
$I^*_1=\frac{(m-\theta\mathcal{P}-s\bar{G}\bar{H})-\sqrt{(m-\theta\mathcal{P}-s\bar{G}\bar{H})^2-4s\theta \mathcal{P}\bar{G}\bar{H}}}{2\mathcal{P}}$, and $E^*=\frac{q\mathcal{R}}{r_N(q+\mathcal{S})}$.



(iii) When condition $\mathcal{P}<0$, $\mathcal{Q}>0$, $\mathcal{R}>0$, $\mathcal{S}>0$, and $m>(\mathcal{P}+s\bar{G})(\theta+\bar{H})$ hold, equation (\ref{E*}) has a unique positive roots and satisfies $N^*>0$, $A^* > 0$ and $E^* > 0$. That is, under these conditions, the system has a unique coexistence equilibrium $\Xi^*_1=(N^*_1,A^*_1,I^*_1,E^*)$, where $N^*_1=\frac{\mathcal{Q} }{p}A^*_1$, $A^*_1=\bar{G}(\bar{H}-I^*_1)$,
$I^*_1=\frac{(m-\theta\mathcal{P}-s\bar{G}\bar{H})-\sqrt{(m-\theta\mathcal{P}-s\bar{G}\bar{H})^2-4s\theta \mathcal{P}\bar{G}\bar{H}}}{2\mathcal{P}}$, and $E^*=\frac{q\mathcal{R}}{r_N(q+\mathcal{S})}$.

(iv) When condition $\mathcal{P}=0$, $\mathcal{Q}>0$, $\mathcal{R}>0$, $\mathcal{S}>0$, and $m>s\bar{G}(\theta+\bar{H})$ hold, equation (\ref{E*}) has a unique positive roots and satisfies $N^*>0$, $A^* > 0$ and $E^* > 0$. That is, under these conditions, the system has a unique coexistence equilibrium $\Xi^*_1=(N^*_1,A^*_1,I^*_1,E^*)$, where $N^*_1=\frac{\mathcal{Q} }{p}A^*_1$, $A^*_1=\bar{G}(\bar{H}-I^*_1)$,
$I^*_1=\frac{s\theta \bar{G}\bar{H}}{m-s\bar{G}\bar{H}}$, and $E^*=\frac{q\mathcal{R}}{r_N(q+\mathcal{S})}$.

(v) When condition $\mathcal{P}<-\frac{s\theta\bar{G}}{\bar{H}}$, $\mathcal{Q}>0$, $\mathcal{R}>0$, $\mathcal{S}>0$, and $m=\theta\mathcal{P}+s\bar{G}\bar{H}>0$ hold, equation (\ref{E*}) has a unique positive roots and satisfies $N^*>0$, $A^* > 0$ and $E^* > 0$. That is, under these conditions, the system has a unique coexistence equilibrium $\Xi^*_1=(N^*_1,A^*_1,I^*_1,E^*)$, where $N^*_1=\frac{\mathcal{Q} }{p}A^*_1$, $A^*_1=\bar{G}(\bar{H}-I^*_1)$,
$I^*_1=\sqrt{-\frac{s\theta\bar{G}\bar{H}}{\mathcal{P}}}$, and $E^*=\frac{q\mathcal{R}}{r_N(q+\mathcal{S})}$.

(vi) When condition $\mathcal{P}>\frac{s\theta\bar{G}}{\bar{H}}$, $\mathcal{Q}>0$, $\mathcal{R}>0$, $\mathcal{S}>0$, and $m=\theta\mathcal{P}+s\bar{G}\bar{H}+2\sqrt{s\theta\mathcal{P}\bar{G}\bar{H}}$ hold, equation (\ref{E*}) has a unique positive roots and satisfies $N^*>0$, $A^* > 0$ and $E^* > 0$. That is, under these conditions, the system has a unique coexistence equilibrium $\Xi^*_1=(N^*_1,A^*_1,I^*_1,E^*)$, where $N^*_1=\frac{\mathcal{Q} }{p}A^*_1$, $A^*_1=\bar{G}(\bar{H}-I^*_1)$,
$I^*_1=\sqrt{\frac{s\theta\bar{G}\bar{H}}{\mathcal{P}}}$, and $E^*=\frac{q\mathcal{R}}{r_N(q+\mathcal{S})}$.

(vii) The system does not exist equilibrium $\Xi^*=(N^*,A^*,I^*,E^*)$ for all conditions except for the five parameter combination cases mentioned above.

\section{The proof of Theorem \ref{sxi}}\label{psxi}
\begin{proof}
    We calculate the Jacobi matrix evaluated around steady state $\Xi^*$, which is expressed as
\begin{eqnarray*}
J|_{\Xi^*}=
\begin{pmatrix}
-r_NN^* & -r_NN^*  & -r_NN^*  & -r_NN^*\\
-r_AA^*+p & -\frac{pN^*}{A^*}-r_AA^*  & -r_AA^*  & -r_AA^*\\
-r_II^* & -r_II^*+s  & -r_II^*-s\frac{A^*}{I^*}+\frac{mI^*}{(\theta+I^*)^2}  & -r_II^*\\
-r_EE^*+q & -r_EE^*+q  & -r_EE^*+q  & -\frac{q(N^*+A^*+I^*)}{E^*}-r_EE^*\\
\end{pmatrix}.
\end{eqnarray*}
Therefore, the characteristic equation for the matrix $J|_{\Xi^*}$ is

\begin{equation}\label{CEs}
\lambda^4+\Upsilon_{1\Xi^*}\lambda^3+\Upsilon_{2\Xi^*}\lambda^2+\Upsilon_{3\Xi^*}\lambda+\Upsilon_{4\Xi^*}=0,
\end{equation}
where
\begin{align}\label{ts}
    \Upsilon_{1\Xi^*}=& \frac{pN^*}{A^*}+\frac{sA^*}{I^*}+r_II^*-\frac{mI^*}{(\theta+I^*)^2}+\frac{q(N^*+A^*+I^*)}{E^*}+r_EE^*+r_AA^*+r_NN^*,\\
    \Upsilon_{2\Xi^*}=& \left(\frac{sA^*}{I^*}+r_II^*-\frac{mI^*}{(\theta+I^*)^2}\right)\left(\frac{q(N^*+A^*+I^*)}{E^*}+q\right)+\left(\frac{sA^*}{I^*}-\frac{mI^*}{(\theta+I^*)^2}\right) \notag \\
    &\left(r_EE^*-q\right)+\frac{pN^*}{A^*}\left(\frac{sA^*}{I^*}+r_II^*-\frac{mI^*}{(\theta+I^*)^2}+\frac{q(N^*+A^*+I^*)}{E^*}+r_EE^*\right), \notag \\
    &r_AA^*\left(\frac{sA^*}{I^*}-\frac{mI^*}{(\theta+I^*)^2}+s+\frac{q(N^*+A^*+I^*)}{E^*}+q\right)+r_NN^*\left(p+\frac{pN^*}{A^*}\right. \notag \\
    &\left.+\frac{sA^*}{I^*}-\frac{mI^*}{(\theta+I^*)^2}+\frac{q(N^*+A^*+I^*)}{E^*}+q\right), \notag \\
    \Upsilon_{3\Xi^*}=& \frac{pN^*}{A^*}\left(\frac{sA^*}{I^*}+r_II^*
    -\frac{mI^*}{(\theta+I^*)^2}\right)\left(\frac{q(N^*+A^*+I^*)}{E^*}+q\right)
    +\frac{pN^*}{A^*}\left(\frac{sA^*}{I^*}\right. \notag \\
    &\left.-\frac{mI^*}{(\theta+I^*)^2}\right)\left(r_EE^*-q\right)+r_AA^*\left(\frac{sA^*}{I^*}
    -\frac{mI^*}{(\theta+I^*)^2}+s\right)\left(\frac{q(N^*+A^*+I^*)}{E^*}\right. \notag \\
    &\left.+q\right)+pr_NN^*\left(\frac{sA^*}{I^*}-\frac{mI^*}{(\theta+I^*)^2}+s+\frac{sN^*}{I^*}-\frac{mN^*I^*}{A^*(\theta+I^*)^2}\right)
    +r_NN^* \notag \\
    &\left(\frac{q(N^*+A^*+I^*)}{E^*}+q\right)\left(p+\frac{pN^*}{A^*}+\frac{sA^*}{I^*}-\frac{mI^*}{(\theta+I^*)^2}\right), \notag \\
    \Upsilon_{4\Xi^*}=& r_NN^*\left(\frac{q(N^*+A^*+I^*)}{E^*}+q\right)\left[p\left(\frac{sA^*}{I^*}-\frac{mI^*}{(\theta+I^*)^2}+s\right)+\frac{pN^*}{A^*}\right. \notag \\
    &\left.\left(\frac{sA^*}{I^*}-\frac{mI^*}{(\theta+I^*)^2}\right)\right]. \notag
\end{align}

According to the Routh-Hurwitz criterion, we know that all roots of (\ref{CEs}) have negative real parts if and only if the conditions $\Upsilon_{1\Xi^*},\Upsilon_{4\Xi^*}>0$,  $\Upsilon_{1\Xi^*}\Upsilon_{2\Xi^*}-\Upsilon_{3\Xi^*}>0$, and $\Upsilon_{3\Xi^*}(\Upsilon_{1\Xi^*}\Upsilon_{2\Xi^*}-\Upsilon_{3\Xi^*})-\Upsilon_{1\Xi^*}^2\Upsilon_{4\Xi^*}>0$ are satisfied. 
That is, the equilibrium $\Xi^*$ is locally asymptotically stable under the conditions $\Upsilon_{1\Xi^*},\Upsilon_{4\Xi^*}>0$,  $\Upsilon_{1\Xi^*}\Upsilon_{2\Xi^*}-\Upsilon_{3\Xi^*}>0$, and $\Upsilon_{3\Xi^*}(\Upsilon_{1\Xi^*}\Upsilon_{2\Xi^*}-\Upsilon_{3\Xi^*})-\Upsilon_{1\Xi^*}^2\Upsilon_{4\Xi^*}>0$.
\end{proof}

\section{The proof of Theorem \ref{unique}}\label{punique}
\begin{proof}
    Since the coefficients of system (\ref{Smodel}) satisfy the local Lipschitz condition, then for any initial value $(N_0, A_0, I_0, E_0 )\in \mathbb{R}^4_+$ there  is a unique local solution $(N(t), A(t), I(t), E(t))$ on $[0, \tau_e)$, where $\tau_e$ is the explosion time. To prove this solution is global, we only need to verify that $\tau_e = \infty$ a.s. To this end, let $n_0 > 0$ be sufficiently  large such that $N_0$, $A_0$, $I_0$ and $E_0$ all lie within the interval $[ \frac{1}{n_0}, n_0]$. For each integer $n > n_0$, define the stopping time
    \begin{small}
    \begin{equation*}
    \tau_n=\inf\{t\in[0,\tau_e):\min\{N(t),A(t),I(t),E(t)\}\leq\frac{1}{n} ~~\text{or}~~\max\{N(t),A(t),I(t),E(t)\}\geq n\}\}.
    \end{equation*}
    \end{small}

    Without loss of generality, define $\inf \emptyset =\infty$. It is easy to see that $\tau_n$ is increasing as $n\rightarrow \infty$. Let $\tau_{\infty} = \lim\limits_{n\rightarrow \infty} \tau_n$, then $\tau_{\infty} \leq \tau_e$ a.s. model (\ref{Smodel}) has a unique global positive solution if $\tau_{\infty} = \infty$  a.s. If this assertion is false, there are constants $\tilde{T}> 0$ and an $ \epsilon  \in(0, 1)$ such that $\mathbb{P}\{\tau_{\infty} \leq \tilde{T}\} >\epsilon $. As a result, there is an integer $n_1 \geq n_0$ such that
    \begin{equation}\label{tau}
        \mathbb{P}\{\tau_{\infty} \leq \tilde{T}\}>\epsilon,~~n\geq n_1. 
    \end{equation}
Define the Lyapunov function $V:\mathbb{R}_+^4\rightarrow \mathbb{R}_+$ as
$$
V=(N-1-\ln N)+(A-1-\ln A)+(I-1-\ln I)+(E-1-\ln E).
$$
The nonnegativity of this function can be seen from  $u-1- \ln u \geq 0$, $\forall~u > 0$. Applying the It$\hat{o}$'s formula to $V$, we have
\begin{equation}\label{dV}
    \mathrm{d}V=  \mathcal{L}V \mathrm{d}t+\sigma_1(N-1)\mathrm{d}B_1(t)+\sigma_2(A-1)\mathrm{d}B_2(t)+\sigma_3(I-1)\mathrm{d}B_3(t)+\sigma_4(E-1)\mathrm{d}B_4(t),
\end{equation}
where
\begin{equation*}
    \begin{aligned}
        \mathcal{L}V=&(N-1)\left[r_N\left(1-N-A-I-E\right)-d_N-p-q\right]+\frac{\sigma_1^2}{2}\\
        &+(A-1)\left[r_A\left(1-N-A-I-E\right)-d_A+p\frac{N}{A}-q-s\right]+\frac{\sigma_2^2}{2}\\
        &+(I-1)\left[r_I\left(1-N-A-I-E\right)-d_I+s\frac{A}{I}-q-\frac{m}{\theta+I}\right]+\frac{\sigma_3^2}{2}\\
        &+(E-1)\left[r_E\left(1-N-A-I-E\right)-d_E+q\frac{N+A+I}{E}\right]+\frac{\sigma_4^2}{2}\\
        \leq & r_NN(1-N)+r_AA(1-A)+r_II(1-I)+r_EE(1-E)\\
        &+(r_N+r_A+r_I+r_E)(N+A+I+E)+d_N+p+q+d_A+q+s\\
        &+d_I+q+\frac{m}{\theta}+d_E+\frac{\sigma_1^2+\sigma_2^2+\sigma_3^2+\sigma_4^2}{2}\\
        \leq & r_N+r_A+r_I+r_E+d_N+p+q+d_A\\
        &+q+s+d_I+q+\frac{m}{\theta}+d_E+\frac{\sigma_1^2+\sigma_2^2+\sigma_3^2+\sigma_4^2}{2}\\
       :=&\mathfrak{D}  
    \end{aligned}
\end{equation*}
where $\mathfrak{D}$ is a positive constant. Integrating both sides of (\ref{dV}) from 0 to $\tau_n \land \tilde{T}$ and taking the expectation lead to
\begin{equation}\label{EV}
\mathbb{E}V[N(\tau_n\land\tilde{T}),A(\tau_n\land\tilde{T}),I(\tau_n\land\tilde{T}),E(\tau_n\land\tilde{T})]\leq V(N_0,A_0,I_0,E_0)+\mathfrak{D}\tilde{T}. 
\end{equation}

Define $\Omega_n=\{\omega\in\Omega:\tau_n\leq\tilde{T}\}$ for $n\geq n_1$. By (\ref{tau}), we get $\mathbb{P}(\Omega_n)\geq \epsilon$. It follows that, for any $\omega\in\Omega_n$, there exists $N(\tau_n, \omega)$ or $A(\tau_n, \omega)$ or $I(\tau_n, \omega)$ or$E(\tau_n, \omega)$ equals either $n$ or $\frac{1}{n}$.  Consequently, we obtain
$$
V(N(\tau_n,\omega),A(\tau_n,\omega),I(\tau_n,\omega),E(\tau_n,\omega))\geq (n-1-\ln n)\land (\frac{1}{n}-1-\ln\frac{1}{n}).
$$
According to (\ref{EV}), one can get that
\begin{equation*}
    \begin{aligned}
        V(N_0,A_0,I_0,E_0)+\mathfrak{D}\tilde{T}\geq & \mathbb{E}[\mathbb{I}_{\Omega_n(\omega)}V(N(\tau_n\land\tilde{T}),A(\tau_n\land\tilde{T}),I(\tau_n\land\tilde{T}),E(\tau_n\land\tilde{T}))]\\
        \geq &\epsilon[(n-1-\ln n)\land (\frac{1}{n}-1-\ln\frac{1}{n})],
    \end{aligned}
\end{equation*}
where $\mathbb{I}_{\Omega_n(\omega)}$ is the indicator function of $\Omega_n$. Letting $n\rightarrow \infty$ leads to the contradiction
$$
\infty >V(N_0,A_0,I_0,E_0)+\mathfrak{D}\tilde{T}= \infty.
$$
Hence, $\tau_{\infty}=\infty$ a.s. This completes the proof.
\end{proof}

\section{The proof of Theorem \ref{moment}} \label{pmoment}

\begin{proof}
    Define a $\mathbb{C}^2$-function $V$: $\mathbb{R}_+^4\rightarrow\mathbb{R}_+$ such that
    \begin{equation}\label{V}
V(N,A,I,E)=e^t(N+A+I+E)^{\theta},
    \end{equation}
where $(N,A,I,E)\in\mathbb{R}_+^4$. Differentiating equation (\ref{V}) by It$\hat{o}$'s formula, we get

\begin{equation*}
    \begin{aligned}
\mathcal{L}V(N,A,I,E)=&\theta e^t (N+A+I+E)^{\theta-2}\left\{(N+A+I+E)\left[\left(1-N-A-I-E\right)\right.\right.\\
&\left.(r_NN+r_AA+r_II+r_EE)-d_NN-d_AA-d_II-d_EE-\frac{mI}{\theta + I}\right]\\
&+\left.\frac{\theta-1}{2}(\sigma_1^2N^2+\sigma_2^2A^2+\sigma_3^2I^2+\sigma_4^2E^2)+\frac{1}{\theta}(N+A+I+E)^2\right\}\\
\leq & \theta e^t (N+A+I+E)^{\theta-2}W(N,A,I,E),
    \end{aligned}
\end{equation*}
where 
\begin{equation*}
    \begin{aligned}
        W(N,A,I,E)= & -r_NN^3+\left(r_N-d_N+\frac{\theta-1}{2}\sigma_1^2+\frac{1}{\theta}\right)N^2+\left(-r_AA^2+r_AA-d_AA\right.\\
        &\left.+\frac{2}{\theta}A-r_II^2+r_II-d_II+\frac{2}{\theta}I-r_EE^2+r_EE-d_EE+\frac{2}{\theta}E\right)N\\
        &-r_AA^3+\left(r_A-d_A+\frac{\theta-1}{2}\sigma_2^2+\frac{1}{\theta}\right)A^2+\left(-r_NN^2+r_NN-d_NN\right.\\
        &\left.-r_II^2+r_II-d_II+\frac{2}{\theta}I-r_EE^2+r_EE-d_EE+\frac{2}{\theta}E\right)A\\
        &-r_II^3+\left(r_I-d_I+\frac{\theta-1}{2}\sigma_3^2+\frac{1}{\theta}\right)I^2+\left(-r_NN^2+r_NN-d_NN\right.\\
        &\left.-r_AA^2+r_AA-d_AA-r_EE^2+r_EE-d_EE+\frac{2}{\theta}E\right)I\\
        &-r_EE^3+\left(r_E-d_E+\frac{\theta-1}{2}\sigma_4^2+\frac{1}{\theta}\right)E^2+\left(-r_NN^2+r_NN-d_NN\right.\\
        &\left.-r_AA^2+r_AA-d_AA-r_II^2+r_II-d_II\right)E.\\
    \end{aligned}
\end{equation*}
One observes that 
\begin{equation*}
    \lim\limits_{N^2+A^2+I^2+E^2}(N+A+I+E)^{\theta-2}W(N,A,I,E)=-\infty,
\end{equation*}
which together with the continuity of $(N+A+I+E)^{\theta-2}W(N,A,I,E)$ in $\mathbb{R}_+^4$ implies that 
\begin{equation}\label{L(theta)}
    L_1(\theta):=\theta\sup\limits_{N,A,I,E\in\mathbb{R}_+}\{(N+A+I+E)^{\theta-2}W(N,A,I,E)\}<+\infty.
\end{equation}
Thus, we have 
$$
\mathcal{L}V(N,A,I,E)\leq L_1(\theta)e^t.
$$

Integrating both sides of the above inequality from 0 to $t_n \land t$ and taking expectation leads to the following inequality
\begin{equation*}
    \mathbb{E}V(N(t_n \land t),A(t_n \land t),I(t_n \land t),E(t_n \land t))\leq V(N_0,A_0,I_0,E_0)+L_1(\theta)\mathbb{E}\int_0^{t_n \land t}e^s\mathrm{d}s.
\end{equation*}
    If we take $n\rightarrow\infty$, then we have
    \begin{equation*}
    \mathbb{E}V(N(t),A(t),I(t),E( t))\leq V(N_0,A_0,I_0,E_0)+L_1(\theta)(e^t-1).
    \end{equation*}
which implies that
    \begin{equation*}
    e^{-t}\mathbb{E}V(N(t),A(t),I(t),E( t))\leq e^{-t}V(N_0,A_0,I_0,E_0)+L_1(\theta).
    \end{equation*}
Letting $t\rightarrow\infty$, we obtain
$$
\limsup\limits_{t\rightarrow\infty}\mathbb{E}[(N(t)+A(t)+I(t)+E(t))^{\theta}]\leq L_1(\theta).
$$
The proof is complete.
\end{proof}

\section{The proof of Theorem \ref{stationary}} \label{pstationary}

\begin{proof}
The diffusion matrix of system (\ref{Smodel}) is
$$
\mathcal{A}=\left(
\begin{array}{cccc}
    \sigma_1^2N^2 & 0 & 0 & 0 \\
    0 &  \sigma_2^2A^2 & 0 & 0\\
    0 & 0 &\sigma_3^2I^2 & 0\\
    0 & 0 & 0 &\sigma_4^2E^2
\end{array}
\right)
$$
Choosing $H=\min\limits_{(N,A,I,E)\in\bar{D}\subset\mathbb{R}_+^4}\{\sigma_1^2N^2,\sigma_2^2A^2,\sigma_3^2I^2,\sigma_4^2E^2\}$, we obtain
$$
\sum_{i,j=1}^4 a_{ij}(N,A,I,E)\xi_i\xi_j=\sigma_1^2N^2+\sigma_2^2A^2+\sigma_3^2I^2+\sigma_4^2E^2\geq H|\xi|^2,
$$
where $\bar{D}=[\epsilon,\frac{1}{\epsilon}]\times [\epsilon,\frac{1}{\epsilon}]\times[\epsilon,\frac{1}{\epsilon}]\times[\epsilon,\frac{1}{\epsilon}]$, then the condition (A1) of Lemma \ref{SD} holds.

Define a $C^2$-function $V_1:\mathbb{R}_+^4\rightarrow\mathbb{R}_+$ by
\begin{equation*}
    V_1=-\frac{1}{N}-c_1\ln A-c_2\ln I-c_3\ln E+(N+A+I+E)
\end{equation*}
where $c_1, c_2, c_3$ are positive constants to be determined later. Making use of It$\hat{o}$’s formula, we get
    \begin{align*}
        \mathcal{L}V_1= & \frac{r_N}{N}-\frac{r_N(N+A+I+E)}{N}-\frac{d_N+p+q+\sigma_1^2}{N}-c_1r_A+c_1r_A(N+A+I+E)\\
        &+c_1(d_A+q+s+\frac{\sigma_2^2}{2})-c_1p\frac{N}{A}-c_2r_I+c_2r_I(N+A+I+E)+c_2(d_I+q\\
        &+\frac{\sigma_3^2}{2})+\frac{c_2m}{\theta+I}-c_2s\frac{A}{I}-c_3r_E+c_3r_E(N+A+I+E)+c_3(d_E+\frac{\sigma_4^2}{2})\\
        &-c_3q\frac{N}{E}-c_3q\frac{A}{E}-c_3q\frac{I}{E}+(r_NN+r_AA+r_II+r_EE)\left(1-N-A-I-E\right)\\
        &-d_NN-d_AA-d_II-d_EE-\frac{mI}{\theta+I}\\
        \leq & -\frac{d_N+p+q+\sigma_1^2}{N}-c_1p\frac{N}{A}-c_2s\frac{A}{I}-c_3q\frac{I}{E}-d_EE+\frac{r_N}{N}+c_1(d_A+q+s\\
        &+\frac{\sigma_2^2}{2})+c_2(d_I+q+\frac{\sigma_3^2}{2}+\frac{m}{\theta})+c_3(d_E+\frac{\sigma_4^2}{2})+\left[c_1r_A+c_2r_I+c_3r_E+\max\{\right.\\
        &\left.r_N,r_A,r_I,r_E\}\right](N+A+I+E)-\min\{r_N,r_A,r_I,r_E\}(N+A+I+E)^2\\
        \leq & -4\sqrt[4]{c_1c_2c_3(d_N+p+q+\sigma_1^2)psqd_E}+\frac{r_N}{N}+c_1\left(d_A+q+s+\frac{\sigma_2^2}{2}\right)+c_2\left(d_I\right.\\
        &\left.+q+\frac{\sigma_3^2}{2}+\frac{m}{\theta}\right)+c_3\left(d_E+\frac{\sigma_4^2}{2}\right)+\mathcal{D}.
    \end{align*}
Let 
\begin{align*}
&c_1\left(d_A+q+s+\frac{\sigma_2^2}{2}\right)
=c_2\left(d_I+q+\frac{\sigma_3^2}{2}+\frac{m}{\theta}\right)
=c_3\left(d_E+\frac{\sigma_4^2}{2}\right)\\
=&\frac{psqd_E(d_N+p+q+\sigma_1^2)}{(d_A+q+s+\frac{\sigma_2^2}{2})(d_I+q+\frac{\sigma_3^2}{2}+\frac{m}{\theta})(d_E+\frac{\sigma_4^2}{2})},    
\end{align*}
then we have
$$c_1=\frac{psqd_E(d_N+p+q+\sigma_1^2)}{(d_A+q+s+\frac{\sigma_2^2}{2})^2(d_I+q+\frac{\sigma_3^2}{2}+\frac{m}{\theta})(d_E+\frac{\sigma_4^2}{2})},$$
$$c_2=\frac{psqd_E(d_N+p+q+\sigma_1^2)}{(d_A+q+s+\frac{\sigma_2^2}{2})(d_I+q+\frac{\sigma_3^2}{2}+\frac{m}{\theta})^2(d_E+\frac{\sigma_4^2}{2})},$$
$$c_3=\frac{psqd_E(d_N+p+q+\sigma_1^2)}{(d_A+q+s+\frac{\sigma_2^2}{2})(d_I+q+\frac{\sigma_3^2}{2}+\frac{m}{\theta})(d_E+\frac{\sigma_4^2}{2})^2}.$$
Hence
\begin{equation*}
    \begin{aligned}
        \mathcal{L}V_1\leq & \frac{psqd_E(d_N+p+q+\sigma_1^2)}{(d_A+q+s+\frac{\sigma_2^2}{2})(d_I+q+\frac{\sigma_3^2}{2}+\frac{m}{\theta})(d_E+\frac{\sigma_4^2}{2})}-\mathcal{D}+\frac{r_N}{N}\\
        :=&-\lambda+\frac{r_N}{N},
    \end{aligned}
\end{equation*}
where 
$$
\lambda=\mathcal{D}(R_s^*-1)>0.
$$
Constructing a $C^2$-function $Q:\mathbb{R}_+^4\rightarrow\mathbb{R}_+$ in the following form
$$
Q=MV_1+V_2+V_3+V_4+V_5,
$$
where $V_2=\frac{1}{\theta+1}(N+A+I+E)^{\theta+1}$, $V_3=-\ln A$, $V_4=-\ln I$, $V_5=-\ln E$, $\theta>0$ is a constant and $M>0$ is a sufficiently large number satisfying the following condition
\begin{equation}
    -M\lambda+B+F+C\leq -2,
\end{equation}
where 
\begin{equation}\label{BCF}
\begin{aligned}
    B = & \sup\limits_{(N,A,I,E)\in\mathbb{R}_+^4}\left\{\left[\max\{r_N,r_A,r_I,r_E\}-\min\{d_N,d_A,d_I,d_E\}+\frac{\theta}{2}\max\{\sigma_1^2,\sigma_2^2,\sigma_3^2,\right.\right.\\
    & \left.\left.\sigma_4^2\}\right](N+A+I+E)^{\theta+1}-\frac{1}{2}\min\{r_N,r_A,r_I,r_E\}(N+A+I+E)^{\theta+2}\right\},  \\
    F = & \sup\limits_{(N,A,I,E)\in\mathbb{R}_+^4}\left\{\left(r_A+r_I+r_E\right)(N+A+I+E)+\frac{Mr_N}{N}-\frac{\min\{r_N,r_A,r_I,r_E\}}{4}\right.\\&\left.(N^{\theta+2}+A^{\theta+2}+I^{\theta+2}+E^{\theta+2})\right\},\\
    C = & d_A+q+s+\frac{\sigma_2^2}{2}+d_I+q+\frac{\sigma_3^2}{2}+\frac{m}{\theta}+d_E+\frac{\sigma_4^2}{2}.
\end{aligned}
\end{equation}

Define a nonnegative $C^2$-function $V:\mathbb{R}_+^4\rightarrow\mathbb{R}_+$ as follows
\begin{equation*}
\begin{aligned}
    V(N,A,I,E)=&Q(N,A,I,E)-Q(\bar{N}_0,\bar{A}_0,\bar{I}_0,\bar{E}_0),\\
    =&MV_1+V_2+V_3+V_4+V_5-Q(\bar{N}_0,\bar{A}_0,\bar{I}_0,\bar{E}_0),
\end{aligned}
\end{equation*}
where $Q(\bar{N}_0,\bar{A}_0,\bar{I}_0,\bar{E}_0)$ is the minimum value of $Q(N,A,I,E)$.

Applying It$\hat{o}$’s formula, we can obtain
\begin{equation*}
    \begin{aligned}
        \mathcal{L}V_2=&(N+A+I+E)^{\theta}\left[(r_NN+r_AA+r_II+r_EE)\left(1-N-A-I-E\right)\right.\\
        &\left.-d_NN-d_AA-d_II-d_EE-\frac{mI}{\theta+I}\right]+\frac{\theta}{2}(N+A+I+E)^{\theta-1}(\sigma_1^2N^2\\
        &+\sigma_2^2A^2+\sigma_3^2I^2+\sigma_4^2E^2)\\
        \leq & \max\{r_N,r_A,r_I,r_E\}(N+A+I+E)^{\theta+1}-\min\{r_N,r_A,r_I,r_E\}\\
        &(N+A+I+E)^{\theta+2}-\min\{d_N,d_A,d_I,d_E\}(N+A+I+E)^{\theta+1}\\
        &+\frac{\theta}{2}\max\{\sigma_1^2,\sigma_2^2,\sigma_3^2,\sigma_4^2\}(N+A+I+E)^{\theta+1}\\
        \leq & B-\frac{1}{2}\min\{r_N,r_A,r_I,r_E\}(N+A+I+E)^{\theta+2}\\
         \leq & B-\frac{1}{2}\min\{r_N,r_A,r_I,r_E\}(N^{\theta+2}+A^{\theta+2}+I^{\theta+2}+E^{\theta+2})
    \end{aligned}
\end{equation*}
By using the It$\hat{o}$’s formula,
\begin{equation*}
    \begin{aligned}
        \mathcal{L}V_3=&-r_A+r_A(N+A+I+E)+d_A+q+s+\frac{\sigma_2^2}{2}-p\frac{N}{A}\\
        \leq & r_A(N+A+I+E)+d_A+q+s+\frac{\sigma_2^2}{2}-p\frac{N}{A},\\
        \mathcal{L}V_4=&-r_I+r_I(N+A+I+E)+d_I+q+\frac{\sigma_3^2}{2}-\frac{mI}{\theta+I}-s\frac{A}{I}\\
        \leq & r_I(N+A+I+E)+d_I+q+\frac{\sigma_3^2}{2}-\frac{m}{\theta}-s\frac{A}{I},\\
         \mathcal{L}V_5=&-r_E+r_E(N+A+I+E)+d_E+\frac{\sigma_4^2}{2}-q\frac{N}{E}-q\frac{A}{E}-q\frac{I}{E}\\
        \leq & r_E(N+A+I+E)+d_E+\frac{\sigma_4^2}{2}-q\frac{N}{E}.
    \end{aligned}
\end{equation*}
Therefore, 
\begin{equation*}
    \begin{aligned}
        \mathcal{L}V \leq & -M\lambda+\frac{Mr_N}{N}+B-\frac{\min\{r_N,r_A,r_I,r_E\}}{2}(N^{\theta+2}+A^{\theta+2}+I^{\theta+2}+E^{\theta+2})\\
        &+(r_A+r_I+r_E)N+(r_A+r_I+r_E)A+(r_A+r_I+r_E)I+(r_A+r_I+r_E)E\\
        &+d_A+q+s+\frac{\sigma_2^2}{2}+d_I+q+\frac{\sigma_3^2}{2}+\frac{m}{\theta}+d_E+\frac{\sigma_4^2}{2}-p\frac{N}{A}\\
        &-s\frac{A}{I}-q\frac{N}{E}\\
        \leq & -M\lambda+B-\frac{\min\{r_N,r_A,r_I,r_E\}}{4}(N^{\theta+2}+A^{\theta+2}+I^{\theta+2}+E^{\theta+2})+F\\
        &-p\frac{N}{A}-s\frac{A}{I}-q\frac{N}{E}+C
    \end{aligned}
\end{equation*}
where $B$, $F$, and $C$ are given in (\ref{BCF}).

Denote
$$
D=\{(N,A,I,E)\in\mathbb{R}_+^4:\epsilon\leq N\leq \frac{1}{\epsilon},\epsilon^2\leq A\leq \frac{1}{\epsilon^2},\epsilon^3\leq I\leq \frac{1}{\epsilon^3},\epsilon^2\leq E\leq \frac{1}{\epsilon^2}\},
$$
where $0<\epsilon<1$ is sufficiently small constant satisfying the following conditions
\begin{align}
    \label{1}&-2+r_A\epsilon\leq -1,\\
     \label{2}&B+F+C-\frac{p}{\epsilon}\leq -1,\\
     \label{3}&B+F+C-\frac{s}{\epsilon}\leq -1,\\
     \label{4}&B+F+C-\frac{q}{\epsilon}\leq -1,\\
     \label{5}&-\frac{\min\{r_N,r_A,r_I,r_E\}}{4}\frac{1}{\epsilon^{\theta+2}}+B+F+C\leq -1,\\
     \label{6}&-\frac{\min\{r_N,r_A,r_I,r_E\}}{4}\frac{1}{\epsilon^{2(\theta+2)}}+B+F+C\leq -1,\\
     \label{7}&-\frac{\min\{r_N,r_A,r_I,r_E\}}{4}\frac{1}{\epsilon^{3(\theta+2)}}+B+F+C\leq -1.
\end{align}
In the following, we will prove $\mathcal{L}V\leq-1$ for any $(N,A,I,E)\in\mathbb{R}_+^4\backslash D$. To achieve that, we can divide $\mathbb{R}_+^4\backslash D$ into the following eight domains,
\begin{align*}
    &D_1=\{(N,A,I,E)\in\mathbb{R}_+^4:0<N<\epsilon\},\\
    &D_2=\{(N,A,I,E)\in\mathbb{R}_+^4:0<A<\epsilon^2, N\geq\epsilon\},\\
    &D_3=\{(N,A,I,E)\in\mathbb{R}_+^4:0<I<\epsilon^3, A\geq\epsilon^2\},\\
    &D_4=\{(N,A,I,E)\in\mathbb{R}_+^4:0<E<\epsilon^2, N\geq\epsilon\},\\
    &D_5=\{(N,A,I,E)\in\mathbb{R}_+^4:N>\frac{1}{\epsilon}\},~~D_6=\{(N,A,I,E)\in\mathbb{R}_+^4:A>\frac{1}{\epsilon^2}\}, \\
    &D_7=\{(N,A,I,E)\in\mathbb{R}_+^4:N>\frac{1}{\epsilon^3}\},~~D_8=\{(N,A,I,E)\in\mathbb{R}_+^4:N>\frac{1}{\epsilon^2}\}.
\end{align*}
Notice that $D^c=D_1\cup D_2\cup D_3\cup D_4 \cup D_5\cup D_6\cup D_7\cup D_8$, then we only need to prove $\mathcal{L}V\leq -1$ on the above eight domains, respectively.

Case 1. If $(N,A,I,E)\in D_1$, i.e., $0<N<\epsilon$,
\begin{equation*}
    \begin{aligned}
        \mathcal{L}V\leq & -M\lambda+B+F+r_AN-r_AN+C\\
        \leq & -2+r_A\epsilon\leq-1,
    \end{aligned}
\end{equation*}
which follows from (\ref{1}).

Case 2. If $(N,A,I,E)\in D_2$, i.e., $0<A<\epsilon^2$ and $N\geq\epsilon$.
\begin{equation*}
    \begin{aligned}
        \mathcal{L}V\leq & B+F+C-p\frac{N}{A}\\
        \leq & B+F+C-\frac{p}{\epsilon}\leq-1,
    \end{aligned}
\end{equation*}
which follows from (\ref{2}).

Case 3. If $(N,A,I,E)\in D_3$, i.e., $0<I<\epsilon^3$ and $A\geq\epsilon^2$.
\begin{equation*}
    \begin{aligned}
        \mathcal{L}V\leq & B+F+C-s\frac{A}{I}\\
        \leq & B+F+C-\frac{s}{\epsilon}\leq-1,
    \end{aligned}
\end{equation*}
which follows from (\ref{3}).

Case 4. If $(N,A,I,E)\in D_4$, i.e., $0<E<\epsilon^2$ and $N\geq\epsilon$.
\begin{equation*}
    \begin{aligned}
        \mathcal{L}V\leq & B+F+C-q\frac{N}{E}\\
        \leq & B+F+C-\frac{q}{\epsilon}\leq-1,
    \end{aligned}
\end{equation*}
which follows from (\ref{4}).

Case 5. If $(N,A,I,E)\in D_5$, i.e., $N>\frac{1}{\epsilon}$.
\begin{equation*}
    \begin{aligned}
        \mathcal{L}V\leq & B+F+C-\frac{\min\{r_N,r_A,r_I,r_E\}}{4}N^{\theta+2}\\
        \leq & B+F+C-\frac{\min\{r_N,r_A,r_I,r_E\}}{4}\frac{1}{\epsilon^{\theta+2}}\leq-1,
    \end{aligned}
\end{equation*}
which follows from (\ref{5}).

Case 6. If $(N,A,I,E)\in D_6$, i.e., $A>\frac{1}{\epsilon^2}$.
\begin{equation*}
    \begin{aligned}
        \mathcal{L}V\leq & B+F+C-\frac{\min\{r_N,r_A,r_I,r_E\}}{4}A^{\theta+2}\\
        \leq & B+F+C-\frac{\min\{r_N,r_A,r_I,r_E\}}{4}\frac{1}{\epsilon^{2(\theta+2)}}\leq-1,
    \end{aligned}
\end{equation*}
which follows from (\ref{6}).

Case 7. If $(N,A,I,E)\in D_7$, i.e., $I>\frac{1}{\epsilon^3}$.
\begin{equation*}
    \begin{aligned}
        \mathcal{L}V\leq & B+F+C-\frac{\min\{r_N,r_A,r_I,r_E\}}{4}I^{\theta+2}\\
        \leq & B+F+C-\frac{\min\{r_N,r_A,r_I,r_E\}}{4}\frac{1}{\epsilon^{3(\theta+2)}}\leq-1,
    \end{aligned}
\end{equation*}
which follows from (\ref{7}).

Case 8. If $(N,A,I,E)\in D_8$, i.e., $E>\frac{1}{\epsilon^2}$.
\begin{equation*}
    \begin{aligned}
        \mathcal{L}V\leq & B+F+C-\frac{\min\{r_N,r_A,r_I,r_E\}}{4}E^{\theta+2}\\
        \leq & B+F+C-\frac{\min\{r_N,r_A,r_I,r_E\}}{4}\frac{1}{\epsilon^{2(\theta+2)}}\leq-1,
    \end{aligned}
\end{equation*}
which follows from (\ref{6}).

To sum up, we can conclude that
$$
\mathcal{L}V\leq -1,
$$
for all $(N,A,I,E)\in D^c$ as long as $\epsilon$ sufficiently small. Hence, the condition (A2) of Lemma \ref{SD} holds. It follows from Lemma \ref{SD} that system (\ref{Smodel}) is ergodic and has a unique stationary distribution. This completes the proof.

\end{proof}

\section{The proof of Theorem \ref{extinction}} \label{pextinction}
\begin{proof}
Let P(t)=N(t)+A(t)+I(t)+E(t). Differentiating $\ln P$ by It$\hat{o}$'s formula, we get

    \begin{align}\label{lnp}
        \mathrm{d}\ln P=&\left\{\frac{1}{N+A+I+E}\left[(r_NN+r_AA+r_II+r_EE)\left(1-N-A-I-E\right)-d_NN\right.\right.\\
        &\left.\left.-d_AA-d_II-d_EE-\frac{mI}{\theta+I}-\frac{1}{2(N+A+I+E)^2}\left(\sigma_1^2N^2+\sigma_2^2A^2+\sigma_3^2I^2\right.\right.\right. \notag \\
        &\left.\left.\left.+\sigma_4^2E^2\right)\right]\right\}\mathrm{d}t+\frac{1}{N+A+I+E}[\sigma_1N\mathrm{d}B_1(t)+\sigma_2A\mathrm{d}B_2(t)+\sigma_3I\mathrm{d}B_3(t) \notag \\
        &+\sigma_4E\mathrm{d}B_4(t)] \notag \\
        \leq & \max\{r_N,r_A,r_I,r_E\}\mathrm{d}t-\frac{1}{(N+A+I+E)^2}\left[\left(d_N+\frac{\sigma_1^2}{2}\right)N^2+\left(d_A+\frac{\sigma_2^2}{2}\right)\right. \notag \\
        &\left.A^2+\left(d_I+\frac{\sigma_3^2}{2}\right)I^2+\left(d_E+\frac{\sigma_4^2}{2}\right)E^2\right]\mathrm{d}t+\frac{1}{N+A+I+E}[\sigma_1N\mathrm{d}B_1(t) \notag \\
   &+\sigma_2A\mathrm{d}B_2(t)+\sigma_3I\mathrm{d}B_3(t)+\sigma_4E\mathrm{d}B_4(t)] \notag \\
        \leq & \left[\max\{r_N,r_A,r_I,r_E\}-\frac{1}{4}\min\left\{d_N+\frac{\sigma_1^2}{2},d_A+\frac{\sigma_2^2}{2},d_I+\frac{\sigma_3^2}{2},d_E+\frac{\sigma_4^2}{2}\right\}\right]\mathrm{d}t \notag \\
        & +\frac{1}{N+A+I+E}[\sigma_1N\mathrm{d}B_1(t)+\sigma_2A\mathrm{d}B_2(t)+\sigma_3I\mathrm{d}B_3(t)+\sigma_4E\mathrm{d}B_4(t)]. \notag
    \end{align}

Integrating (\ref{lnp}) from 0 to $t$ and then dividing by $t$ on both sides, we have
\begin{small}
\begin{equation}\label{intlnp}
\begin{aligned}
    &\frac{\ln P(t)}{t}-\frac{\ln P(0)}{t}\\
    \leq & \max\{r_N,r_A,r_I,r_E\}-\frac{1}{4}\min\left\{d_N+\frac{\sigma_1^2}{2},d_A+\frac{\sigma_2^2}{2},d_I+\frac{\sigma_3^2}{2},d_E+\frac{\sigma_4^2}{2}\right\}\\
    & +\frac{\sigma_1}{t}\int_0^t\frac{N(s)}{N(s)+A(s)+I(s)+E(s)}\mathrm{d}B_1(s)+\frac{\sigma_2}{t}\int_0^t\frac{A(s)}{N(s)+A(s)+I(s)+E(s)}\mathrm{d}B_2(s)\\
    &+\frac{\sigma_3}{t}\int_0^t\frac{I(s)}{N(s)+A(s)+I(s)+E(s)}\mathrm{d}B_3(s) +\frac{\sigma_4}{t}\int_0^t\frac{E(s)}{N(s)+A(s)+I(s)+E(s)}\mathrm{d}B_4(s) 
\end{aligned}
\end{equation}
\end{small}
Taking the superior limit on both sides of (\ref{intlnp}) and combining with (\ref{17}), (\ref{18}) and noting that $R_0^s<1$, one can see that

\begin{equation*}
\begin{aligned}
&\limsup\limits_{t\rightarrow\infty}\frac{\ln P(t)}{t}\\
&\leq \max\{r_N,r_A,r_I,r_E\}-\frac{1}{4}\min\left(d_N+\frac{\sigma_1^2}{2},d_A+\frac{\sigma_2^2}{2},d_I+\frac{\sigma_3^2}{2},d_E+\frac{\sigma_4^2}{2}\right)<0~~a.s.
\end{aligned}
\end{equation*}
which implies that
$$
\lim_{t\rightarrow\infty}N(t)=\lim_{t\rightarrow\infty}A(t)=\lim_{t\rightarrow\infty}I(t)=\lim_{t\rightarrow\infty}E(t)=0~~a.s.
$$

\end{proof}

\section{Parameter values for model (\ref{nonmodel})}

\begin{table}[h]
\centering
\caption{\centering Default value of all nondimensionalized parameters in the model (\ref{nonmodel})}\label{smtab}
\begin{tabular}{ccc}
\toprule
Parameters  & Default Value  & References\\
\midrule
$r_N$    &   $1$            & \cite{lakatos2020evolutionary} \\
$d_N$    &   $0.001$        & \cite{chen2025modeling}\\
$p$      &   $0.15$         &  \cite{aguade2022transition}\\
$q$      &   $10^{-5}$      & \cite{chen2025modeling} \\
$r_A$    &   $0.5$          &  \cite{benzekry2014classical,aguade2022transition}\\
$d_A$    &   $0.001$        & \cite{chen2025modeling}\\
$s$      &   $0.15$         &  \cite{aguade2022transition}\\
$r_I$    &   $0.45$         &  \cite{benzekry2014classical,aguade2022transition}\\
$d_I$    &   $0.001$        & \cite{chen2025modeling}\\
$m$      &  $3.2\times10^{-5}$    & Estimation \\
$\theta$ &  $4\times10^{-4}$     & Estimation\\
$r_E$    &   $0.4$         &  \cite{benzekry2014classical,aguade2022transition} \\
$d_E$    &   $0.001$       & \cite{chen2025modeling}\\
\hline
\end{tabular}
\end{table}

\section{Parameter sensitive analysis}\label{PSA}

This section uses the Latin hypercube sampling (LHS) method to sample parameters, generating 1,000 samples to calculate the partial rank correlation coefficient (PRCC) of the proportions of the four cell types. A quantitative analysis is performed on the sensitivity of the four cell proportions to changes in the input parameters.

\begin{figure}[H]
    \centering
\subfigure[]{\includegraphics[width=0.48\linewidth]{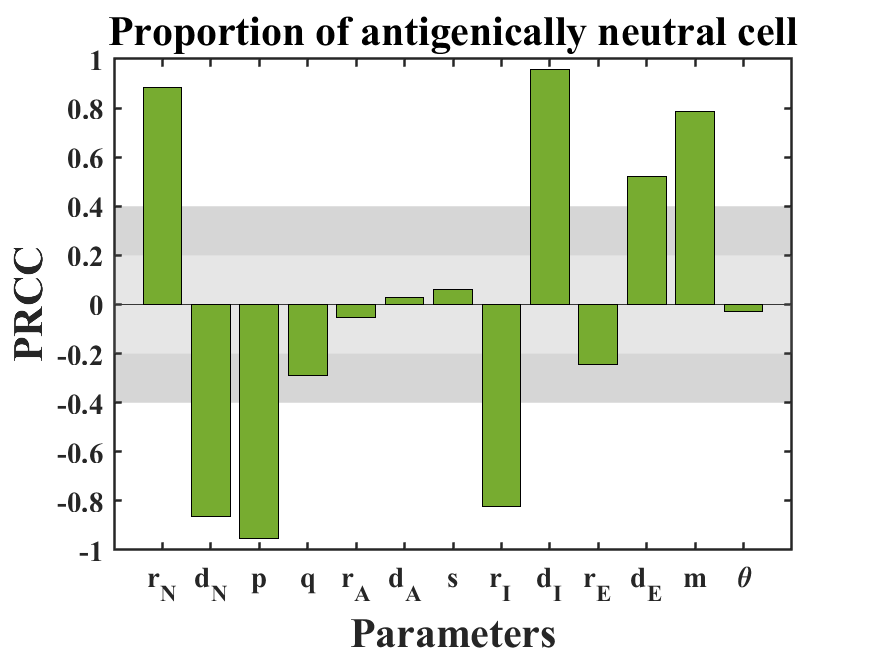}}
\subfigure[]{\includegraphics[width=0.48\linewidth]{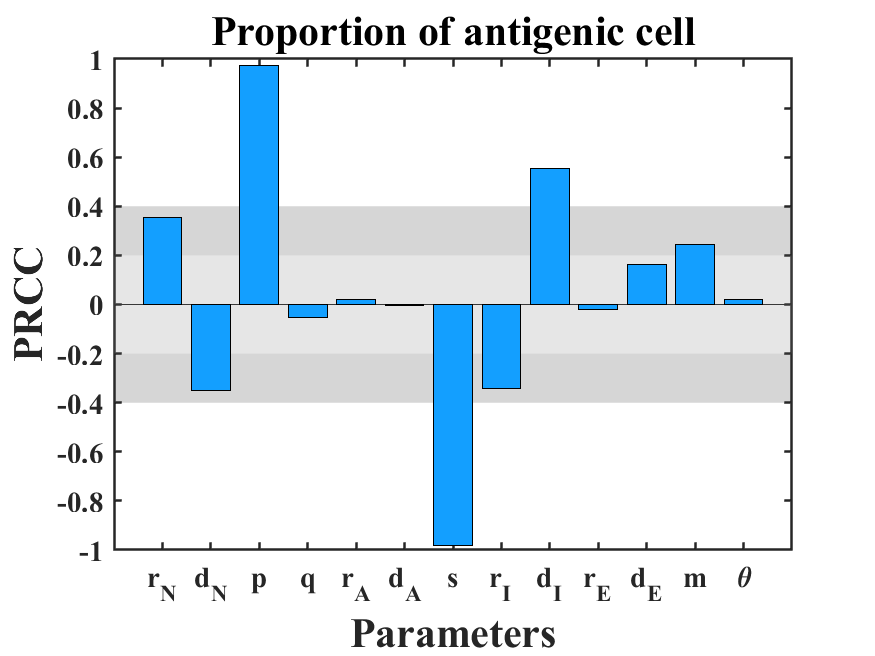}}\\
\subfigure[]{\includegraphics[width=0.48\linewidth]{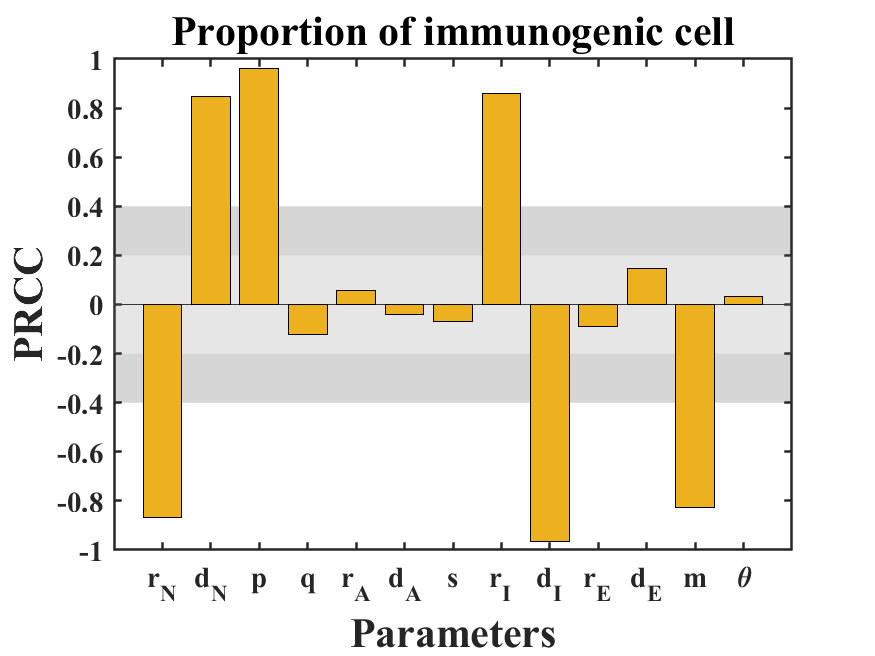}}
\subfigure[]{\includegraphics[width=0.48\linewidth]{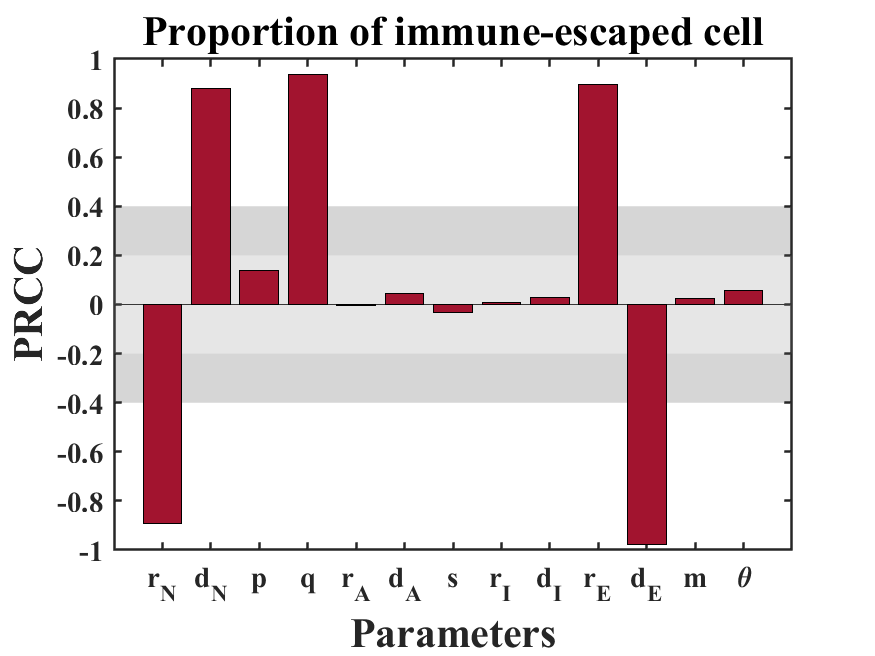}}\\	
    \caption{ Results of sensitivity analysis of four cell ratios. The light gray areas, dark gray areas, and white areas correspond to weak, moderate, and strong correlations, respectively.
    The baseline value of the parameter $p$ is taken as $0.00008$ and the baseline values of the other parameters are given in Table \ref{tab1}.}
    \label{fig5}
\end{figure}

Assuming that all input parameters follow  uniform distribution, with a $10\%$ floating value as an optional range, the baseline values of each parameter are shown in Table \ref{tab1}. The initial values of the model are $N(0)=1,~A(0)=I(0)=E(0)=0$.
The sensitivity index of PRCC ranges from $-1$ to $+1$, indicating the strength of the correlation between the proportion of related cells and parameters. We denote the value of PRCC as the $\tilde{p}$. Then, when $0\leq\tilde{p}<0.2$, the correlation between the input parameters and the output variables is not significant; when $0.2\leq\tilde{p}<0.4$, the input parameters are moderately correlated with the output variables; when $0.4\leq\tilde{p}< 1$, the input parameters are highly correlated with the output variables. The results of the parameter sensitivity analysis are shown in Fig. \ref{fig5}.

In the tumor immune escape kinetic model, global sensitivity analysis is performed using PRCCs to investigate the nonlinear relationship between parameters and cell proportions, revealing the following core regulatory mechanisms.
The proportion of antigenically neutral cell ($N$) is significantly and positively affected by the proliferation rate of antigenically neutral cell ($r_N$), death rate of immunogenic cell  ($d_I$), immune system-mediated death rate ($m$), and death rate of immune-escaped cell ($d_E$). 
This means that these parameters promote the survival of antigenically neutral cell through direct or indirect pathways (such as reducing the density of competing cells). 
However, the conversion from antigenically neutral cell to antigenic cell, increased apoptosis of antigenically neutral cell, or resource competition caused by the proliferation of immunogenic cell can have negative effects.
The positive regulation of the antigenic cell ratio depends on $p$ and $d_I$, which are achieved through the conversion of $N$ to $A$ or the apoptosis of $I$  to release resources.
The proportion of immunogenic cell is positively driven by $p$, $d_N$, and $r_I$, but the mortality rate of immunogenic cells ($d_I$), the mortality rate mediated by the immune system ($m$), and the proliferation rate of antigenically neutral cells ($r_N$) have a negative impact.
Positive regulation of the proportion of immune-escaped cell ($E$) is dominated by the immune escape rate ($q$), mortality rate  of antigenically neutral cell ($d_N$), and proliferation rate of immune-escaped cell ($r_E$), which reflects the acceleration of $N/A/I\rightarrow E$ transformation or resource release, whereas the negative effects of the mortality rate of immune-escaped cell ($d_E$) and the proliferation rate of antigen-neutral cell  ($r_N$) inhibit their growth through direct clearance of $E$ or competition. 


\begin{figure}[H]
    \centering
\subfigure[]{\includegraphics[width=0.48\linewidth]{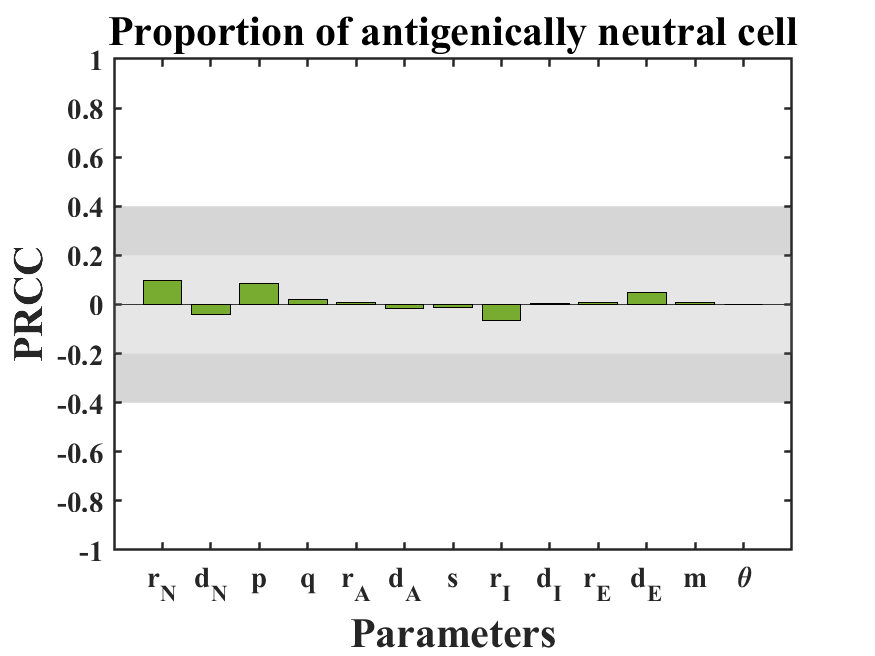}}
\subfigure[]{\includegraphics[width=0.48\linewidth]{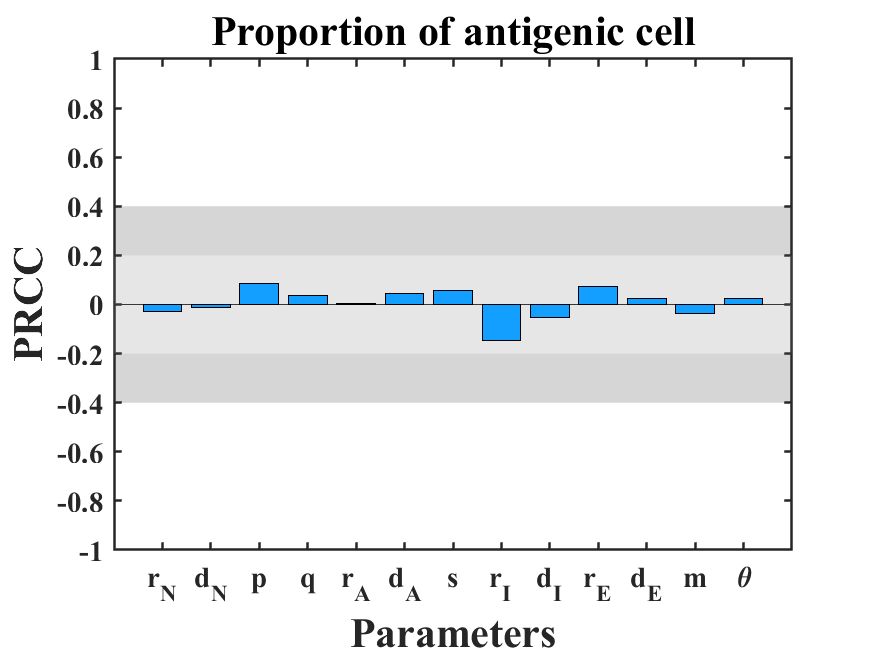}}\\
\subfigure[]{\includegraphics[width=0.48\linewidth]{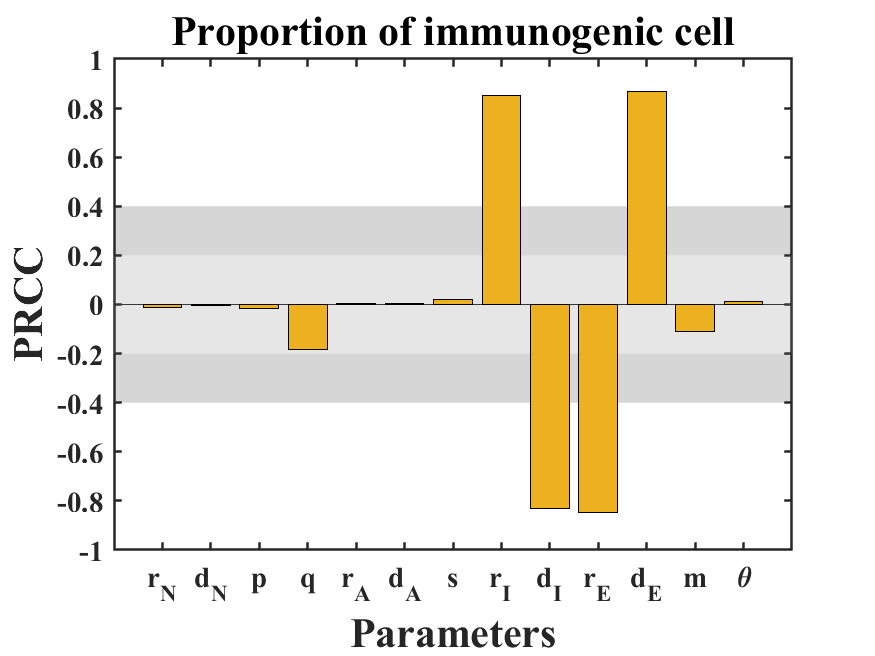}}
\subfigure[]{\includegraphics[width=0.48\linewidth]{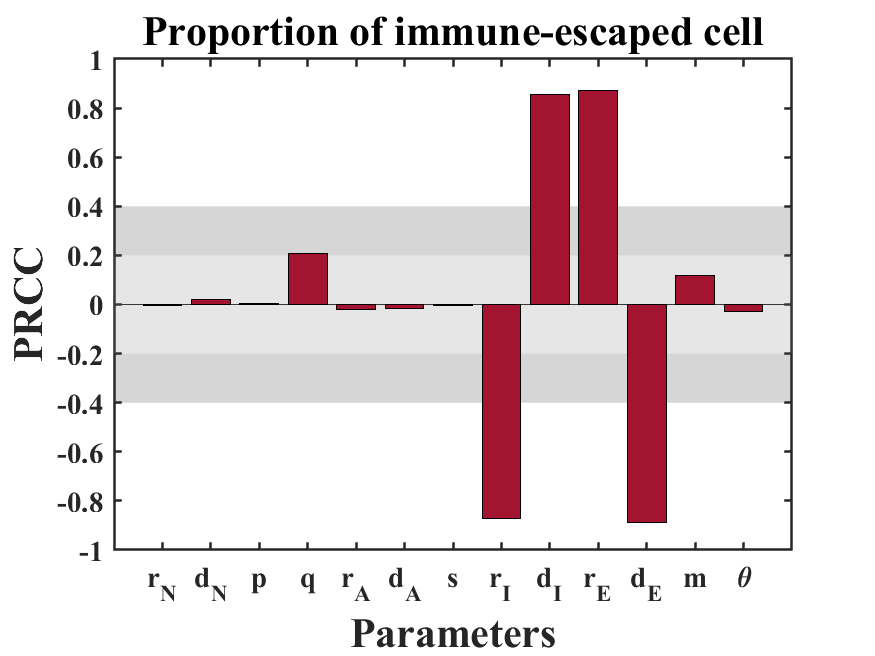}}\\	
    \caption{ Results of sensitivity analysis of four cell ratios. The light gray areas, dark gray areas, and white areas correspond to weak, moderate, and strong correlations, respectively.
    The baseline value of the parameter $p$ is taken as $0.15$ and the baseline values of the other parameters are given in Table \ref{tab1}.}
    \label{fig51}
\end{figure}

Within the bistable region of antigenic mutation rate $p=0.15$, the system dynamics are dominated by the dynamic equilibrium of immunogenic cell $I$ and immune-escaped cell $E$, whereas the proportions of antigenically neutral cell $N$ and antigenic cell $A$ are significantly less sensitive to system parameters. Sensitivity analyses show that the proportion of immunogenic cell is positively regulated by $r_I$ (the proliferation rate of immunogenic cell) and $d_E$ (the mortality rate of immune-escaped cell), whereas the proportion of immune-escaped cell is positively driven by $d_I$ (the  mortality rate of immunogenic cell) and $r_E$ (the proliferation rate of immune-escaped cell), see Fig. \ref{fig51} for more details.

In the region of bistability, the response of the system to external interventions is highly dependent on the initial conditions, and the parameters of self-growth and death ($r_I,d_I,r_E,d_E$) become the key in regulating homeostasis. This phenomenon suggests that the competition in the tumor microenvironment has shifted from antigenic diversity to a direct game between immunogenic cell and immune-escaped cell when $p$ exceeds a threshold. 
In addition, by comparing the analysis results of Figs. \ref{fig5} and \ref{fig51}, it can be seen that simply increasing the mortality rate mediated by the immune system ($m$) has a limited effect on breaking the bistable balance between immunogenic cell and immune-escaped cell. This indicates that the therapeutic strategy needs to break through the limitations of traditional immune enhancement methods, and shift to the combined regulation of the autonomous proliferation and death parameters of immunogenic cell and immune-escapes cell.

\section{Stable manifold computation}\label{App1}

First, we consider the following nonlinear system: 
\begin{equation}\label{A1}
    \dot{x} =f(x,p)
\end{equation}
 where $x\in \mathbb{R}^n$ are the state variables, $p\in\mathbb{R}^m$ are model parameters, and $f:\mathbb{R}^n  \times \mathbb{R}^m\rightarrow\mathbb{R}^n$ is assumed to be smooth. Let $x_e$ be a hyperbolic equilibrium of (\ref{A1}), its stable and unstable manifolds $W^s(x_e)$ and $W^u(x_e)$ are defined as follows:
 \begin{equation}
 \begin{aligned}
     W^s(x_e)=&\{x\in\mathbb{R}^n:\lim_{t\rightarrow +\infty}\Phi(t,x)=x_e\}\\
     W^u(x_e)=&\{x\in\mathbb{R}^n:\lim_{t\rightarrow -\infty}\Phi(t,x)=x_e\}
 \end{aligned}
 \end{equation}
 where $\Phi(t,x)$ is the trajectory of (\ref{A1}) through point $x$.

 If $x_s$ is a stable equilibrium, the stability region of $x_s$ is defined as $\tilde{A}(x_s)=W^s(x_s)\in\mathbb{R}^n$, and its boundary is denoted as $\partial \tilde{A}(x_s)\in \mathbb{R}^{n-1}$. It is proved in \cite{chiang1988stability} that under the following conditions:\\
 (1) All the equilibrium points on $\partial \tilde{A}(x_s)$ are hyperbolic;\\
 (2) The stable and unstable manifolds of equilibrium points on $\partial \tilde{A}(x_s)$ satisfy the transversality condition; \\
 (3) Every trajectory on $\partial \tilde{A}(x_s)$ converges to an equilibrium point,\\
 the stability boundary $\partial \tilde{A}(x_s)$ is the union of the stable manifold of the unstable equilibrium points on the boundary. That is, $\partial \tilde{A}(x_s)=\bigcup_i W^s(x_i)$, where $x_i, ~i=1,2,\cdots$ be the equilibrium points on the stability boundary $\partial \tilde{A}(x_s)$. However, $x_i\in \partial \tilde{A}(x_s)$ if and only if $W^u(x_i)\cap \tilde{A}(x_s) \ne  \emptyset $ and $W^s(x_i)\subseteq \partial \tilde{A}(x_s)$.
 Therefore, we need to calculate $W^s(x_i)$ to find $\partial \tilde{A}(x_s)$. 

For the immune escape model (\ref{model}), the parameter values are all taken from Table \ref{tab1}, which yields two stable equilibria $\Xi_1=(0,0,0,0.9975)$ and $\Xi_{21}=(0,0,0.87238,0.12$ $528)$. Moreover, we have a unsatble equilibrium $x_u=\Xi_{22}=(0,0,0.28945,0.70806)$ and its corresponding eigenvalues $\bar{\lambda}=(-0.41344,-0.14977,-0.14852,0.00007)$. Then we apply a linear transformation 
$$\begin{bmatrix} N  \\ A \\ I \\ E \end{bmatrix}=Q\begin{bmatrix} u_1 \\ u_2 \\ u_3  \\ v \end{bmatrix}+x_u$$
with
$$
Q=\begin{bmatrix} 
0 & 0 & 0.0058 & 0  \\ 
0 & 0.7070 & 0.7040 & 0\\
-0.4177 & -0.7073 & -0.7102 &-0.7072\\
-0.9086 & 0.0006 & 0.0007 & 0.7070
\end{bmatrix}.
$$
Model (\ref{model}) is transformed to
$$\dot{U}=J_gU+g(U),$$
where $U=(u_1,u_2,u_3,v)^{\mathrm{T}}$, $J_g=diag\{\bar{\lambda}\}$, $g=(g_1,g_2,g_3,g_4)^{\mathrm{T}}=O(||u||^2)$. 

Denote
$$P=Q^{-1}=\begin{pmatrix}
    -0.7501 & -0.7535 &-0.7538 & -0.7540\\
    -170.4381 & 1.4145 & 0 & 0\\
    171.1536 & 0 & 0 & 0\\
    -0.9785 & -0.9695 & -0.9687 & 0.4454
\end{pmatrix},$$
let $P_4$ be the $4$-th row of $P$, and $\tilde{s}(x)=\sum_{i=1}^{4}P_{4i}f_i(x)$, which $f_i$ is the $i$-th component of $f$. Then the Hessian matrix of $\tilde{s}(x)$ by $$A_2=\begin{pmatrix}
    1.9570 & 1.4632 & 1.4144 & 0.8003\\
    1.4632 & 0.9695 & 0.9207 & 0.3066\\
    1.4144 & 0.9207 & 0.8719 & 0.2578\\
    0.8003 & 0.3066 & 0.2578 & -0.3563
\end{pmatrix}.$$ Next compute the matrix  
$$
G_0=\frac{1}{2}Q^{\mathrm{T}}A_2Q
=\begin{pmatrix}
    0.0268 & -0.0227 & -0.0247 & 0.2880\\
    -0.0227 & 0.00001 & 0.00001 & -0.00008\\
    -0.0247 & 0.00001 & 0.00001 & -0.00008\\
    0.2880 & -0.00008 & -0.00008 & 0.00009
\end{pmatrix}.
$$
Let $G$ be the upper left sub-matrix of $G_0$. Compute $H_2$ from 
$$
\lambda_uH_2-(H_2+H_2^{\mathrm{T}})A_s=-G.
$$
where
$$
H_2=\begin{pmatrix}
    -0.03251 & 149.78311 & 164.03071\\
    -149.70237 & -0.00004 & -0.00075 \\
     -163.94271 & 0.00067 & -0.00004 
\end{pmatrix}.
$$
The local quadratic stable manifold approximation  is given by
$$
v=(u_1~u_2~u_3)H_2\begin{pmatrix}
    u_1\\
    u_2\\
    u_3
\end{pmatrix}.
$$
Substituting $U = P(x-x_u)$ back to the original variables, where $x=(N,A,I,E)^{\mathrm{T}}$, we have
\begin{equation*}
\begin{aligned}
h_2(N,A,I,E)=&-0.01845158767E^2 + (-0.4085921830 - 1.01785353N - 0.1229938\\
&539A- 0.03689208021I)E - 1.00404343N^2 + (1.993707422 - 1.1044 \\
&5836A- 1.01754749I)N - 0.1045504614A^2 + (1.092217394 - 0.1229\\
&568752I)A+ 1.005531645I - 0.01844049420I^2 + 0.01660999215. 
\end{aligned}
\end{equation*}
Hence $h_2(N,A,I,E)=0$ is the quadratic approximation of $W^s(x_u)$ for model (\ref{model}).

\bibliographystyle{siamplain}
\bibliography{references}